\documentclass[12pt]{article}

\usepackage{amssymb}
\usepackage{textcomp}
\usepackage{tikz}
\usepackage{etoolbox}
\usepackage{bm}
\usepackage{color}
\usepackage{MnSymbol}
\definecolor{orange}{rgb}{1.0,0.4,0.0}
\definecolor{blueviolet}{rgb}{0.3,0,0.7}
\pdfoutput=1
\newcommand{\fbcrcl}{
\begin{tikzpicture}
\filldraw[fill=black,draw=green] circle (3pt);
\end{tikzpicture}
}
\newcommand{\frcrcl}{
\begin{tikzpicture}
\filldraw[fill=red,draw=green] circle (3pt);
\end{tikzpicture}
}

\newrobustcmd*{\mytriangle}[1]{\tikz{\filldraw[draw=green,fill=#1] (0,0) --
(0.2cm,0) -- (0.1cm,0.2cm);}}

\newrobustcmd*{\mybarredtriangle}[1]{\tikz{\draw[draw=#1] (0,0) --
(0.2cm,0) -- (0.1cm,0.2cm) -- (0cm,0cm); \draw[draw=#1] (-0.1cm, 0.07cm) -- (0.3cm, 0.07cm)}}

\newrobustcmd*{\mybarredsquare}[1]{\tikz{\draw[draw=#1] (0,0)
rectangle (0.2cm,0.2cm); \draw[draw=#1] (-0.1cm, 0.1cm) -- (0.3cm, 0.1cm)}}

\newrobustcmd*{\mythickbarredsquare}[1]{\tikz{\draw[line width=0.4mm,draw=#1] (0,0)
rectangle (0.2cm,0.2cm); \draw[draw=#1] (-0.1cm, 0.1cm) -- (0.3cm, 0.1cm)}}

\newrobustcmd*{\mybarredcircle}[1]{\tikz{\draw[draw=#1] (0,0)
circle (0.1cm); \draw[draw=#1] (-0.2cm, 0.0cm) -- (0.2cm, 0.0cm)}}

\newrobustcmd*{\mythickbarredcircle}[1]{\tikz{\draw[line width=0.4mm,draw=#1] (0,0)
circle (0.1cm); \draw[draw=#1] (-0.2cm, 0.0cm) -- (0.2cm, 0.0cm)}}

\newrobustcmd*{\mydashedline}[1]{\tikz{\draw[draw=#1] (-0.2cm, 0.2cm) -- (-0.1cm, 0.2cm); \draw[draw=#1] (-0.0cm, 0.2cm) -- (0.1cm, 0.2cm)}}

\newrobustcmd*{\mybarredcross}[1]{\tikz{\draw[line width=0.4mm,draw=#1] (0,0) --
(0.2cm,0); \draw[line width=0.4mm,draw=#1] (0.1cm,-0.1cm) -- (0.1cm,0.1cm); \draw[draw=#1] (-0.1cm,0) -- (0.3cm,0);}}

\newrobustcmd*{\mythickline}[1]{\tikz{\draw[line width=0.4mm,draw=#1] (-0.15cm, 0.1cm) -- (0.15cm, 0.1cm);\draw[line width=0.4mm,draw=#1] (-0.0cm, 0.0cm);}}

\newrobustcmd*{\mythickdashedline}[1]{\tikz{\draw[line width=0.4mm,draw=#1] (-0.2, 0.1cm) -- (-0.1cm, 0.1cm); \draw[line width=0.4mm,draw=#1] (-0.0cm, 0.1cm) -- (0.1cm, 0.1cm); \draw[line width=0.4mm,draw=#1] (-0.0cm, 0.0cm);}}

\newrobustcmd*{\mycircle}[1]{\tikz{\draw[draw=#1] (0,0)
circle (0.1cm);}}

\begin{document}

%\begin{frontmatter}

\begin{center}
{\bf \Large A minimally-dissipative low-Mach number solver for complex reacting flows in OpenFOAM \vspace{0.2in}\\}
%\tnotetext[label0]{This is only an example}
{\large Malik Hassanaly, Heeseok Koo, Christopher Lietz, Shao Teng Chong, and Venkat Raman \vspace{0.1in} \\}
{\large Department of Aerospace Engineering, University of Michigan \\}
\end{center}

%
%\author[label1]{Malik Hassanaly\corref{cor1}}
%\address[label1]{Department of Aerospace Engineering, University of Michigan, Ann Arbor, MI 48109, United States}
%\ead{malik.hassanaly@gmail.com}
%
%\author[label1]{Heeseok Koo}
%%\address[label2]{Sandia National Laboratory}
%\ead{heeseokkoo@gmail.com}
%
%\author[label3]{Christopher F. Lietz}
%\address[label3]{SLI, Inc., Air Force Research Lab, Edwards AFB, CA 93524, United States}
%\ead{cflietz@gmail.com}
%
%\author[label1]{Shao Teng Chong}
%\ead{chongshaoteng@gmail.com}
%
%\author[label1]{Venkat Raman}
%\ead{ramanvr@umich.edu}

\begin{abstract}

Large eddy simulation (LES) has become the de-facto computational tool for modeling complex reacting flows, especially in gas turbine applications. However, readily usable general-purpose LES codes for complex geometries are typically academic or proprietary/commercial in nature. The objective of this work is to develop and disseminate an open source LES tool for low-Mach number turbulent combustion using the OpenFOAM framework. In particular, a collocated-mesh approach suited for unstructured grid formulation is provided. Unlike other fluid dynamics models, LES accuracy is intricately linked to so-called primary and secondary conservation properties of the numerical discretization schemes. This implies that although the solver only evolves equations for mass, momentum, and energy, the implied discrete equation for kinetic energy (square of velocity) should be minimally-dissipative. Here, a specific spatial and temporal discretization is imposed such that this kinetic energy dissipation is minimized. The method is demonstrated using manufactured solutions approach on regular and skewed meshes, a canonical flow problem, and a turbulent sooting flame in a complex domain relevant to gas turbines applications. %Finally, a discussion on code scalability is provided. {\color{red} Note this last part - can you guys add something on how this scales on cores - both weak and strong scaling.}
\end{abstract}

%\begin{keyword}
%Low-Mach number solver \sep Kinetic energy conservation \sep OpenFOAM
%\end{keyword}

%\end{frontmatter}

% \linenumbers

Large eddy simulation (LES) has become the de-facto computational tool for modeling complex reacting flows, especially in gas turbine applications. However, readily usable general-purpose LES codes for complex geometries are typically academic or proprietary/commercial in nature. The objective of this work is to develop and disseminate an open source LES tool for low-Mach number turbulent combustion using the OpenFOAM framework. In particular, a collocated-mesh approach suited for unstructured grid formulation is provided. Unlike other fluid dynamics models, LES accuracy is intricately linked to so-called primary and secondary conservation properties of the numerical discretization schemes. This implies that although the solver only evolves equations for mass, momentum, and energy, the implied discrete equation for kinetic energy (square of velocity) should be minimally-dissipative. Here, a specific spatial and temporal discretization is imposed such that this kinetic energy dissipation is minimized. The method is demonstrated using manufactured solutions approach on regular and skewed meshes, a canonical flow problem, and a turbulent sooting flame in a complex domain relevant to gas turbines applications.

\section{Introduction}

The use of large eddy simulation in complex turbulent flows has increased substantially in the last decade. This advance is being driven by the rapid growth in computational power, as well as advances in numerical algorithms for such complex flows. This has led to the development of a number of LES solvers that are routinely deployed in industrial applications \cite{charles,cdp,pyfr,openfoam,fluent}. Nevertheless, many of these solvers remain proprietary and not open to the research community. This deficiency is particularly challenging for LES due to the sensitivity of LES results to numerical accuracy. For instance, when head-to-head comparison of LES models are made, such studies rely on different LES solvers and numerical methods. Consequently, the conclusions are highly sensitive to these numerical details, and may not even be relevant to model comparison \cite{kaul_ctm}. Furthermore, many combustion models and subfilter closures are developed for canonical flow problems and are not exercised in full-scale geometries or application-relevant flow problems. This sparsity in full-scale validation is at least partially due to the lack of an easy approach to porting models to complex flow problems. As LES models mature, there is a clear need for a robust open source platform to demonstrate their performance for practical configurations. In this context, OpenFOAM \cite{openfoam} is a prime candidate for such a framework. Based on a field operation approach \cite{jasakthesis}, OpenFOAM provides a convenient code-base for numerically solving partial differential equations. Moreover, OpenFOAM has developed a broad community of developers and users, who have added valuable tools and methods to the base solver \cite{OFworkshop}. This robust development ecosystem has been leveraged extensively in the turbulent flow and turbulent combustion research and application communities \cite{firefoam,fureby,diesel-openfoam,edm-openfoam}. The current work intends to upgrade the implementation of LES algorithms that are indispensable in solving variable density reacting flows of interest to gas turbine applications. With the will of gradually moving combustion research applications to OpenFOAM, key issues regarding LES numerical accuracy are addressed.

Many gas turbine applications, especially related to the combustor section, operate in the low-Mach number regime, characterized by velocities smaller than 0.3 Ma, where the Mach number is defined based on local fluid properties. In this regime, the acoustic component of the Navier-Stokes equations is decoupled from the basic flow physics. Consequently, by reformulating the governing equations for this low-Ma regime, it is possible to go beyond the CFL restriction imposed by the acoustic wave speed (i.e., local sound speed). In many applications, this can amount to an order of magnitude increase in timestep used. Although OpenFOAM has been used in combustion applications, many of these are related to the compressible flow regime (see, for instance \cite{fureby}). In the base distribution of OpenFOAM, the variable density solver the closest to a low-Mach number solver uses an all-Mach approach, where some level of compressibility-related coupling of the governing equations is retained even in the limit of zero Mach number \cite{ferziger}. Hence, developing a robust low-Ma solver for variable density flows will be of interest to combustion applications.

Another issue of importance to LES is the coupling between numerical discretization and modeling errors, as discussed in \cite{katopodes,bose,kravmoin-leserrors,ghosal-jcp,kaul-pof1,kaul-pof2}. Briefly, there are two types of numerical errors. First, the spatial discretization of derivative operators for fields containing high wave number components can be highly erroneous \cite{moin-book}. In practical LES, features comparable to the filter size can thus be contaminated by discretization errors \cite{ghosal,katopodes,kaul-pof1}. In turbulent combustion applications, where small-scale models are critical for capturing the mixing and reactions processes, this discretization error represents a major concern. The most comprehensive solution is to use an explicit filtering technique, where the small-scales are removed through a filter during the simulation in order to prevent deposition of energy at such scales \cite{vasilyev,katopodes,bose,youExpl}. However, extending these techniques to variable density flows has not been achieved yet \cite{heye-tfsp}. Several approaches \cite{kaul-pof2,seqmom} have been proposed to mitigate these numerical effects. The second issue concerns the so-called conservation of secondary quantities. Finite-volume approaches discretely conserve primary quantities such as mass and momentum. With LES, the accurate representation of the turbulent energy spectrum is important the validity of the modeling assumptions. Therefore, minimizing numerical dissipation of kinetic energy is a key point of the LES solvers. In low-Mach number incompressible flows, and in the absence of viscous dissipation, kinetic energy should be exactly conserved. In this case, the spatial and temporal discretization should ensure that such conservation or, at the least, a minimization of dissipation is achieved \cite{ferziger}. In this regard, there has been considerable progress in the design of energy-conserving numerical schemes \cite{morinishi-skew,felten-jcp,morinishi,hamiac-kecons,mahesh-jcp}. Again, a comprehensive implementation of such tools in the OpenFOAM framework is not available.

With this introduction, the objectives of this work are as follows: 1) Analyze the variable density solvers in OpenFOAM, and implement a consistent low-Ma solver that preserves spatial and temporal accuracy in the limit of zero Mach number, and 2) evaluate the energy conserving properties of existing OpenFOAM solvers, and implement a minimally-dissipative approach. In Sec.~\ref{sec:lowMachOFintro}, a variable density low-Mach number solver is designed and its implementation in OpenFOAM is described. In particular, it is stressed that the baseline implementation of variable density flow solver in OpenFOAM was previously only allowed for compressible flow cases. The solver is then tested using the method of manufactured solutions. Sec.~\ref{sec:dissipationOFintro} provides the necessary theoretical background for the design of energy conservative solvers. In Sec.~\ref{sec:KeImpl}, the implementation of a minimally-dissipative solver is described for the OpenFOAM framework. Finally in Sec.~\ref{sec:lowdissVerif}, a set of verification and validation cases are used to demonstrate the capabilities of the new solver. Sec.~\ref{sec:dissipationOFintro}, energy conservation properties of OpenFOAM discretization operators are studied, followed by the implementation of a fully-conservative scheme.

\section{Low-Mach number solvers in OpenFOAM}
\label{sec:lowMachOFintro}

The governing equations of fluid flow of interest here are written as
\begin{equation}
\label{eq:contNSmass}
\frac{\partial \rho}{\partial t} + \nabla \cdot (\rho \boldsymbol{u} ) = 0,
\end{equation}
\begin{equation}
\label{eq:contNSmom}
\frac{\partial \rho {\boldsymbol{u}}}{\partial t} + \nabla \cdot (\rho {\boldsymbol{u}} {\boldsymbol{u}}) = - \nabla p + \nabla \cdot \boldsymbol{\overline{\sigma}},
\end{equation}

where $\rho$ is the flow density, $\boldsymbol{u}$ is the local gas phase velocity vector, $p$ is the mechanical pressure and $\overline{\boldsymbol{\sigma}}$ is the viscous stress tensor.

 When considering chemical reactions, additional equations that describe transport of chemical species and a formulation that couples any heat addition to the density term need to be included. A generic scalar transport equation that may be used for this purpose is given by:
  
\begin{equation}
    \label{eq:scalarcont}
    \frac{\partial \rho \phi}{\partial t} + \nabla \cdot (\rho \phi \boldsymbol{u}) = \nabla \cdot (D \nabla \phi) + \dot{\omega},
\end{equation} 

where $\phi$ denotes the transported scalar and $D$ denotes mass diffusivity and $\dot{\omega}$ is some volumetric source term. 
 
 To solve this system of equations, a numerical approach is used. Three different types of flow solvers are defined here for the sake of future discussion. An \textit{incompressible solver} is defined as a solver that does not take into account any density change. In other words, the density field is treated as a constant value. A \textit{compressible solver} is defined as a solver that takes into account the dynamic coupling between the pressure and density fields. In other words, this solver allows mechanical energy (kinetic energy noted KE) to be converted to thermal energy (through pressure). A \textit{low-Mach number solver} is defined as a solver that does not couple density changes with instantaneous pressure changes. In particular, the pressure field is split into a thermodynamic pressure and a mechanical pressure, with the former held constant while the latter is allowed to vary through velocity changes in the flow. The thermodynamic pressure is used in the equation of state, while the mechanical pressure appears in the momentum transport equation (Eq.~\ref{eq:contNSmom}). Density variations in low-Mach number solvers occur through heat addition or removal, for instance due to chemical reactions. Such low-Ma number flows and associated solvers are the focus of this work.

\subsection{Hybrid solvers in OpenFOAM for variable density flows}
\label{sec:hybridpress}

The base distribution of OpenFOAM \cite{openfoam} contains a suite of solvers targeting reacting flows that involve density changes. Although they nominally fall under the compressible solver definition provided above, some aspects of the low-Mach representation is also included. The main choice is as follows: all variable density solvers use the equation of state to reduce the number of partial differential equations needed (for the thermochemical part) to two of the three variables (pressure, temperature, density). Assuming that an energy equation that provides temperature is available, the choice is then to solve for density and obtain pressure from equation of state or vice-versa. These two types of compressible solvers are: the \textit{density-based solvers} where the density is transported using the continuity equation, and the \textit{pressure-based solvers} where the thermodynamical pressure is computed using a pressure correction method. In the pressure-based solvers handling variable density available in OpenFOAM, a generic pressure correction procedure \cite[Ch.\ 10.2]{ferziger} is used. This approach is easily applicable even with unstructured and complex grids. The pressure correction equation contains an incompressible divergence term (correcting the mass fluxes) and a compressible convective term (correcting the density). Each one of these term becomes dominant when the flow is largely incompressible or compressible, respectively. This pressure-correction strategy is referred to as a \textit{hybrid} approach in the remainder of the paper. This type of pressure correction has been used to simulate combustion cases at high and low-Mach numbers \cite{furebysymp, firefoam}.

Since the pressure correction has an impact on the density field, this procedure involving momentum, scalar transport and pressure correction can be repeated in order to fully couple all transported variables. These iterations are called \textit{outer-iterations} as opposed to \textit{iterations} which denote the time advancement of the variables. Within each outer-iteration (denoted by $(.)_m$), the momentum field is first advanced from the old timestep (denoted by $(.)^n$ below) to an intermediate \textit{fractional time} (denoted by $(.)^*$ below) without taking into account the pressure gradient.

\begin{equation}
    \label{eq:fractionaltime}
    \frac{\rho \boldsymbol{u}^* -  \rho \boldsymbol{u}^n}{\Delta t} + \boldsymbol{\mathcal{C}} = \boldsymbol{\mathcal{D}},
\end{equation}

where $\boldsymbol{\mathcal{C}}$ denotes the convective term, $\boldsymbol{\mathcal{D}}$ the diffusion term and $\Delta t$ is the flow timestep.

The time at which $\boldsymbol{\mathcal{C}}$ and $\boldsymbol{\mathcal{D}}$ is evaluated in Eq.~\ref{eq:fractionaltime} is not specified since the following discussion is about completing the time derivative term with a pressure correction. The field $\boldsymbol{u}^*$ does not necessarily respect the integral continuity equation. It is therefore corrected with a pressure gradient which is tailored to enforce mass conservation in the domain. Because this correction is done through pressure, the thermodynamical state also depends on this correction procedure. This pressure correction equation is therefore solved to determine the correction to apply to the mass fluxes as obtained from the intermediate field $\boldsymbol{u}^*$.

In pressure-based compressible solvers, the pressure correction equation is formulated by replacing density changes by pressure changes under the assumption that temperature is held constant:
\begin{equation}
  \label{eq:compressFact}
  \rho' \approx \Big(\frac{\partial \rho}{\partial p}\Big)_{T} p',
\end{equation}
where $\rho'$ denotes a density change and $p'$ denotes a pressure change. In the notations adopted in OpenFOAM 4.1, the multiplicative factor on the right-hand-side (RHS) of Eq.~\ref{eq:compressFact} is called \verb|psi|. The derivation of this method is covered in detail in \cite[Ch.\ 10.2]{ferziger}. Similar to all predictor corrector methods, the pressure field is used as a corrector which adjusts the mass fluxes in order to ensure conservation of mass at the new timestep (denoted by $(.)^{n+1}$). Unlike a low-Mach implementation where the pressure gradient only corrects the flow velocity, the pressure correction also acts on the density in pressure-based compressible solvers. The goal of the procedure is to find the mass flux correction $\boldsymbol{\phi'_f}$ such that

\begin{equation}
\label{eq:continuityPressureBased}
V_p \frac{\rho^{n+1}-\rho^n}{\Delta t} + \sum_f  \boldsymbol{S_f} \cdot (\rho  \boldsymbol{u}^*_f +  \boldsymbol{\phi'_f})  = 0, 
\end{equation}
leading to
\begin{equation}
\label{eq:pcorrcomp}
     \boldsymbol{\phi'_f} = (\rho^{n+1}  \boldsymbol{u'})_f +  ( \boldsymbol{u}^*_{m} \rho')_f ,
\end{equation}
where $V_p$ is the cell volume, $(.)_f$ denotes a variable defined at a cell face, $\boldsymbol{S_f}$ denotes the cell face normal vector which norm is the cell face surface area. The first term on the RHS of Eq.~\ref{eq:pcorrcomp} comes from the additional pressure gradient due to the computed pressure correction. The second term of the RHS comes from the correction of density which goes to zero in a low-Mach number case \cite[Ch.\ 10.2]{ferziger}. Whenever the hybrid solver is compared to a low-Mach number solver, only the first RHS term is therefore included. Injecting Eq.~\ref{eq:pcorrcomp} into Eq.~\ref{eq:continuityPressureBased}, and using the compressibility factor to transform the derivation of density into pressure, the pressure equation can be formulated for the hybrid solver. Only the time derivative term is discretized in Eq.~\ref{eq:hybridpress} since it is where the numerical differences with a strict low-Mach number solver become apparent.

\begin{equation}
 \label{eq:hybridpress}
 \nabla^2 p_m = \frac{1}{\Delta t}(\frac{\frac{\rho^{n+1}}{p_{m-1}}p_m - \frac{\rho^{n}}{p^n}p^n}{\Delta t} + \nabla \cdot (\rho \boldsymbol{u}^*)).
\end{equation}

The above equation requires a solution to the elliptic Poisson equation system at each outer-iteration step, which is typically computationally expensive. Hence, the cost of the simulation is tied to the number of outer-iterations used in each timestep.

\subsection{Strict low-Mach number solvers for variable density flows}
\label{sec:procedure}

The hybrid approach used in the pressure-based solver of OpenFOAM offers flexibility and allows to deal with flows spanning a large range of Mach numbers. This method is based on the assumption that the compressible and the incompressible terms of the pressure correction equation will be effectively negligible in the right context. However, even in the limit of a low-Mach number flow, some compressibility is retained, and the solver cannot be considered as a strictly acoustically decoupled formulation.

In the context of direct and large eddy simulations, several low-Mach formulations have been developed in the past \cite{pierce_thesis,desjardins-jcp,kimmoin-projection,ham_ctr2006}, which are generally semi-implicit in the temporal direction. The objective of this work is to develop a similar approach for implementation in OpenFOAM. In these methods, the solver is segregated in that the different partial differential equations are solved sequentially within each outer-iteration, and multiple outer-iterations are used to couple the scalar and the velocity fields \cite{pierce_thesis}. Within each outer-iteration, the scalar fields are first advanced. Second, the impact of the scalar fields on the flow variables (density or viscosity for example) are then computed and these variables are updated. Third, the fractional timestep method is used to first advance the momentum equation without the pressure gradient. For an implicit Euler method, the momentum equation discretized for time only takes the form shown in Eq.~\ref{eq:fractionaltime}.

Similar to the pressure based solver presented in Sec.~\ref{sec:hybridpress}, the momentum equation is updated using the pressure gradient obtained from a pressure-correction equation. As opposed to the hybrid solver, which links the pressure to the thermodynamical variables in the pressure equation, the sole role of pressure in a low-Mach number solver is to enforce mass conservation. Hence, this pressure variable is referred to as the mechanical pressure, which is different from the thermodynamical pressure that determined the thermophysical quantities such as density and temperature. For a low-Mach number solver, the pressure equation is derived from the continuity equation and takes the form

\begin{equation}
    \label{eq:lowmachPress}
    \nabla^2 p = \frac{1}{\Delta t} (\frac{ \rho_m  - \rho^{n-1} }{\Delta t} + \nabla \cdot (\rho \boldsymbol{u}^*)).
\end{equation}

Finally, the velocity field is corrected using the newly computed pressure gradient. At the start of the first outer-iteration, the most updated fields available still correspond to the previous timestep. Any term that is required to be defined at the next timestep is estimated using an explicit extrapolation in time. Typically, the mass flux at the face needs to be obtained in this way for the convective terms.

\begin{equation}
    \rho_f \boldsymbol{u_f}^{n+1} \cdot \boldsymbol{S_f} = 2 \rho_f \boldsymbol{u_f}^{n} \cdot \boldsymbol{S_f} - \rho_f \boldsymbol{u_f}^{n-1} \cdot \boldsymbol{S_f}.
\end{equation}

A similar requirement holds for density in variable density cases in order to accurately compute the time derivative terms.

\subsection{Implementation of a low-Mach number solver for combustion applications}
\label{sec:lowMachImpl}

In this section, the details pertaining to the combustion model that is necessary for implementing the low-Mach number solver in OpenFOAM are discussed. 

\subsubsection{Flow-combustion coupling through density field}

Here, it is important to consider the nature of combustion models for low-Mach number flows. While different combustion models are available \cite{ramanpitsch-bluff1, pitsch-arfm,kronenburg-cmc,pierce,menon-lem}, the coupling procedure between reacting scalars and the density/momentum equation set is approximately similar. 
In general, a set of scalars is used to obtain the gas-phase density, viscosity and individual species diffusivity. When the scalars are species mass fractions, the local thermodynamical properties of the mixture can be directly extracted. When the scalars are not species mass fractions, a mapping between these scalars and the thermodynamical properties is first established and used throughout the domain. In some implementations \cite{wang-cf, ramanpitsch-sandia1}, an energy equation is solved, for which the source term due to chemical reactions is provided based on the scalar fields. The density field is then obtained from the energy field, thermodynamic pressure, and scalar fields through a non-linear inversion process. 

For variable density unsteady flows (called \verb|compressible| in OpenFOAM 4.1), the existing turbulence models in OpenFOAM take as an input a correlation between density and pressure. In order to decouple the variations in thermophysical quantities (density, diffusivity, viscosity) from pressure fluctuations, it is necessary to formulate a turbulence model which does not take the thermodynamical model as an input. Instead it takes the thermophysical variables obtained from the combustion model. For this purpose, a new turbulence model \verb|compSansThermo| has been developed. The full source code can be obtained as a supplementary material of this paper.

\subsubsection{Pressure solver}

%\subsubsection{Equation}
\label{sec:presseq_lowmach}
As mentioned in Sec.~\ref{sec:hybridpress}, the pressure correction method used in pressure-based solvers is written to generically handle compressible or incompressible flows. Although similar to Eq.~\ref{eq:lowmachPress}, the hybrid pressure equation given by Eq.~\ref{sec:hybridpress} will always result in a different correction for the mass fluxes from the correction that would be provided by low-Mach number solvers. In the hybrid formulation available in the base OpenFOAM distribution, the pressure equation reduces to the low-Mach number pressure equation if and only if the pressure does not change between two subsequent outer-iterations. In practice this is never exactly satisfied, and this results in numerical compressibility effects, even for zero Mach number flows (Sec.~\ref{sec:veriflowmach}).

%\subsubsection{Boundary Conditions}

The pressure solver ensures discrete mass conservation by re-arranging velocities throughout the domain. A pre-requisite for this solver to function is that the initial discretely non-conserving field satisfies global mass conservation. In other words, the integral of the continuity equation over the computational domain should be satisfied. Here, this is achieved by first altering the outflow surface fluxes to match the sum of density changes in each control volume as well as the fluxes across the boundary surfaces. The global continuity equation can then be written as
\begin{equation}
\label{eq:adjust}
  \sum_{f,in} \rho_f \boldsymbol{u_f} \cdot \boldsymbol{S_f} - \sum_{f,out} \rho_f \boldsymbol{u_f} \cdot \boldsymbol{S_f} + \sum_{cell} \frac{\delta \rho}{\delta t} \delta V_{cell} = 0,
\end{equation}

where $\frac{\delta}{\delta t}$ denotes some time discretization that is not of interest for now.

In order to use the low-Mach number solver in OpenFOAM, the boundary conditions of the pressure fields need to be adequately chosen. Neumann boundary conditions should be applied where velocity must not be corrected, typically at the inlet and at the walls. At the outlet of the domain, the user has the choice between Neumann and Dirichlet boundary conditions. With a Neumann boundary conditions, the outlet velocity adjusted to ensure global mass conservation through the function \verb|adjustPhi| in OpenFOAM 4.1 is preserved. In the original function \verb|adjustPhi| of OpenFOAM 4.1, the last term of the left-hand side of Eq.~\ref{eq:adjust} is missing which is expected since incompressible solvers were not designed to handle variations of density. With a Dirichlet boundary condition, the outlet velocity is not preserved anymore and the global mass conservation is only enforced up to the addition of the mass conservation errors in each cell of the domain (this is the telescopic property of mass conservation). Although the Dirichlet boundary conditions can lead to non-exact global mass conservation, it was observed to speed up the calculations, especially in open geometries.

\subsection{Low-Mach number solver verification and comparison with baseline hybrid solver}
\label{sec:veriflowmach}

\begin{figure}
\center
% left bot right top
\includegraphics[width=0.9\textwidth,trim={0cm 0cm 0cm 0cm},clip]{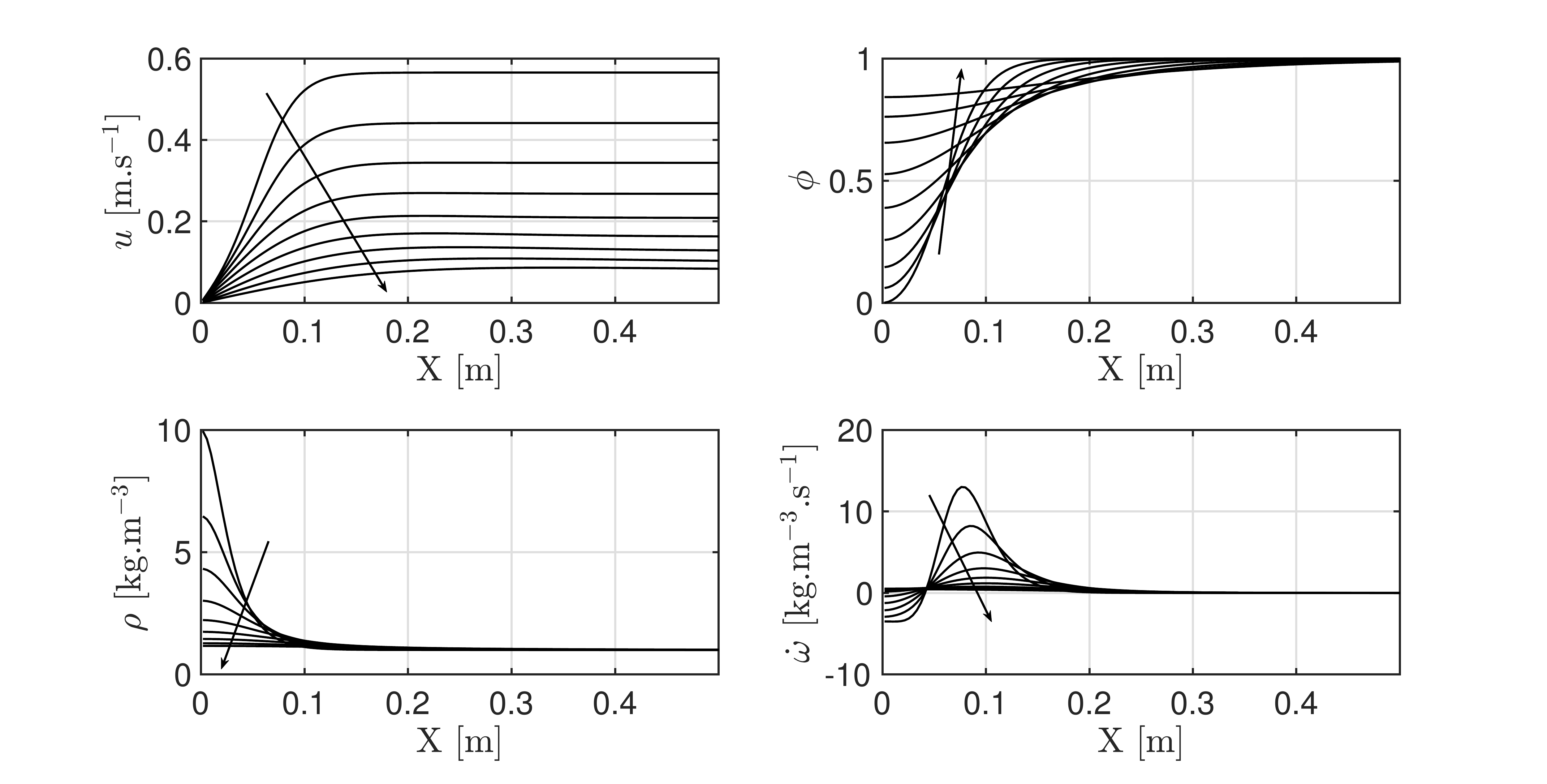}
\caption{Analytical solution of the 1D MMS case. The time evolution is indicated by the arrow. Top left: velocity. Top right: scalar field. Bottom left: density. Bottom right: scalar source term.}
\label{fig:Anal1d}
\end{figure}

The method of manufactured solutions (MMS) is used to verify the solvers and analyze their performance. The one-dimensional variable density flow problem of Shunn et al.\,\cite{shunn-MMS} is studied. Since many of the changes implemented for the low-Mach number solver are done for the mass fluxes and the pressure correction, this test case clearly highlights the difference with the hybrid procedure. The 1D case is computed with the following parameters (see \cite{shunn-MMS} for details of the flow): $\rho_0 = 10; \rho_1 = 1; \mu = 0.03; k_1=4; k_2=2; w_0=50$. The density field is obtained analytically from the value of the transported scalar field $\phi$. Note that no equation of state link the pressure and the density. The analytical solution is shown in Fig.~\ref{fig:Anal1d}.

%The grid sizes are varied from $64$ to $512$ cells for a 2m domain. The timestep is held constant to $6.25\mu s$.% in order to reduce time discretization error and assess only the spatial convergence. In Fig.~\ref{fig:Linferr}, the L$_{2}$-error of the scalar field $\phi$, which is analytically related to the density, is plotted. It can be seen that a second order convergence in space is achieved for both the scalar and the velocity field.

%\begin{figure}
%\center
% left bot right top
%\includegraphics[width=0.45\textwidth,trim={0cm 0cm 0cm 0cm},clip]{./Figures/UerrConv-eps-converted-to.pdf}
%\includegraphics[width=0.45\textwidth,trim={0cm 0cm 0cm 0cm},clip]{./Figures/phiErrConv-eps-converted-to.pdf}
%\includegraphics[width=0.7\textwidth,trim={0cm 0cm 0cm 0cm},clip]{./Figures/1D_spaceConv-eps-converted-to.pdf}
%\caption{Analytical solution of the 1D MMS case. The time evolution is indicated by the arrow. Top left: velocity. Top right: scalar field. Bottom left: density. Bottom right: scalalr source term.}
%\label{fig:Conv1d}
%\end{figure}

\begin{figure}
\center
% left bot right top
\includegraphics[width=0.45\textwidth,trim={0cm 0cm 0cm 0cm},clip]{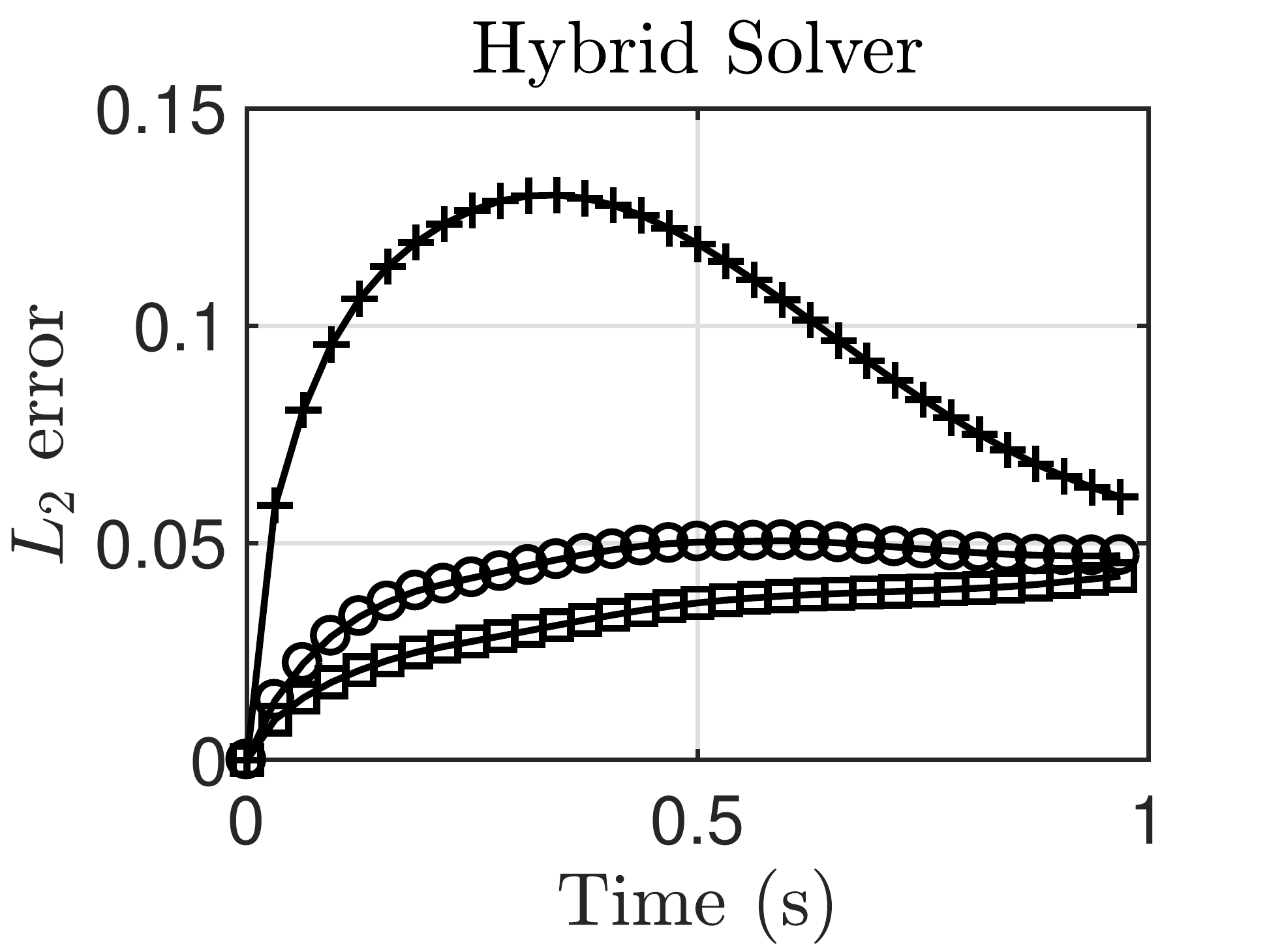}
\includegraphics[width=0.45\textwidth,trim={0cm 0cm 0cm 0cm},clip]{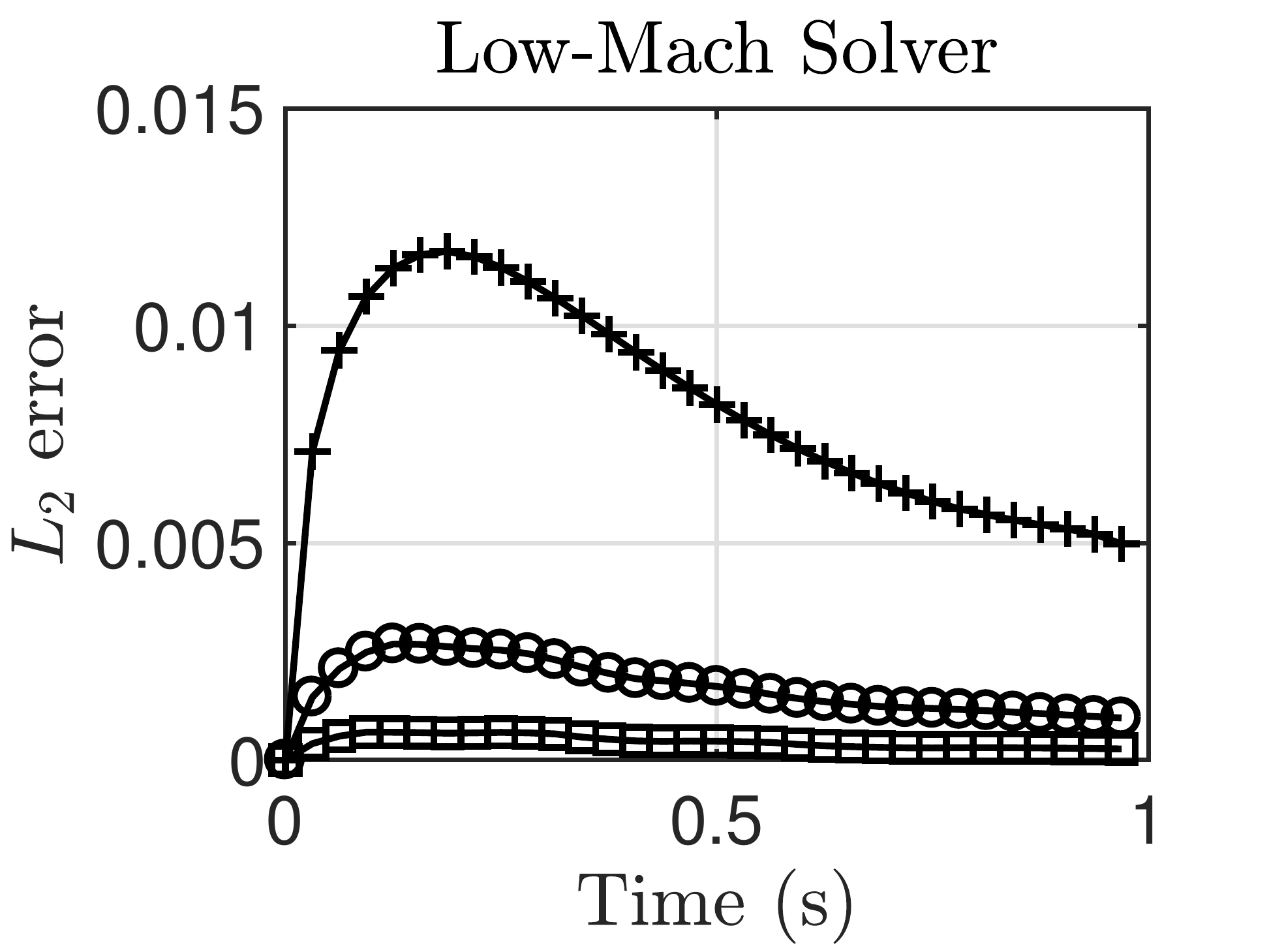}
\caption{L$_2$-error between the analytical transported scalar field $\phi$ and the numerical solution obtained using the hybrid solver (left) and the low-Mach solver (right). Cases were run with 64 cells (\mybarredcross{black}), 128 cells (\mythickbarredcircle{black}) and 256 cells (\mythickbarredsquare{black}).}
\label{fig:Linferr}
\end{figure}

The grid sizes are varied from $64$ to $512$ cells for a 2 m domain. The timestep is held constant to $6.25$ $\mu$s. These cases are used to compare how the original mixed compressible and incompressible approach (referred to as hybrid in the following) performs in OpenFOAM compares to the strict low-Mach approach. The results shown in Fig.~\ref{fig:Linferr} as the L$_{2}$-error to the analytical solution integrated over the domain and plotted against time.

%{\color{red} Where is the arrow?}

% Alex: Did you define L infinity before this? Is it the max of residual?
%Malik : No need, the definition of L infinity is standard

The hybrid pressure correction consistently leads to an increased L$_2$-error for the considered test case, often exceeding by at least one order magnitude. It is likely that the error comes from the approximation of the time derivative of density by the left hand side (LHS) of Eq.~\ref{eq:hybridpress}.
%The shape of the L$_\infty$-error in time also shows oscillation of variables which are absent in the strict low-Mach number case. This is due to artificial pressure wave convected in the hybrid formulation.

\begin{figure}
\center
% left bot right top
\includegraphics[width=0.55\textwidth,trim={1.5cm 0cm 4cm 0cm},clip]{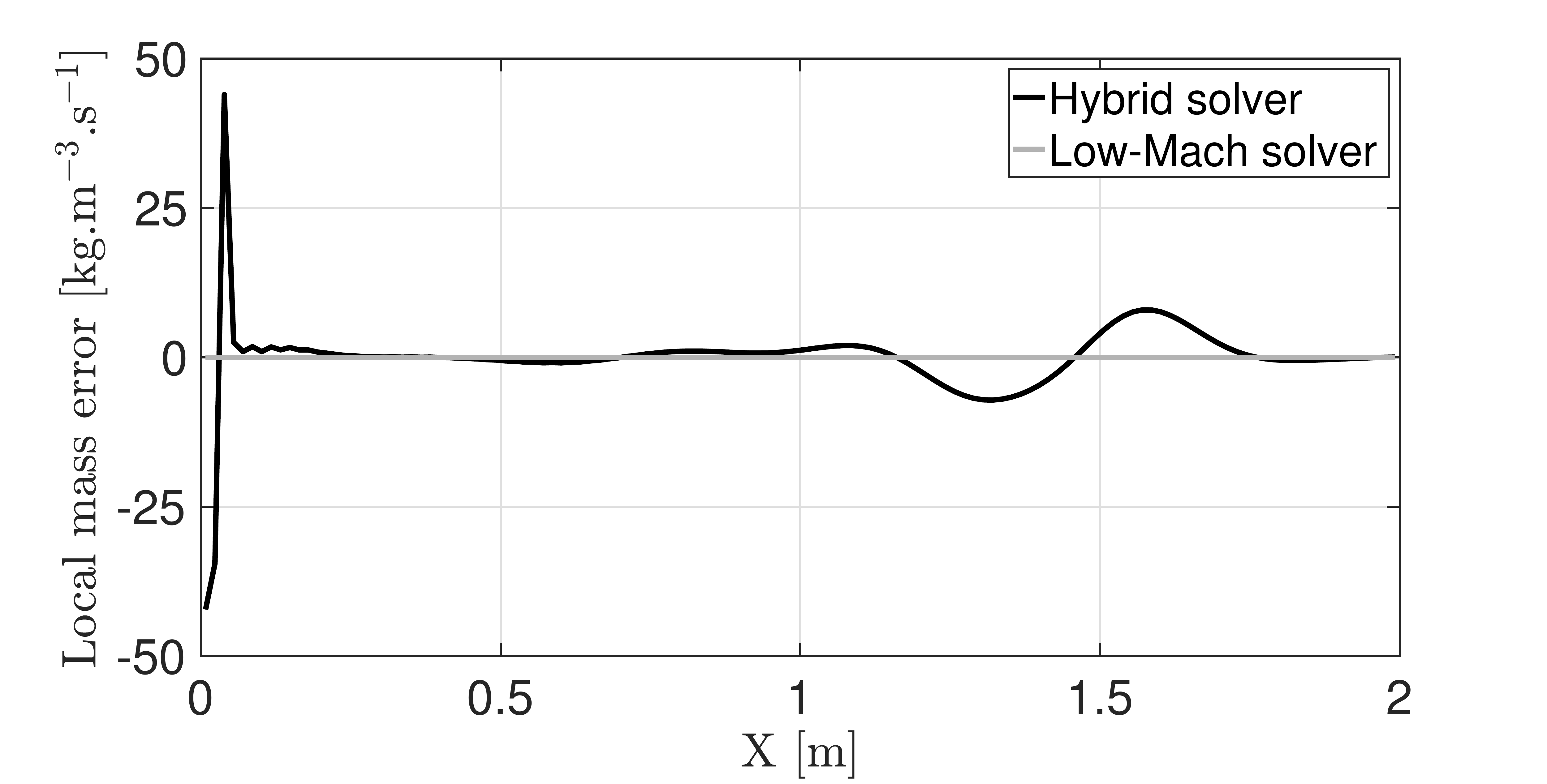}
\caption{Local mass conservation error calculated using Eq.~\ref{eq:masserr} for the hybrid solver and low-Mach number solvers. The number of grid cells is set to 128 for both solvers.}
\label{fig:masserr1d}
\end{figure}

Finally, the impact of the hybrid formulation on mass conservation is investigated. In Fig.~\ref{fig:masserr1d}, the mass balance at every cell is plotted for the hybrid and the low-Mach number solver. The mass balance is computed using mass fluxes obtained from the velocity interpolations at each face (Eq.~\ref{eq:masserr}) as follows:

\begin{equation}
\label{eq:masserr}
  mass_{err}(x,t) = V_p \frac{\rho^{n+1}-\rho^n}{\Delta t} + \sum_f \rho_f \boldsymbol{u_f}(x,t) \cdot \boldsymbol{S_f}. 
\end{equation}

While the low-Mach number solver ensures mass conservation up to machine precision, the hybrid solver generates an error proportional to $\frac{\rho^{n+1}-\rho^n}{\Delta t} - \frac{\frac{p_m}{p_{m-1}}\rho^{n+1}-\rho^n}{\Delta t}$. This difference numerically comes from the difference between the pressure at two successive outer-iterations. This is particularly prominent at locations where the density varies the most, which in this problem is near the left boundary of the 1D domain.

\section{Non-dissipative solvers}
\label{sec:dissipationOFintro}

\subsection{Motivation for reducing numerical dissipation in OpenFOAM}
\label{sec:motiv}

When simulating reacting flows using large eddy simulation (LES), there has to be a special focus on numerical errors due to the strong coupling between discretization schemes and the notion of filtering in this technique \cite{kravmoin-leserrors,kaul-pof1,kaul-pof2,kaul-tfsp}. Unlike in Reynolds-averaged Navier Stokes (RANS) formulations, the simulated field is not smooth with respect to the mesh spacing. More specifically, numerical dispersion errors arising from Taylor series based finite-difference/finite-volume methods introduce scale-dependent errors, with the highest numerical errors occurring at scales close to the filter size \cite{kaul-pof1}. In order to assess and maintain accuracy of solvers, the concept of secondary conservation has been considered. Here, errors in quantities that are not directly solved for, but obtained from output fields, are minimized by appropriate choice of numerical schemes. In the context of turbulent flows, one such key quantity is kinetic energy (KE), which dictates the level of resolved turbulence captured by the numerical approach.

\begin{figure}
\center
% left bot right top
%\includegraphics[width=0.30\textwidth,trim={0cm 0cm 0cm 0cm},clip]{./Figures/BluffBodyConfig.png}
\includegraphics[width=0.32\textwidth,trim={0cm 0cm 0cm 0cm},clip]{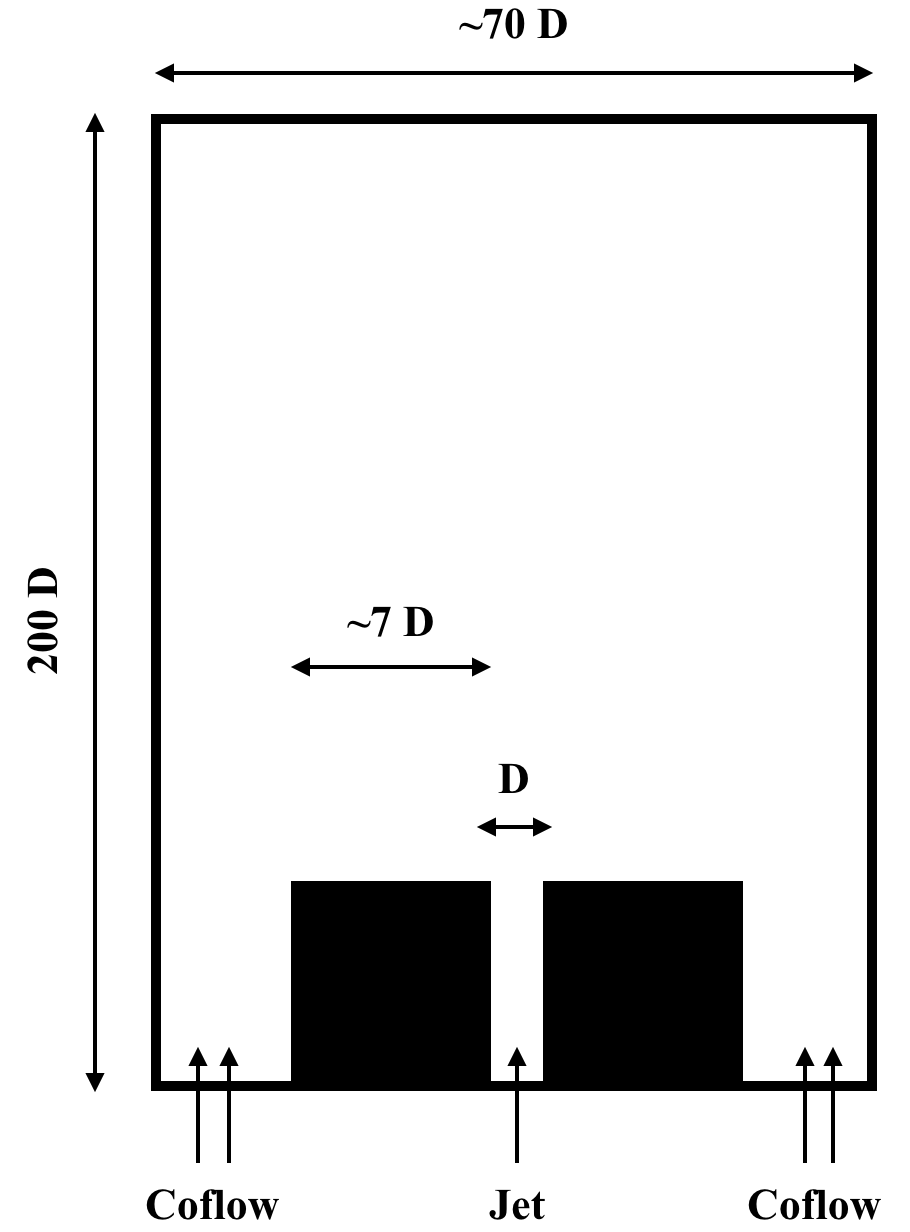}
\includegraphics[width=0.45\textwidth,trim={0cm 0.5cm 0cm 0cm},clip]{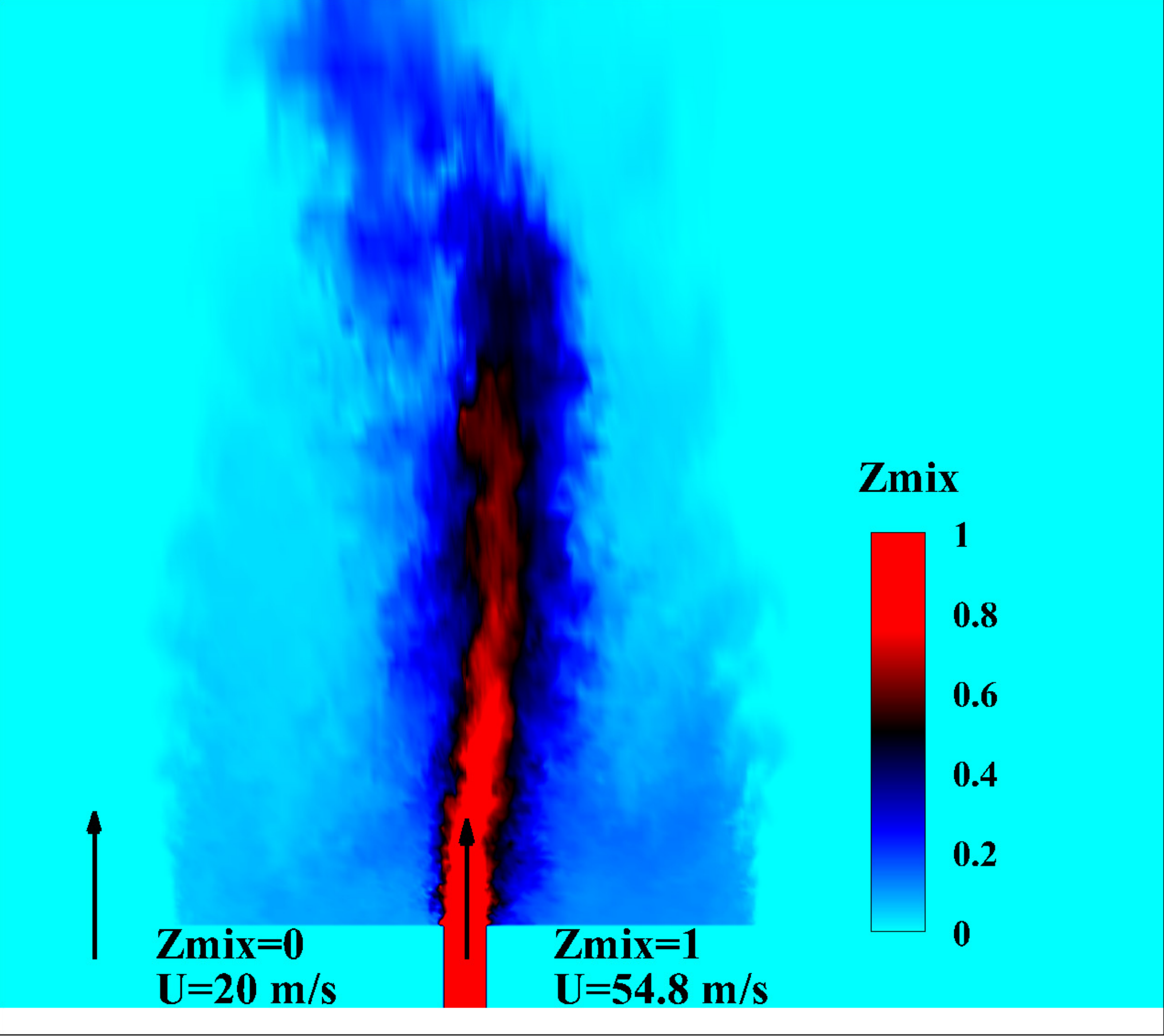}
\caption{Schematic of the bluff-body flame configurations (left) and instantaneous LES field of mixture fraction obtained with NGA (right).}
\label{fig:bbconfig}
\end{figure}

To illustrate the numerical dissipation problem, an constant density bluff-body flow simulation is carried out using two codes: the \verb|pisoFoam| solver used with linear interpolations in OpenFOAM, and the NGA solver used with second order schemes \cite{desjardins-jcp}. The \verb|pisoFoam| solver does not include variable density \textit{i.e.} it is written using an incompressible pressure correction equation and is therefore a low-Mach number solver. Thus, the differences observed between both codes are not due to the numerical compressibility effect noted in Sec.~\ref{sec:lowMachOFintro}. The flow configuration is illustrated in Fig.~\ref{fig:bbconfig}. When used for reacting flows, fuel is injected through the central pipe while air-coflow is injected through the outer annulus. The bluff body acts as a flame stabilization device because the flow recirculation that it creates. In the inert configuration constitutes an interesting test case for assessing the capability to capture recirculating flows.

\begin{figure}
\center
% left bot right top
\includegraphics[width=0.3\textwidth,trim={0cm 0cm 0cm 0cm},clip]{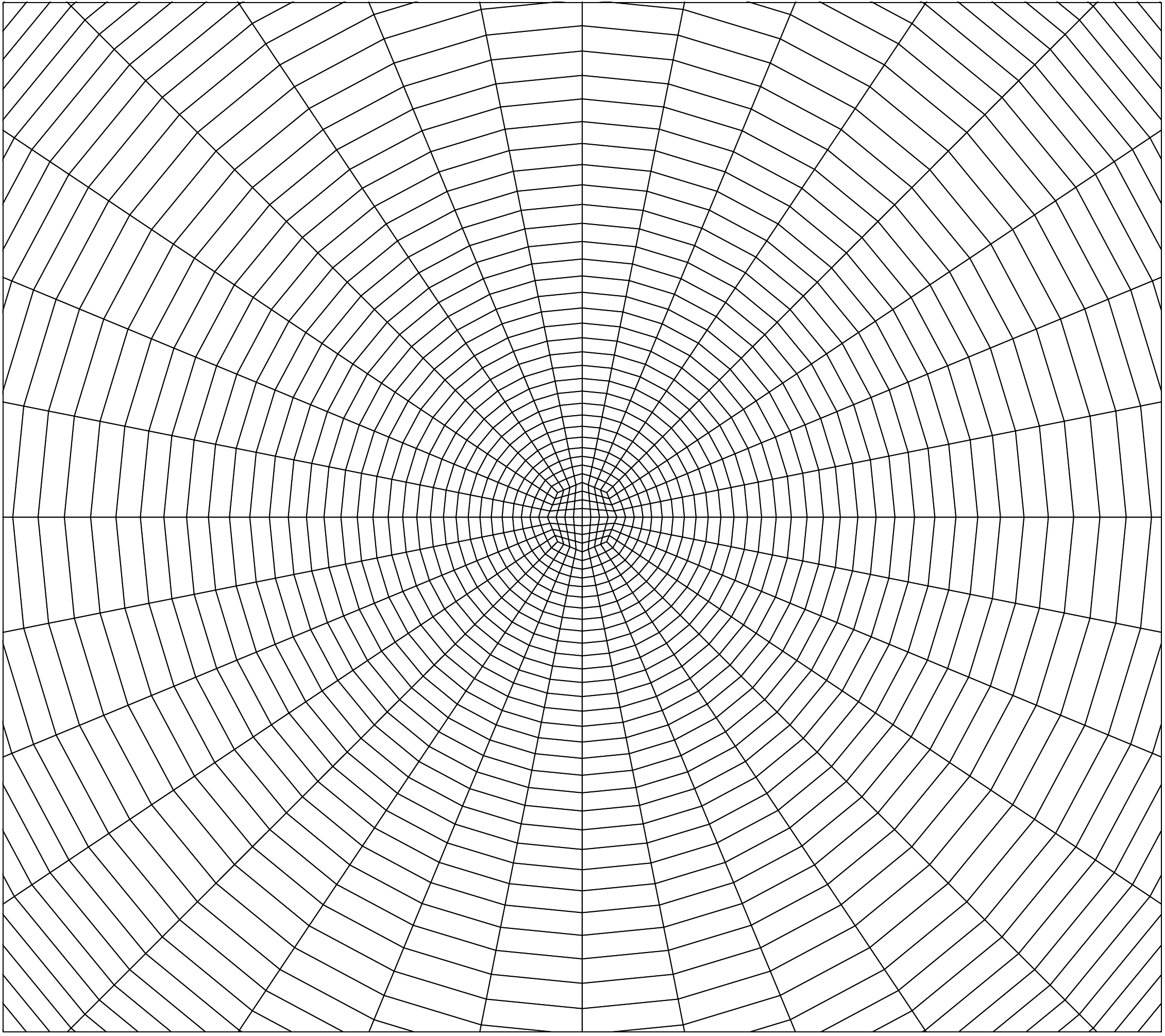}
\includegraphics[width=0.60\textwidth,trim={0cm 0cm 0cm 0cm},clip]{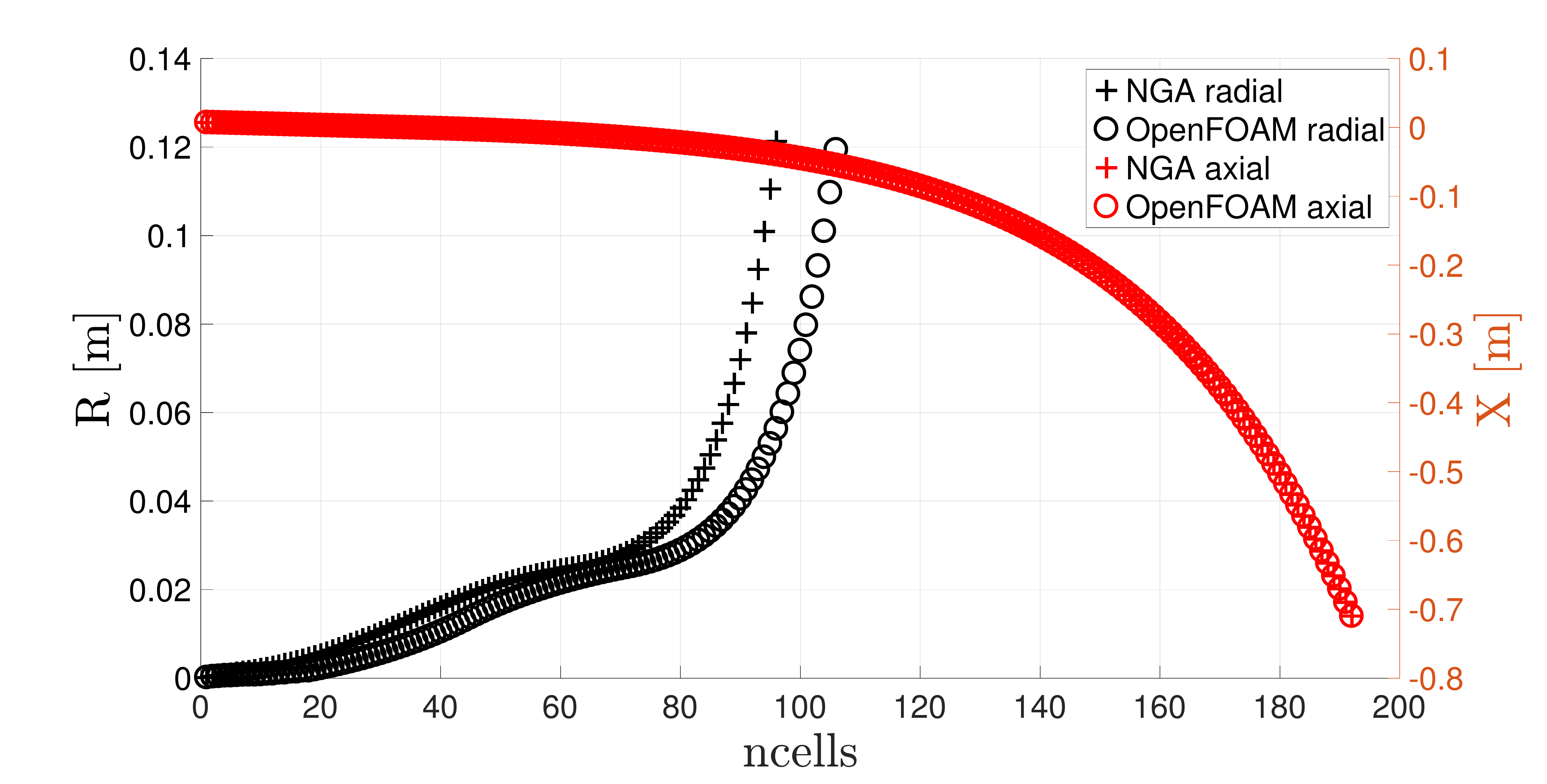}
\caption{Cylindrical grid used in OpenFOAM (left) and distribution of grid points in the radial and axial direction in OpenFOAM and NGA (right).}
\label{fig:meshOF}
\end{figure}

A structured cylindrical grid is used in NGA with a resolution of 192 $\times$ 92 $\times$ 32 in the axial, radial and azimuthal direction. The same grid is used in OpenFOAM except near the centerline. The mesh used in OpenFOAM is illustrated in Figure~\ref{fig:meshOF}. While the axial location of the cell exactly matches (Figure~\ref{fig:meshOF}), the radial location is slightly shifted for the OpenFOAM case because of the particular treatment at the centerline. These differences are deemed minor and the grids are considered identical for following discussion.

The spatial and temporal schemes used in NGA and OpenFOAM are of the same order. However, as opposed to OpenFOAM, NGA uses a staggered arrangement of variables \cite{desjardins-jcp} which ensures exact kinetic energy conservation when the mesh is uniform. In OpenFOAM, numerical dissipation of KE occurs because of the collocated arrangement of variables and because of the numerical discretization of the momentum equation. The latter will be addressed in Sec.~\ref{sec:KeImpl}. The baseline OpenFOAM solver is non-kinetic energy conservative (non-KEcons solver).

To illustrate the difference in the mixing process, a mixture fraction variable \cite{peters} is transported using the following equation.

\begin{equation}
 \frac{\partial Z_{mix}}{\partial t} + \nabla \cdot (\boldsymbol{u} Z_{mix}) = \nabla \cdot (D \nabla Z_{mix}).
\end{equation}

Due to the absence of source terms and the uniform mass diffusivity, the mixture fraction is a conserved variable.

At the fuel inlet boundary, a mixture fraction $Z_{mix}$ is set to 1 and the axial velocity $U$ to 54.8 m/s. At the coflow boundary, the mixture fraction is set to 0 and the axial velocity is set to 20 m/s. The timestep is held constant at 5e-7 s and the turbulence model is disabled in order to remove any differences which would stem from the subgrid scale (SGS) model. The dynamic viscosity and the mass diffusivity are held constant to a molecular value set to $1 \times 10^{-5} $ kg.m$^{-1}$.s$^{-1}$. The jet diameter is set to 3.6 mm.

Figure ~\ref{fig:MeanComp} shows the comparison of the mean mixture fraction fields and the root mean square (RMS) axial velocity, which is a direct measurement of the resolved turbulent kinetic energy. Although the mean structures are recovered, it is apparent that the RMS velocity is overall underpredicted compared to NGA, implying that the turbulent fluctuations are numerically damped more in OpenFOAM. The mean mixture also shows similar pattern as NGA, but the jet seems to mix at a slower rate.

\subsection{Theoretical background}
\label{sec:theory}

\subsubsection{Continuous and discrete operations}

While kinetic energy and mass conservation are uniquely defined in the context of continuous fields, such conservation is valid only in a specified sense in discrete representations. In order to prescribe the notion of conservation, a description of the discrete operation is necessary.

The ultimate goal of a numerical simulation is to be able to approximate the continuous functions $\boldsymbol{u}(\boldsymbol{x},t)$ and $p(\boldsymbol{x},t)$ which satisfy Eq.~\ref{eq:contNSmass} and Eq.~\ref{eq:contNSmom}. The continuous solutions also satisfy other equations obtained from the manipulation of the mass and momentum equation. For example, applying the curl operator to the momentum equation results in the vorticity conservation equation. In this work the dot product of the momentum equation with the velocity is considered. It results in a continuous sense in the KE transport equation written below

\begin{equation}
\label{eq:contKE}
\frac{\partial \rho u^2/2}{\partial t} + \nabla \cdot (\rho \boldsymbol{u} u^2/2)  = -\nabla \cdot (p \boldsymbol{u}) + p \nabla \cdot \boldsymbol{u} + \boldsymbol{u} \cdot (\nabla \cdot \overline{\sigma}).
\end{equation}

\begin{figure}
\center
% left bot right top
\includegraphics[width=0.45\textwidth,trim={0cm 0cm 11cm 0cm},clip]{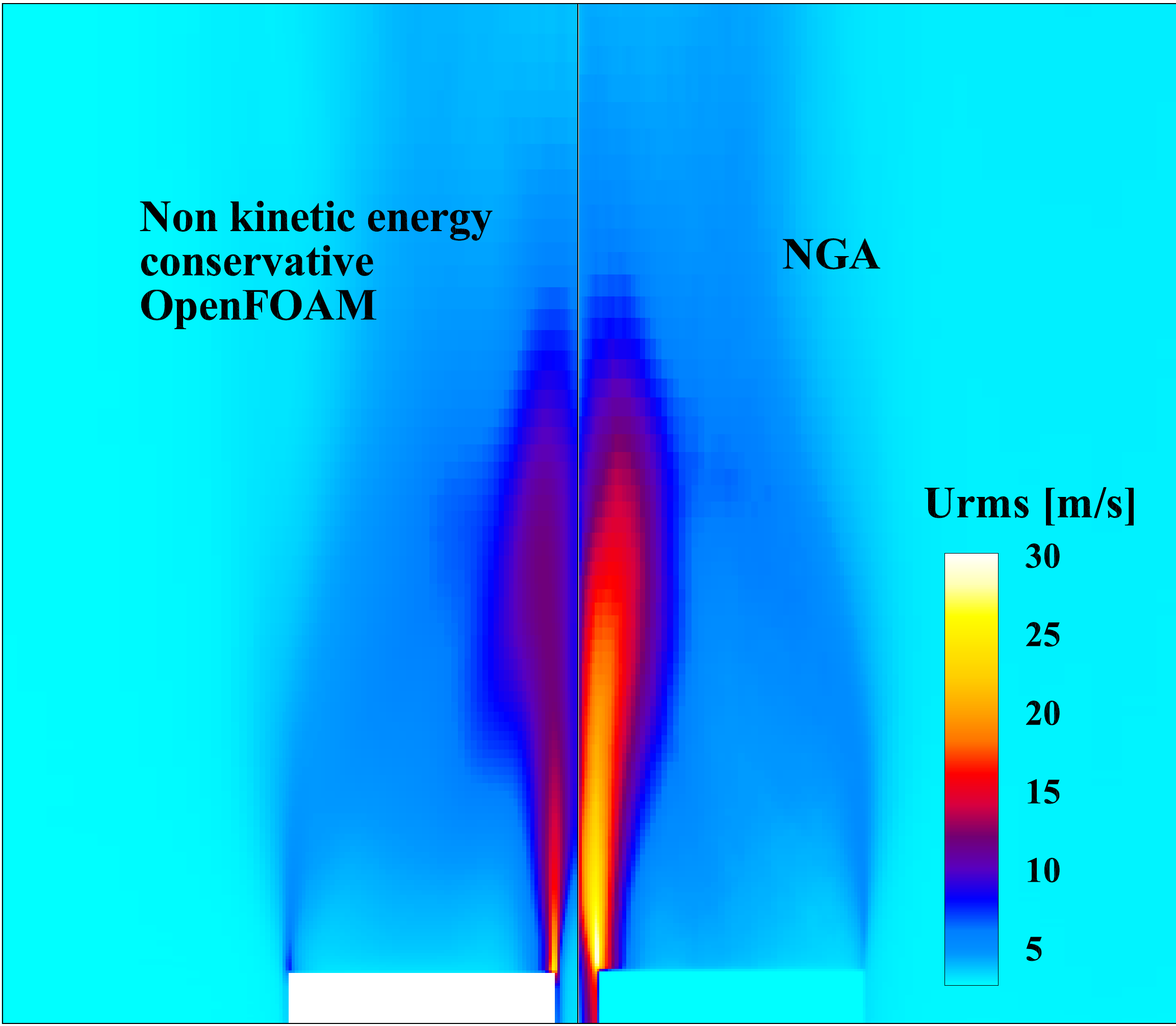}
\includegraphics[width=0.45\textwidth,trim={0cm 0cm 11cm 0cm},clip]{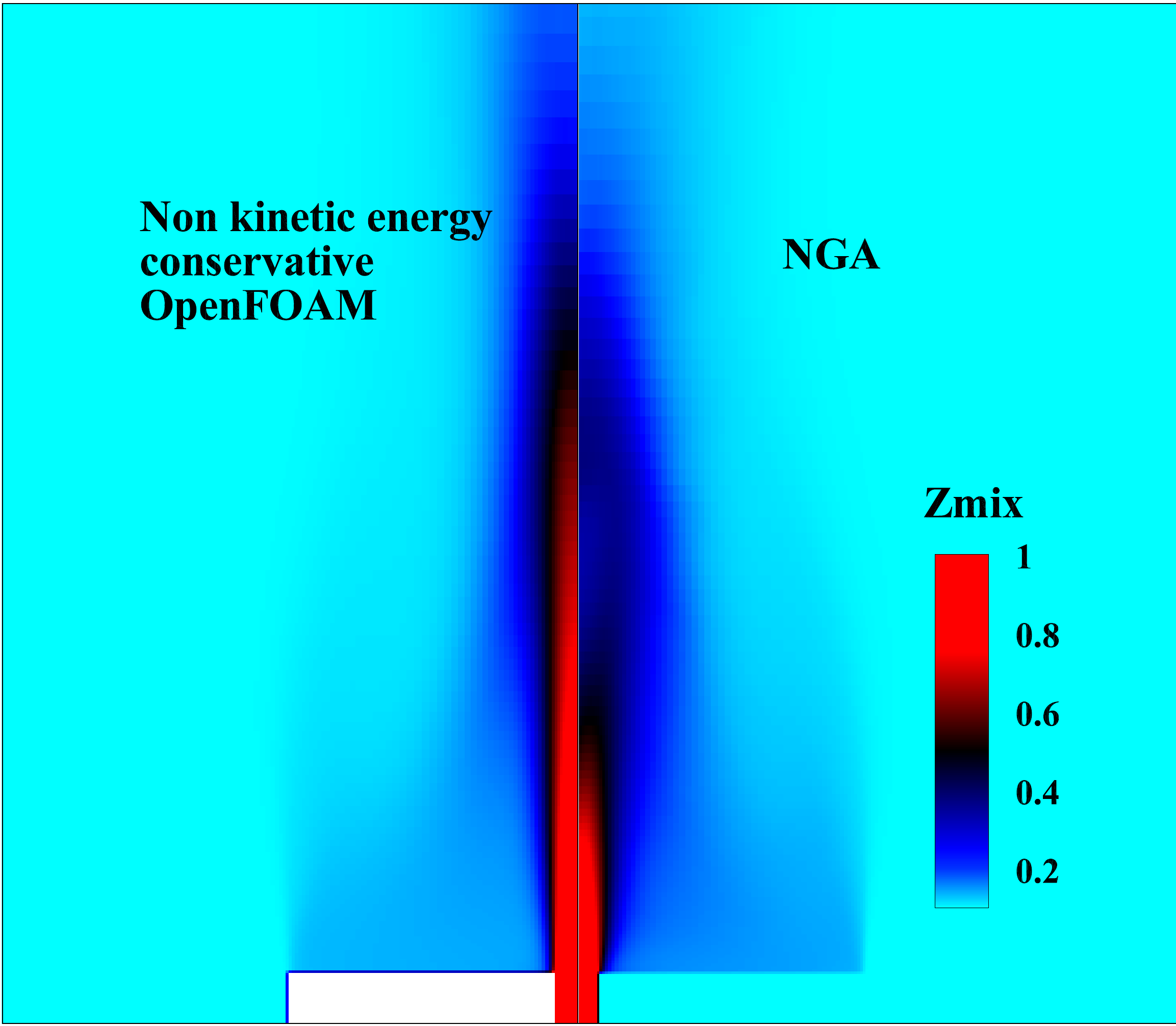}
\caption{Comparison of the axial RMS velocity between the non-KEcons OpenFOAM solver and NGA (left) and comparison of the mean mixture fraction between the non-KEcons OpenFOAM solver and NGA (right).}
\label{fig:MeanComp}
\end{figure}

The discrete fields obtained numerically satisfy the volume integrated momentum conservation equation and mass conservation equation, according to the numerical procedure outlined in Sec.~\ref{sec:procedure}. The velocity at the cell centers and the pressure at the cell centers are defined as fields respecting the momentum conservation equation and the mass conservation equation.

Using a similar strategy as the one used for the continuous fields, the discrete KE equation could be derived from the discrete momentum equation. It can be found that the continuous operations (like summation by parts) used to derive Eq.~\ref{eq:contKE} are not necessarily valid discretely and lead to \textit{extra-terms} which act as sink terms or source terms in each cell of the domain \cite[Ch.\ 10.2]{ferziger}. Two observations can be made: 1) satisfying the discrete momentum equation does not ensure that the KE equation is satisfied 2) the fact that velocity and pressure were solely defined as satisfying momentum and mass conservation does not leave room for introducing another conservation equation. Therefore, the conservation of other quantities like KE must come from an appropriate choice of the discretization schemes, and should be indirectly enforced.

%\subsubsection{Meaning of non-energy-conservative schemes}

Let $K$ be one form of KE for which a transport equation can be deduced from the already enforced momentum and mass conservation equations. An \textit{energy-conservative} scheme guarantees that for each internal face (as opposed to boundary face) of the domain, the amount of $K$ being shared from a computational cell to a neighbour is the exact opposite of the one being shared from this neighbour to this cell. 

As opposed to this definition, with \textit{non-energy-conservative} schemes, one can find an internal face for which the property cited previously is not ensured. Two types of non-energy-conservative schemes can be then defined. A \textit{dissipative} scheme is a scheme which removes KE from a control volume. An \textit{unstable} scheme is a scheme which creates KE in a control volume. Therefore, \textit{non-energy-conservative} schemes do not necessarily lead to more numerical stability through dissipation. Demonstrating that a scheme always dissipate KE (\verb|upwind| schemes for example or certain discrete pressure gradient formulations \cite{hamiac-kecons}) is a considerable achievement in ensuring numerical stability.

The ability to conserve quadratic properties does not imply better accuracy in predicting primary quantities (velocity, pressure etc.). In fact, it will be seen in Sec.~\ref{sec:KeImpl} that some first order interpolation schemes are better suited than second order centered schemes to ensure KE conservation. Hence, a trade-off between accuracy and dissipation needs to be made. Prior implementations have used both high-order spatial schemes and lower order energy conservative spatial schemes \cite{colin_avbp,moureau-yales,ham-ctr-fvm-unstruc}. For example, the compressible AVBP code from the CERFACS research group \cite{moureau_avbp,colin_avbp} has second and third order schemes capabilities, but high-order schemes are usually preferred \cite{garciathesis}. For low-Mach number solvers, high-order capabilities are sparse (for example YALES2 solver from the CORIA research group \cite{moureau-yales,yales2_nambully}) and also exist alongside energy conservative schemes (See \cite{vantieghem_thesis} for the YALES2 solver, and \cite{hamiac-kecons,ham-ctr-fvm-unstruc} for the CDP solver from the CTR research group).

With this background, the goal now is to develop a solver in the OpenFOAM framework that is able to minimize KE losses in the sense described above. 

\section{Minimal KE dissipation in collocated mesh formulations}
\label{sec:KeImpl}

The OpenFOAM solver, similar to be many unstructured mesh formulations, involves collocated variables. Here, all variables (pressure, velocity, density and scalar) are located at the center of a control volume. The alternative approach is the staggered grid technique \cite{harlow_stagg}, where the velocity components are located on the surfaces of the control volume. In collocated mesh solvers, complete KE conservation is not feasible due to irreducible interpolation errors (described below and in \cite{hamiac-kecons,jofre_cons}). Consequently, the objective here is to implement schemes that minimize energy dissipation. 

Schemes that conserve KE are designed such that by taking the dot product of the discrete momentum equation, a discrete KE equation can be obtained with minimal artificial source terms. Here the schemes derived in \cite{morinishi-skew} are chosen for the time derivative and the convective term of the momentum equation. The justification of this choice is provided in Sec.~\ref{sec:convterm_noreac}. As for the pressure gradient term, the scheme introduced in \cite{mahesh-jcp} for collocated arrangement of variables is used. Finally the pressure Laplacian is treated following \cite{hamiac-kecons}. 

%{\color{red} You have to describe in more detail what is the different between each of these papers, and which paper you will follow here. Start with energy conservation in constant density flows and follow the entire set of studies.}

%{\color{red} The discussion below does not say where you get all this material from - is this new or is it from a paper. If this is from some other work, why do we need to discuss this in so much detail? Cant we just provide the scheme and move on?}

%\subsection{Derivation of KEcons schemes}
%\label{sec:derivnoreac}

%In order to minimize dissipation, the discrete operator for each term in the momentum equation needs to be considered separately. In the following, the momentum equation is written between the timestep $n$ and the timestep $n+1$. Unless specified, the velocities are written at the timestep $n+1/2$ using an arithmetic averaging between the timestep $n$ and $n+1$.

\subsection{Convective term and time-derivative term}
\label{sec:convterm_noreac}

The energy conservation statement for the time-derivative and convective term can be written in the continuous form as:
\begin{equation}
\boldsymbol{u} \cdot \{\frac{\partial \rho \boldsymbol{u}}{\partial t} \}+ \boldsymbol{u} \cdot \{ \nabla \cdot (\rho \boldsymbol{u} \boldsymbol{u} )) \} = \frac{\partial \rho u^2/2}{\partial t} + \nabla \cdot (\rho \boldsymbol{u} u^2/2 ).
\end{equation}
The goal then is to obtain a discrete approximation that is consistent with this continuous representation. In this regard, skew-symmetric formulations have been widely used \cite{kravmoin-leserrors,ducros2000}. Achieving discrete energy conservation for constant density flows was done using first interpolations on non-uniform grids (interpolation denoted as midpoint in the following) \cite{verstappen,hamiac-kecons,mahesh-jcp}. A trade-off is therefore made on accuracy to attain energy conservation. In the case variable density flows, the derivations referenced above are not applicable anymore because they require the velocity field to be solenoidal. As a result, extra-terms proportional to the density change appear in the KE equation \cite{pierce_thesis}. Recently, the skew-symmetric schemes were again used to ensure KE conservation of variable density flows in collocated formulations \cite{morinishi-skew}. This formulation removed the requirement of the velocity field to be solenoidal. It should be noted that this formulation still relies on midpoint interpolations of the velocity from the cell centers to the cell faces.

The skew-symmetric form of the momentum convection term used for the derivations of energy conservative properties is written as follows: 

\begin{equation}
\label{eq:skewsymform}
\overline{\sqrt{\overline{\rho}^t}}^t \frac{\delta_1 \sqrt{\overline{\rho}^t} \boldsymbol{u}}{\delta_1 t} + \frac{1}{2} (\frac{\delta_1 \overline{\boldsymbol{\phi}_f}^t \overline{\boldsymbol{\hat{u}}}^x}{\delta_1 x} + \overline{\overline{\boldsymbol{\phi}_f}^t \frac{\delta_1 \boldsymbol{\hat{u}}}{\delta_1 x}}^x),
\end{equation}

where 
\begin{equation}
\label{eq:uhatDef}
 \boldsymbol{\hat{u}} = \frac{\overline{\sqrt{\overline{\rho}^t} u}^t}{\overline{\sqrt{\overline{\rho}^t}}^t},
\end{equation}

with $\frac{\delta_1}{\delta_1 x}$ the second order finite volume approximation of gradients, $\frac{\delta_1}{\delta_1 t}$ the Euler scheme, $\overline{(.)}^t$ the midpoint time interpolation and $\overline{(.)}^x$ the midpoint spatial interpolation as defined in \cite{morinishi-skew}.

In this approach, a divergence form equivalent to the skew-symmetric form can also be derived which makes this scheme well suited for finite volume methods. The equivalence between the divergence and the skew symmetric form relies on the enforcement of the discrete continuity equation given by:
\begin{equation}
\label{eq:cont_final}
\sum_f \overline{\boldsymbol{\phi_f}}^t \cdot \boldsymbol{S_f} + \frac{\delta_1 \overline{\rho}^t}{\delta_1 t} = 0,
\end{equation} 
which is used to formulate the pressure-correction Poisson system (see~\ref{sec:appPoisson}). 

As a side note, in many combustion applications, the presence of liquid (e.g., spray combustion) or solid (e.g., soot formation) phases lead to source/sink terms in the continuity equation representing exchange of mass between phases. It is shown in~\ref{sec:nonmasscons} that the above formulation remains energy-conservative for such modified continuity equations as well.

The form of KE conserved by the schemes is written as
\begin{equation}
\label{eq:timeder_fin}
\frac{1}{2} (\sqrt{\overline{\rho}^t} \boldsymbol{u})^2.
\end{equation}
%\begin{equation}
%\label{eq:timeder_fin}
%\boldsymbol{\hat{u}} \frac{\delta \overline{\rho}^t \boldsymbol{u}}{\delta t}. = \frac{1}{2} \frac{\delta (\sqrt{\overline{\rho}^t} \boldsymbol{u})^2}{\delta t}.
%\end{equation}

 As stated in Sec.~\ref{sec:theory}, this is a non-trivial form of KE which needs to be precisely defined when deriving or describing a solver that conserves secondary properties.

\subsection{Pressure term}
\label{sec:press_noreac}

%{\color{red} This section is not clear: 1) why discuss constant density formulation, 2) what happens to del.u term in 12 - why is the other term discussed but not this one? Needs to be rewritten} 
In the context of minimizing energy dissipation for variable density flows, the pressure gradient term of the momentum equation has not received particular attention. The approach used here relies on the derivations done for constant density flows \cite{mahesh-jcp,hamiac-kecons}. Therefore these schemes ensure that far from density variations in the domain, the KE dissipation is minimized. As noted elsewhere \cite{hamiac-kecons,mahesh-jcp,jofre_cons}, collocated formulation always introduce an extra-term in the KE budget. The only way of dealing with this term is to make it as small as possible. A least square formulation has been proposed by Mahesh \cite{mahesh-jcp} in the computation of the pressure gradient and the application of the pressure gradient to the cell center. However it resulted in instabilities in the solver \cite{hamiac-kecons,mahesh-jcp}. Other methods have been proposed recently but require intrusive modifications to the solver \cite{shashank}. Here, the approach of Ham and Iaccarino \cite{hamiac-kecons} is employed. The discretization of the pressure laplacian is tailored to ensure that the energy dissipation scales as $\mathcal{O}(\Delta t \Delta x^2)$ for skewed meshes. This derivation remains valid in the context of variable density flows. In order to apply the aforementioned methods, the pressure gradient term should be discretized using a midpoint scheme.

\subsection{Scalar transport equation schemes}

Similar to KE for velocity, a quadratic conservation property can be defined for scalars based on the square of the scalar variables. However, scalar equations do not have an accompanying pressure equation. It is known that nonlinear convection terms can lead to the formation of dispersive waves for any discretization scheme of order more than 1. Consequently, the use of midpoint scheme, which will reduce to a second-order central scheme, will cause the generation of location maxima and minima outside the bounds of the scalar \cite{bquick}. In general, some form of numerical dissipation is necessary to ensure stability of the solution. While several techniques have been formulated \cite{wenoorig,eno,quick,tvd,bquick}, none of these schemes will preserve the quadratic properties associated with scalars. 

%{\color{red} It is not clear what is meant by "written at ...". Are you saying that the discrete equations contain $n+1/2$ - then what is this term. For instance, if $n$ and $n+1$ are the solvable terms, then how is $n+1/2$ evaluated?}

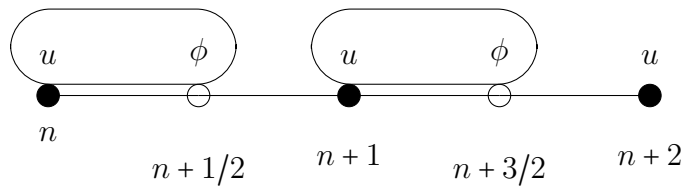
\begin{figure}
\label{fig:timestaggering}
    \centering

\begin{tikzpicture}
\tikzstyle{every node}=[draw,shape=circle]

\draw(0,0) node[circle,fill,inner sep=3pt,label=above:$u$,label=below:$n$](u){} -- (2,0);
\draw(2,0) node[circle,inner sep=3pt,label=above:$\phi$,label=below:$n+1/2$](b){} -- (4,0);
\draw(4,0) node[circle,fill,inner sep=3pt,label=above:$u$,label=below:$n+1$](d){} -- (6,0);
\draw(6,0) node[circle,inner sep=3pt,label=above:$\phi$,label=below:$n+3/2$](d){} -- (8,0);
\draw(8,0) node[circle,fill, inner sep=3pt,label=above:$u$,label=below:$n+2$](d){};
 %\draw
 % (0,0) {[rounded corners=15pt] --
 % ++(2,0)  -- 
 % ++(0,1)} --
%  ++(-2,0) --
%  cycle;
\draw[rounded corners=15pt]
  (-0.5,0.15) rectangle ++(3,1);
\draw[rounded corners=15pt]
  (3.5,0.15) rectangle ++(3,1);
%\draw (b) to[out=110,in=70] ++(2,0);
%\draw (c) to[out=110,in=70] ++(2,0);

\end{tikzpicture}
\caption{Illustration of the time-staggering scheme implemented for the variable density solver. The rounded rectangle shows which terms are advanced together.}
\end{figure}

Here, the focus is on the temporal scheme used for the discretization of Eq.~\ref{eq:scalarcont}. In the solver description, the momentum equation advances the velocity field from the time $n$ to the time $n+1$. The solver procedure involves outer-iterations to couple the velocity and the scalars like mixture fraction or density. In order to make the procedure more implicit, it is useful to use the newly estimated velocity in the formulation of the scalar transport equation. Therefore the scalars are fully transported with the most updated velocity field. Using a centered time-scheme, the scalars are advanced from the time $n+1/2$ to the time $n+3/2$ as shown in Fig.~\ref{fig:timestaggering}. Since the velocity is known at the time $n+1$, this allows to use a central time scheme for the scalar transport equation. The solver is said to be \textit{time-staggered}.

In variable density cases, a time-staggered solver allows the use of a second order scheme for time variation of density field. Without time-staggering, the density would be evaluated only at time $n+1$ and $n$. The pressure-Poisson equation is derived using the $\nabla \cdot (\boldsymbol{u}^{n+1})$ which involves $\frac{\delta_1 \rho^{n+1}}{\delta_1 t}$ (See Sec.~\ref{sec:pressureeqverif} for a complete derivation). In a non time-staggered scheme $\frac{\delta_1 \rho^{n+1}}{\delta_1 t} =  \frac{\rho^{n+1}-\rho^n}{\Delta t}$ which is a 1$^{st}$ order approximation in time. In the time-staggered version $\frac{\delta_1 \rho^{n+1}}{\delta_1 t} =  \frac{\rho^{n+3/2}-\rho^{n+1/2}}{\Delta t}$, which is a $2^{nd}$ order approximation in time.

%\subsubsubsection{Scalar source term}
Similar to the other terms of the scalar transport equation, the scalar source term $\dot{\omega}$ is written using a time interpolation. Since the source term is a function of scalar values, it is written as:

\begin{equation}
\dot{\omega}^{n+1} = \frac{\dot{\omega}(\phi^{n+3/2})+\dot{\omega}(\phi^{n+1/2})}{2}.
\end{equation}

\subsection{Implementation in OpenFOAM}
\label{sec:impl_consdens}

\subsubsection{Spatial schemes}

As highlighted in Sec.~\ref{sec:convterm_noreac} and Sec.~\ref{sec:press_noreac}, midpoint schemes (\verb|Gauss midPoint| in OpenFOAM) are better suited than second order central schemes (\verb|Gauss linear| in OpenFOAM) for the convective term of the momentum equation and the pressure gradient. In addition, in order for the above derivation to be valid, the mass flux used in the convective momentum term should satisfy the continuity equation in an integral sense.%In a low-Mach number solver using a pressure correction method, the velocity obtained after solving the momentum transport equation can be far from the velocity obtained after the pressure correction is applied. In other words, the correction velocity can be large, which might cause stability issues. In order to mitigate this effect, one can either use a time-explicit method for the momentum transport, or use the previously computed gradient of pressure as a source term in the momentum equation. A more systematical approach is to use a PISO (Pressure Implicit with Splitting of Operator) \cite{issa-piso} algorithm which converges to a zero pressure gradient correction with sufficient number of inner iterations. 

%Eq.~\ref{eq:uhatDef} can be used and gives the following form for the convective term:

%\begin{equation}
%\boldsymbol{\hat{u}} =  \frac{\sqrt{\overline{\rho}^t}^{n+1} \boldsymbol{u}^{n+1}}{\overline{\sqrt{\overline{\rho}^t}}^t} - \frac{\sqrt{\overline{\rho}^t}^{n} \boldsymbol{u}^{n}}{\overline{\sqrt{\overline{\rho}^t}}^t}.
%\end{equation}

%The first term of the RHS can be written implicitly (involves time $(.)^{n+1}$) while the second can be written explicitly ($(.)^{n}$).

The Laplacian computation of the pressure follows the approach of Ham and Iaccarino \cite{hamiac-kecons}. To take into account non-orthogonality of the computational mesh, the cell-face pressure gradient is obtained using a combination of cell-center pressure values, and cell center pressure gradient values. This family of laplacian schemes is the \verb|corrected| laplacian schemes in OpenFOAM. The ratio of orthogonal to non-orthogonal contribution can be adjusted using the \verb|limited| parameter \cite[Ch.\ 3.3.1.3]{jasakthesis}. Since the non-orthogonal part of the laplacian is treated explicitly, the pressure equation is repeated \verb|nNonOrthogonalCorrectors|$ +1$ times to account for this explicit part. %Therefore, it is strongly suggested to set this parameter the number of orthogonal correction to more than zero. %  A number of tests has shown that $\phi$=0.7 is a suitable choice . {\color{red} this is just the input command - do we need this?}The laplacian scheme that provides the best results in OpenFOAM is \verb|Gauss midPoint limited 0.7|.

The viscosity terms in the momentum equations dissipate KE. Hence, their discretization is governed primarily by truncation errors in spatial and temporal directions. To ensure temporal consistency, these terms are evaluated at $n+1/2$ in the time-staggered scheme, using a midpoint interpolation in time.

In OpenFOAM syntax, the final momentum equation takes the form.
\begin{verbatim}
rhoTimeInterp = 0.5*(rho+rho.oldTime());
rhoTimeInterp.oldTime() = 0.5*(rho.oldTime()+rho.oldTime().oldTime());

sqrt_rhoTimeInterp = sqrt(rhoTimeInterp);
sqrt_rhoTimeInterp.oldTime() = sqrt(rhoTimeInterp.oldTime());
sqrt_rhoTimeInterp_timeInterp = 
0.5*(sqrt_rhoTimeInterp+sqrt_rhoTimeInterp.oldTime());


fvVectorMatrix UEqn
(
    fvm::ddt(rhoTimeInterp,U)
  + (sqrt_rhoTimeInterp/sqrt_rhoTimeInterp_timeInterp)*
  fvm::div((phi+phi.oldTime())/4.0, U)
  + (sqrt_rhoTimeInterp.oldTime()/sqrt_rhoTimeInterp_timeInterp)*
  fvc::div((phi+phi.oldTime())/4.0, U.oldTime())
  - fvm::laplacian(0.5*turbulence->muEff(),U)
  - fvc::laplacian(0.5*turbulence->muEff()*U.oldTime())
  - fvc::div(turbulence->muEff()*0.5*dev2(Foam::T(fvc::grad(U))))
  - fvc::div(turbulence->muEff()*0.5*dev2(Foam::T(fvc::grad(U.oldTime()))))
);
\end{verbatim}

\subsubsection{Temporal schemes}

Regarding energy dissipation, there exist a constraint on the convective term: conservation is fully achieved only when the velocity convected is the same as the velocity updated. This can be satisfied using some semi-implicit treatment in the algorithm (through outer-iterations). Given that this treatment is correctly implemented, the solver structure discussed above is equally valid for implicit or explicit Euler timestepping. 

The temporal treatment of the pressure formulation needs to be further discussed, since this can affect the computational expense. More precisely, the choice is between the PISO procedure, which is already implemented in OpenFOAM, and the fractional timestep approach that is used by most research codes \cite{desjardins-jcp,ham_ctr2006}\cite[Ch.\ 3.2.2.1]{kraushaar}. Numerical differences between PISO and the fractional timestep method are briefly described in~\ref{sec:pisoFts}.

The PISO procedure involves twice the number of pressure-Poisson solutions as the fractional timestep method, and is therefore twice as expensive. However, the PISO procedure allows the momentum equations to be treated fully implicitly, which allows the use of larger timesteps. 
It has been argued \cite{ferziger} that the PISO procedure was tailored for steady-state problems. This has to do with the better stability properties of the PISO algorithm. When large CFL numbers can be used (for instance, in time-marching to steady-state), the PISO procedure becomes attractive. However, for transient problems where time accuracy is important, a large timestep is not useful and the PISO procedure has no added value compared to the fractional timestep method. Furthermore, for reacting cases, outer-iterations that couple scalar equations are used and are likely to be redundant with the pressure iterations involved in the PISO algorithm. Since their impact on the KE conservation property is not significant (Sec.~\ref{sec:resultstg}), the final application will determine the approach needed. The PISO implementation is described here and can be easily extended to fractional timestep using details in~\ref{sec:pisoFts} and~\ref{sec:appPoisson}.

\subsubsection{Pressure Poisson equation}
\label{sec:pressureeqverif}

The final piece of this low-Mach number solver is the solution of the pressure-Poisson equation. Based on the description in Sec.~\ref{sec:press_noreac}, the pressure equation has to be written such that the discrete continuity equation is enforced after the correction step. The derivation of the Poisson equation is detailed in~\ref{sec:appPoisson}. The consistent pressure equation can be written as
\begin{equation}
\label{eq:pressequform}
    \frac{\rho^{n+3/2}-\rho^{n+1/2}}{\Delta t} + \nabla \cdot (\boldsymbol{\phi_f}^*)= \nabla \cdot(a^{-1}. \overline{\rho}^t \nabla p^{n+1/2}),
\end{equation} 
where $a^{-1}$ is the component at the cell considered of the inverse of the split operator $[A]$. In the OpenFOAM framework it translates to
\begin{verbatim}
phi = fvc::interpolate(rhoTimeInterp*U) & mesh.Sf();

fvScalarMatrix pEqn
(
     fvm::laplacian(rhoTimeInterp*rAU,p)
   - fvc::ddt(rho)
   - fvc::div(phi)
);
\end{verbatim}

%{\color{red} Change this - add the equation that says how mass flux is corrected or refer to the right equation}

%The mass flux which is corrected when \verb|pEqn.flux()| is called will discretely satisfy the requirement which guarantees the equivalence between the skew-symmetric and the divergence form of the momentum equation. 

%Note also that the \verb|dmdt| in the \verb|adjustPhi| subroutine should be calculated with \verb|rho| consistently with the equation above.

\section{Verification of the minimally KE dissipative solver implementation}
\label{sec:lowdissVerif}
The solver verification is conducted in various configurations with increasing complexity. First, the minimal dissipation property of the solver is assessed using an inert periodic configuration. Using the upgraded solver, the inert bluff-body flame configuration which was presented in Sec.~\ref{sec:motiv} as a motivation for improving the KE conservation property of the solver is revisited. This test is used to assess the conservation properties in flows with unsteady inflow, non-periodic boundary conditions, and complex flow patterns with a range of length scales. Then a variable density case using periodic boundaries is studied. Finally a variable density case in a complex geometry using multiphase physics is studied.

\subsection{Constant density Taylor-Green (TG) Vortex Case}
\label{sec:verifconsdens}

\begin{figure}
\center
% left bot right top
%\includegraphics[width=0.9\textwidth,trim={0cm 0cm 0cm 0cm},clip]{./Figures/KooTG.png}
\includegraphics[width=0.49\textwidth,trim={1.5cm 0cm 8cm 0.5cm},clip]{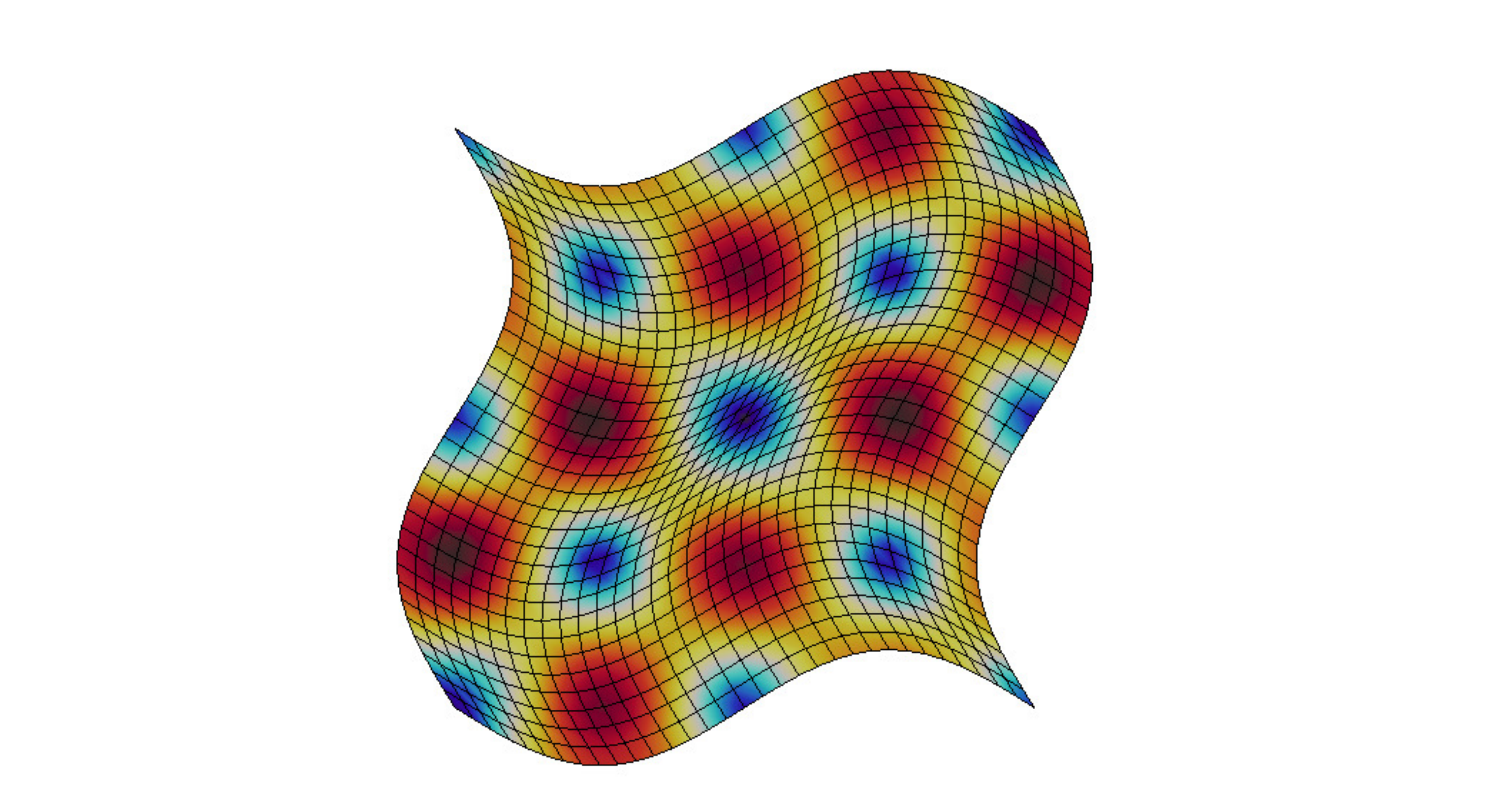}
\includegraphics[width=0.49\textwidth,trim={6.5cm 0cm 3cm 0.5cm},clip]{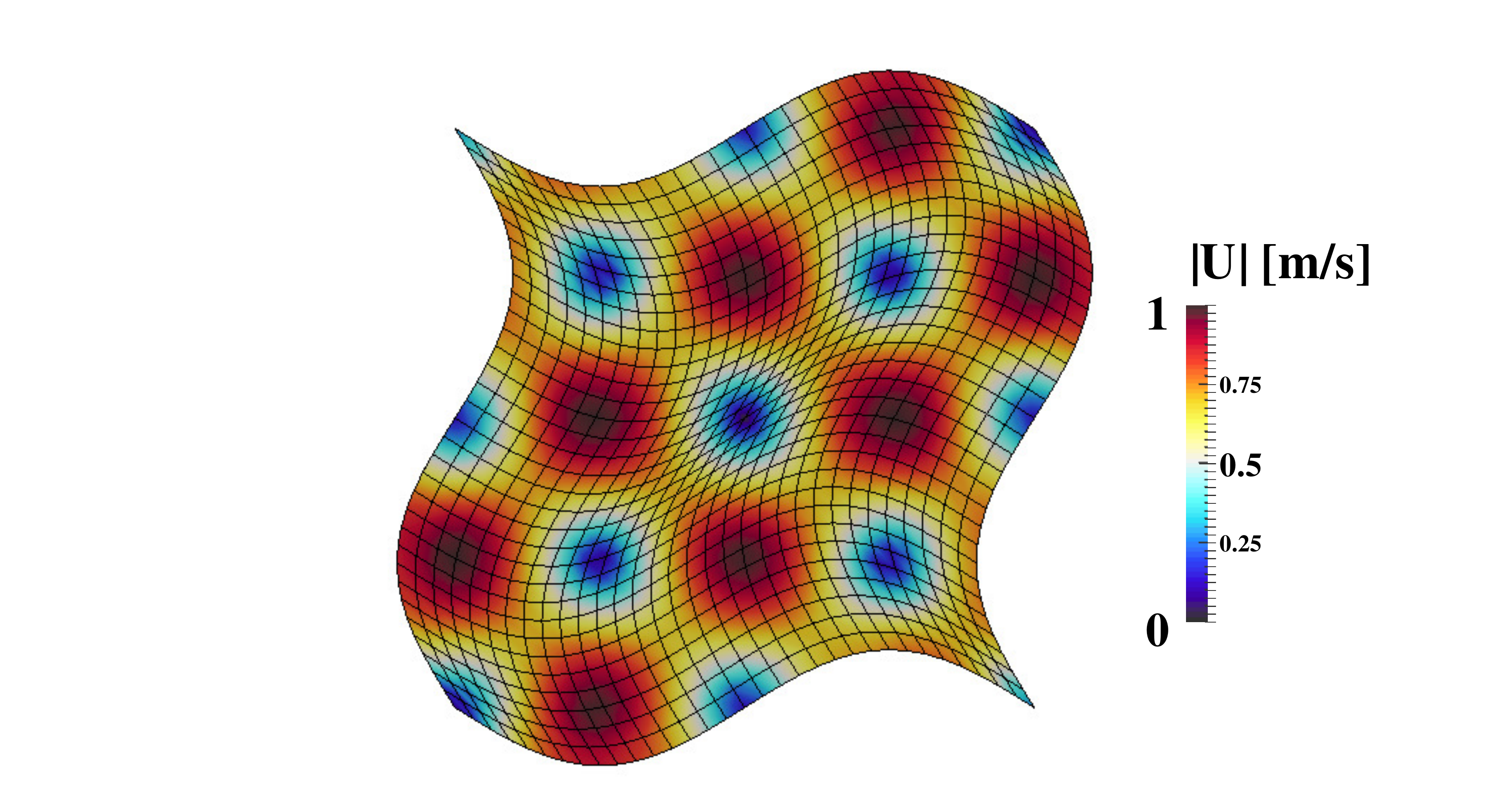}
\caption{Velocity magnitude at $t=1$ s for the constant density Taylor-Green vortex flow using a skewed mesh for the non-KEcons solver (left) and the KEcons solver (right).}
\label{fig:TG}
\end{figure}

The steady Taylor-Green vortex problem is used to test KE conservation properties of the new solver. This solver is called KEcons solver as opposed to the baseline non-KEcons solver available in OpenFOAM. The initial flow field consists in a series of vortices inside a periodic domain. The viscosity is set to zero so that any dissipation is purely numerical. Two cases are investigated, one with a uniform and isotropic grid, and another with a skewed mesh. The parameters chosen for the size and timestepping of the problem can be found in \cite{hamiac-kecons}. The flow described by the following set of equations:
\begin{equation}
u_x(x,y) = - cos(\pi x) sin(\pi y),
\end{equation}
\begin{equation}
u_y(x,y) = sin(\pi x) cos(\pi y),
\end{equation}
\begin{equation}
p(x,y) = -\frac{1}{4} (cos (2\pi x)+cos(2 \pi y)).
\end{equation}
for $x \in [-1;1]$ and $y \in [-1;1]$. For the first test case, the grid is regular and structured. Similar to \cite{hamiac-kecons}, a second case is investigated where the grid has undergone a skewing operation defined by
\begin{equation}
x' = x+ 0.2 sin(\pi y),
\end{equation}
\begin{equation}
y' = y+ 0.2 sin(\pi x).
\end{equation}
The skewed grid is shown in Fig.~\ref{fig:TG}.

%{\bf Alex: Perhaps instead of plotting Kin energy in the y-axis, we could plot percentage of Kin Energy dissipated}
%{\bf Malik: Ok for plots that don't contain comparison with Ham}

%\subsubsection{Results}
\label{sec:resultstg}

Since the solution is time-invariant, KE should be preserved. However, errors in discretization will lead to change in KE. The proposed method leads to decay of KE, albeit at a slow rate, compared to existing OpenFOAM formulations without the pressure correction and interpolation schemes. The rate of decay is presented against results from Ham et al.~\cite{hamiac-kecons}, where a similar collocated solver was used for the same test case.

%\begin{figure}
%\center
% left bot right top
%\includegraphics[width=0.49\textwidth,trim={1.5cm 0cm 10cm 0.5cm},clip]{./Figures/structSpace-eps-converted-to.pdf}
%\includegraphics[width=0.49\textwidth,trim={1.5cm 0cm 10cm 0.5cm},clip]{./Figures/skewedSpace-eps-converted-to.pdf}
%\caption{Global evolution and illustration of the spatial convergence of the KE  using fractional timestep method in the KEcons solver and a mesh size $\Delta x = 62.5mm$ (blue triangles); fractional timestep method in the KEcons solver and mesh size $\Delta x = 31.25mm$ (red triangles); PISO in the KEcons solver and a mesh size $\Delta x = 62.5mm$ (blue circles); PISO in the KEcons solver and a mesh size $\Delta x = 31.25mm$ (red circles)}
%\label{fig:tgspaceconv}
%\end{figure}

\begin{figure}
\center
% left bot right top 2.5 for right
%\includegraphics[width=0.49\textwidth,trim={1.5cm 0cm 9cm 0cm},clip]{./Figures/orthTG-eps-converted-to.pdf}
\includegraphics[width=0.49\textwidth,trim={0cm 0cm 4cm 0cm},clip]{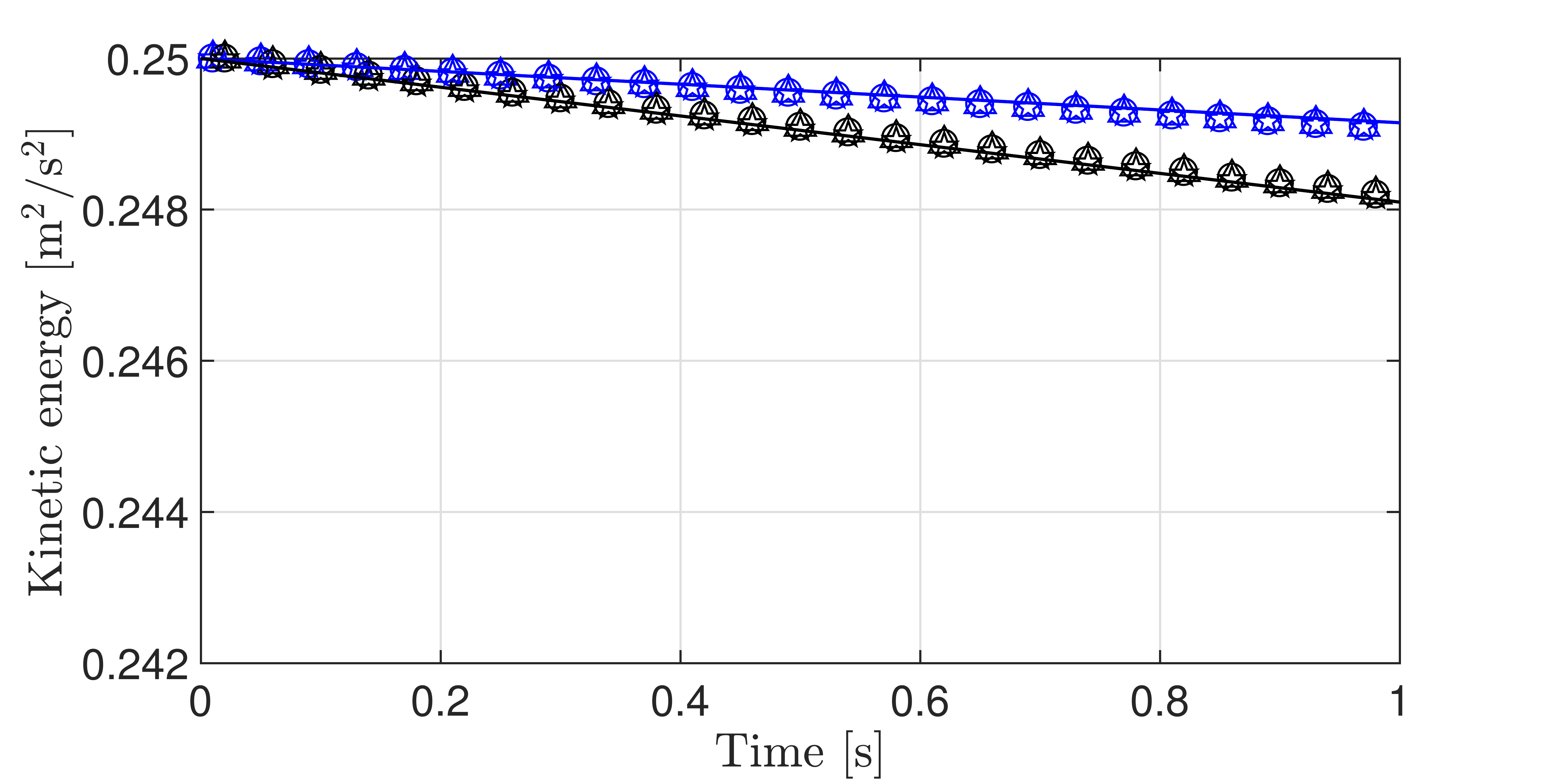}
\includegraphics[width=0.49\textwidth,trim={0cm 0cm 4cm 0cm},clip]{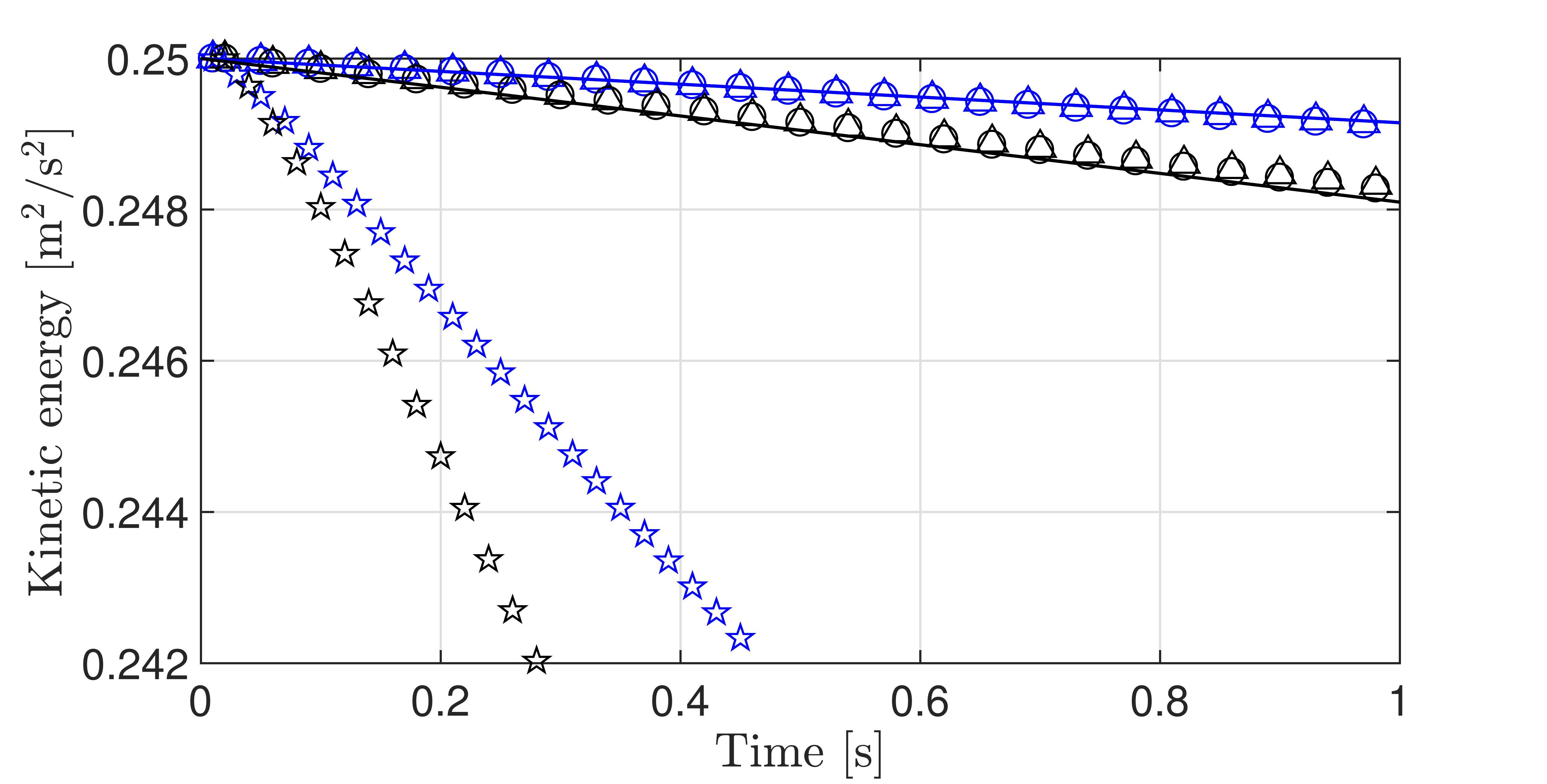}
%\caption{Global evolution and illustration of the temporal convergence of KE, using fractional timestep method in the KEcons solver and a timestep $\Delta t = 10$ ms (blue triangles); fractional timestep method in the KEcons solver and a timestep $\Delta t = 20$ ms (black triangles); PISO in the KEcons solver and a timestep $\Delta t = 10$ ms (blue circles); PISO in the KEcons solver and a timestep $\Delta t = 20$ ms (black circles); icoFoam with $\Delta t = 10$ ms (blue stars); icoFoam with $\Delta t = 20$ ms (black stars); Ham et al. \cite{hamiac-kecons} $\Delta t = 10$ ms (blue line); Ham et al. \cite{hamiac-kecons} $\Delta t = 20$ ms (black line)}
\caption{Global evolution and illustration of the temporal convergence of KE for the structured grid (left) and skewed grid (right) cases. \textcolor{blue}{$\medtriangleup$} fractional timestep method in the KEcons solver and a timestep $\Delta t = 10$ ms; \textcolor{black}{$\medtriangleup$} fractional timestep method in the KEcons solver and a timestep $\Delta t = 20$ ms; \textcolor{blue}{$\medcircle$} PISO in the KEcons solver and a timestep $\Delta t = 10$ ms; \textcolor{black}{$\medcircle$} PISO in the KEcons solver and a timestep $\Delta t = 20$ ms; \textcolor{blue}{$\medstar$} icoFoam with $\Delta t = 10$ ms; \textcolor{black}{$\medstar$} icoFoam with $\Delta t = 20$ ms; \mythickline{blue} Ham et al. \cite{hamiac-kecons} $\Delta t = 10$ ms; \mythickline{black} Ham et al. \cite{hamiac-kecons} $\Delta t = 20$ ms }
\label{fig:tgtimeconv}
\end{figure}

The structured grid solution shows little difference between the conservative and non-conservative formulations (Fig.~\ref{fig:tgtimeconv}). This is expected, since energy loss is primarily associated with the interpolation schemes of the convection term and the pressure laplacian formulation. Uniform structured grids minimize such interpolation errors, especially when the flow is well-resolved using small computational cells. In this case, linear interpolation adequately captures spatial variation. However, skewed mesh calculations show larger difference between conservative and non-conservative solvers (Fig.~\ref{fig:tgtimeconv}). It is seen that the OpenFOAM formulation produces results consistent with the collocated arrangement solver of Ham et al.~\cite{hamiac-kecons}. It is also seen that non-conservative approaches such as that implemented in the OpenFOAM \verb|icoFoam| solver with no pressure correction schemes, lead to much higher KE loss.

%{\color{red} Fix caption - grammar and ms}

\begin{figure}
\center
% left bot right top
%\includegraphics[width=0.49\textwidth,trim={1.5cm 0cm 10cm 0cm},clip]{./Figures/BudgetStruct-eps-converted-to.pdf}
%\includegraphics[width=0.49\textwidth,trim={1.5cm 0cm 10cm 0cm},clip]{./Figures/BudgetSkewed-eps-converted-to.pdf}
\includegraphics[width=0.55\textwidth,trim={1.5cm 0cm 4cm 0cm},clip]{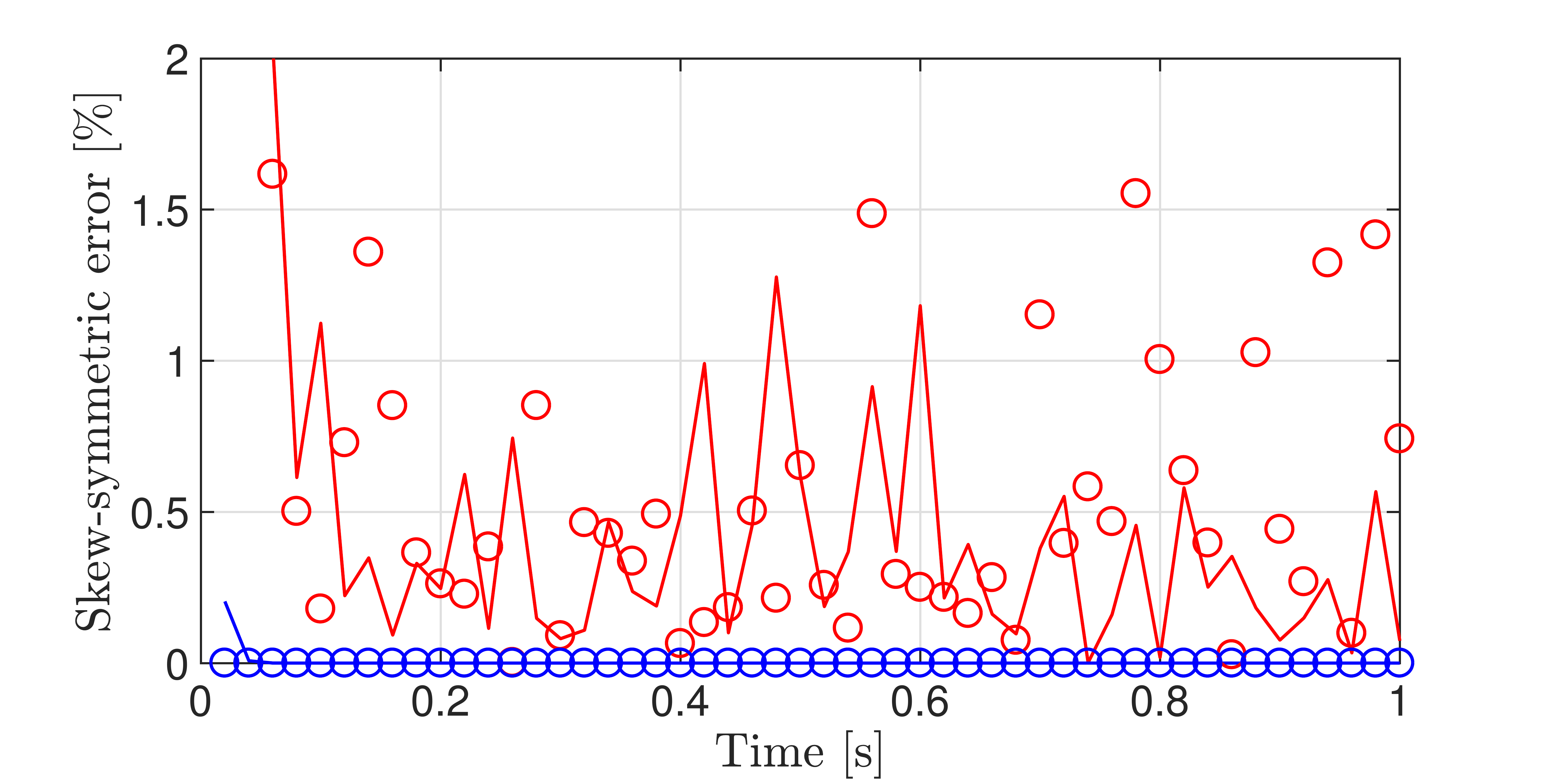}
\caption{Contribution of the convective term to the total energy dissipation with $\Delta t = 20$ ms, $\Delta x = 0.03125$ m. The convection term is defined as $\sum_{P} V_P (\hat{u}_i \cdot \frac{\delta u_i u_j}{\delta x_j})$ for the fractional timestep. \mythickline{blue} structured mesh with fractional timestep; \mythickline{red} skewed mesh with fractional timestep;  \textcolor{blue}{$\medcircle$} structured mesh with PISO; \textcolor{red}{$\medcircle$} skewed mesh with PISO.}
\label{fig:tgbudget}
\end{figure}
%\caption{KE budget contribution to the energy dissipation with $\Delta t = 20 ms, \Delta x = 0.03125m$ for the structured grid case (left) and the skewed grid case (right). The pressure term contribution to dissipation is defined as $\sum_{P} V_P (\hat{u}_i \cdot \frac{\delta P}{\delta x_i})$ for the fractional timestep (blue line) and PISO (blue circles). The convection term is defined as $\sum_{P} V_P (\hat{u}_i \cdot \frac{\delta u_i u_j}{\delta x_j})$ for the fractional timestep (red line) and PISO (red circles). The time term is defined as $\sum_{P} V_P (\hat{u}_i \cdot \frac{\delta u_i}{\delta t} - \frac{u_i^{2,n+1}/2 - u_i^{2,n}/2}{\Delta t})$. $V_P$ is the local cell volume and $P$ is the identifier of the cell.}

This test case is also used to assess the influence of the timestepping technique (PISO or fractional timestep. See~\ref{sec:pisoFts} for details about these methods). Figure ~\ref{fig:tgtimeconv} shows that both techniques lead to the same amount of numerical dissipation. %In the fractional timestep formulation, the accuracy of the estimation of $\boldsymbol{u}^{n+1/2}$ improves with the number of outer-iterations, making the contribution of the convective term to the KE error smaller and smaller. The effect is not as evident for the PISO procedure, which is an implicit approach.

For a constant density case, the definition of $\boldsymbol{\hat{u}}$ in Sec.~\ref{eq:uhatDef} reduces to $\frac{\boldsymbol{u}^n+\boldsymbol{u}^{n+1}}{2}$. The definition of  KE at time $n$ becomes $\frac{\boldsymbol{u}^n\cdot \boldsymbol{u}^n}{2}$. In this case, the budget of the discrete KE equation integrated over the simulation domain can be used to analyze the contribution of the individual terms. The contribution of the convective term to the energy dissipation budget is plotted in Fig.~\ref{fig:tgbudget}. This KE error is called the \textit{skew-symmetric error}. It is first observed that in both the PISO and the fractional timestep methods, the contribution of the convective term to the total energy dissipation is negligible and is therefore dominated by the pressure term $\sum_{P} V_P (\hat{u}_i \cdot \frac{\delta_1 P}{\delta_1 x_i})$. The time derivative term  $(\hat{u}_i \cdot \frac{\delta_1 u_i}{\delta_1 t} - \frac{u_i^{2,n+1}/2 - u_i^{2,n}/2}{\Delta t})$ is exactly zero by construction. Here, $V_P$ is the local cell volume and $P$ is the identifier of the cell. To have complete cancellation of the KE dissipation due to convection, it is essential to ensure that mass is exactly conserved to a high precision in the entire domain. In the skewed case, using the same tolerances for the pressure equation, the mass conservation error was higher.  %{\color{red} You should say what is convection term - which equation, which term - same for pressure term}

\begin{figure}
\center
% left bot right top
\includegraphics[width=0.44\textwidth,trim={9cm 0cm 3cm 0cm},clip]{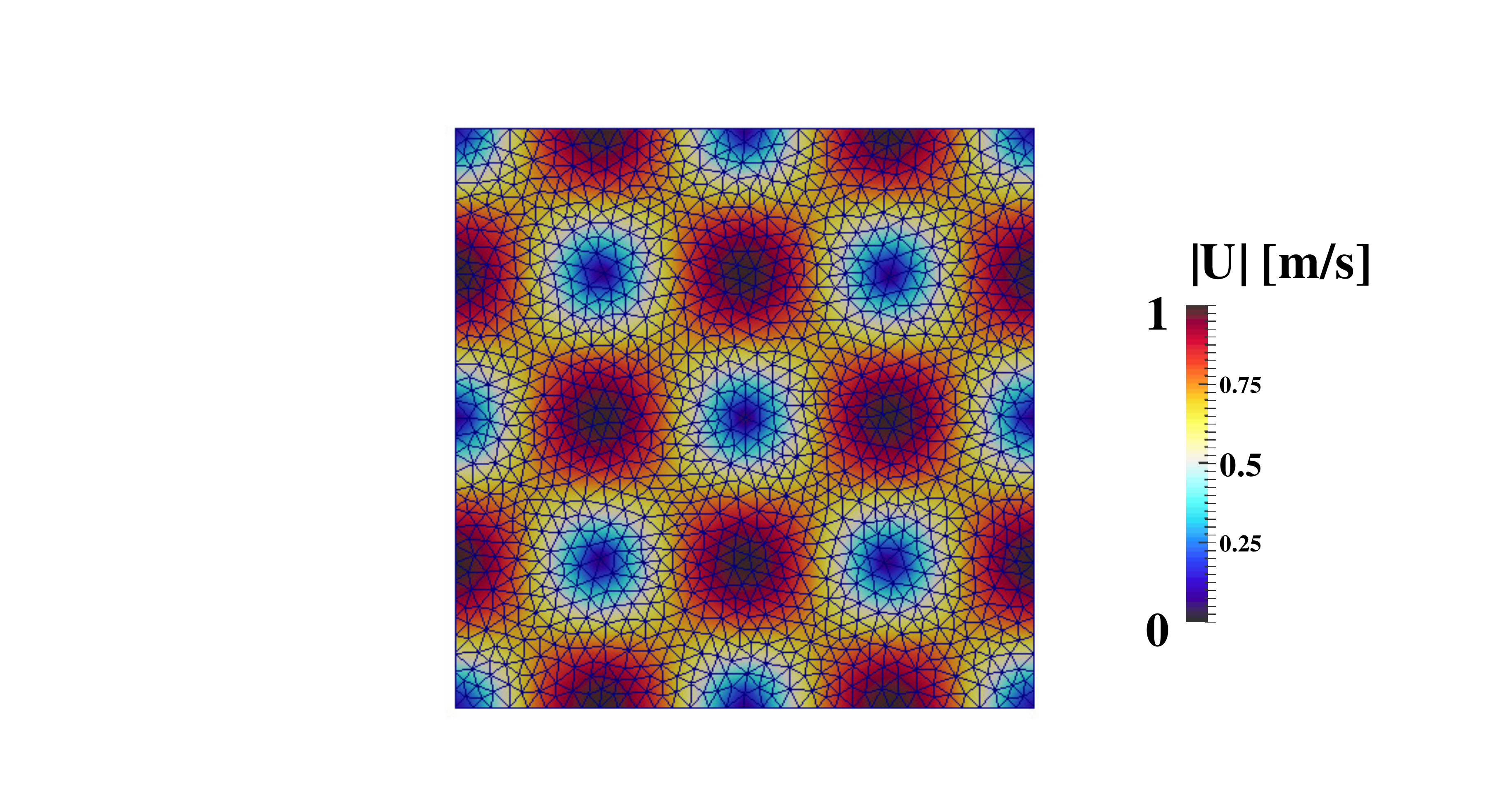}
\includegraphics[width=0.54\textwidth,trim={0cm 0cm 4cm 0cm},clip]{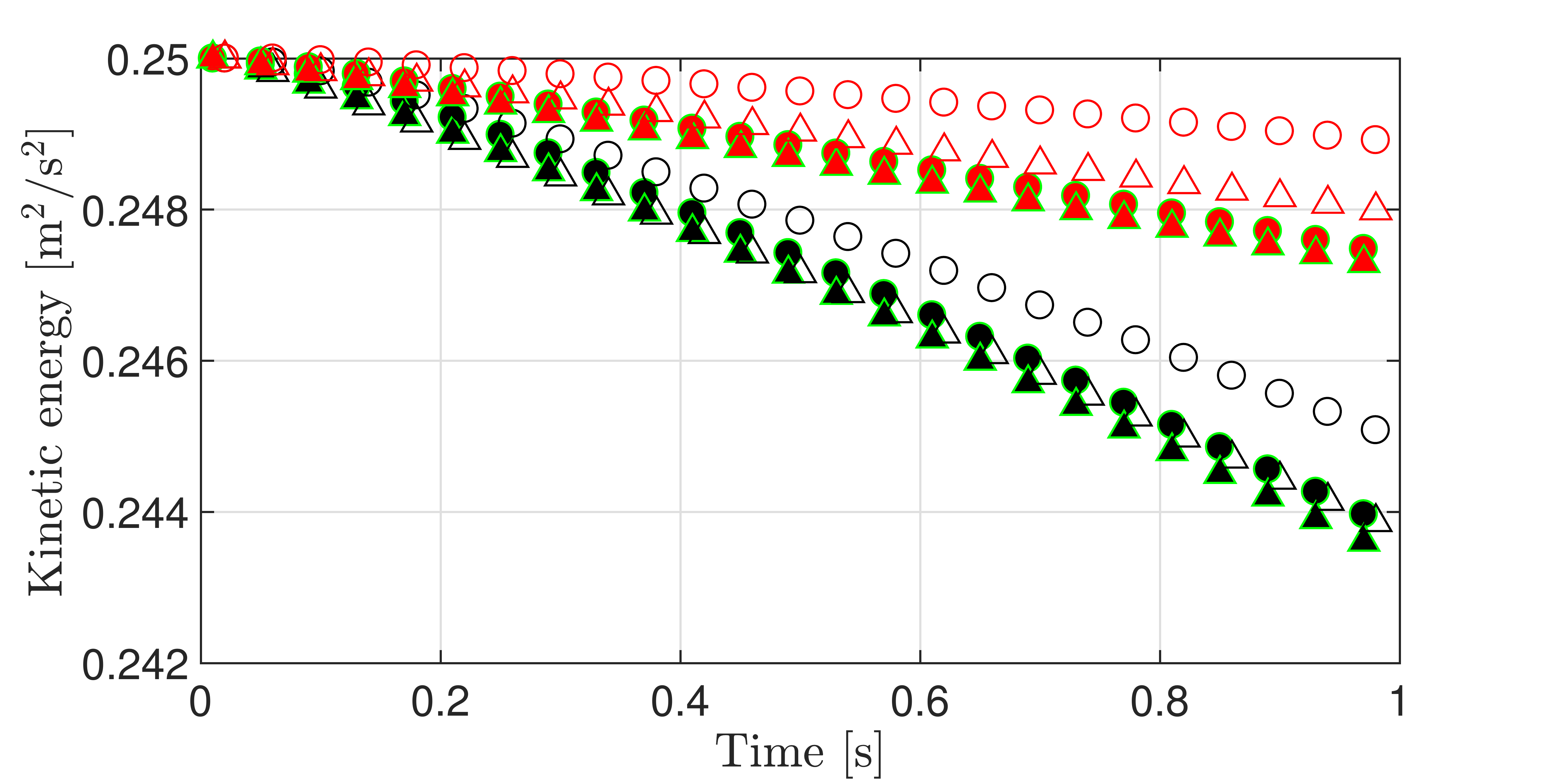}
\caption{Illustration of the unstructured mesh case $\Delta x = 0.0625$ m (left) and global evolution with spatial and temporal convergence of KE (right). \protect\fbcrcl PISO with KEcons solver $\Delta t = 10$ ms and $\Delta x = 0.0625$ m; \mytriangle{black} icoFoam $\Delta t = 10$ ms and $\Delta x = 0.0625$ m; \textcolor{black}{$\medcircle$} PISO with KEcons solver $\Delta t = 20$ ms and $\Delta x = 0.0625$ m; \textcolor{black}{$\medtriangleup$} icoFoam $\Delta t = 20$ ms and $\Delta x = 0.0625$ m; \protect\frcrcl PISO with KEcons solver $\Delta t = 10$ ms and $\Delta x = 0.03125$ m; \mytriangle{red} icoFoam $\Delta t = 10$ ms and $\Delta x = 0.03125$ m; \textcolor{red}{$\medcircle$} PISO with KEcons solver $\Delta t = 20$ ms and $\Delta x = 0.03125$ m; \textcolor{red}{$\medtriangleup$} icoFoam $\Delta t = 20$ ms and $\Delta x = 0.03125$ m.}
\label{fig:unstructTG}
\end{figure}

The TG test case is also solved using an unstructured grid, which is representative of the practical configurations that will be simulated using this solver. To ensure direct comparison, the number of triangles in the tetrahedral case is set equal to the number of hexahedrons on the boundaries. The results from these simulations are shown in Fig.~\ref{fig:unstructTG}. It is observed that KE dissipation is considerably larger than in the structured and skewed mesh cases (Fig.~\ref{fig:tgtimeconv}). In fact, the rate of dissipation increases with time initially, but reaches a constant value after this initial transient. This leads to a concave shaped dissipation plot. These results were also observed elsewhere \cite{benhamadouche}, and suggests accumulation of numerical error. It is interesting to note that the dissipation of energy does not improve with a reduction in timestep, implying that the dissipation error obtained from the interpolation of the pressure gradient at the cell faces to the cell centers is $\mathcal{O}(\frac{1}{\Delta t})$. However, the results are improved compared to the original \verb|icoFoam| solver.

%\caption{Illustration of the unstructured mesh case $\Delta x = 0.0625$ m (left) Global evolution with spatial and temporal convergence of KE (right) for PISO in the KEcons solver and a timestep $\Delta t = 10$ ms and $\Delta x = 0.0625$ m (filled black circles); icoFoam and a timestep $\Delta t = 10$ ms and $\Delta x = 0.0625m$ (filled black triangles); PISO in the KEcons solver and a timestep $\Delta t = 20$ ms and $\Delta x = 0.0625$ m (black circles); icoFoam and a timestep $\Delta t = 20$ ms and $\Delta x = 0.0625$ m (black triangles); PISO in the KEcons solver and a timestep $\Delta t = 10$ ms and $\Delta x = 0.03125$ m (filled red circles); icoFoam and a timestep $\Delta t = 10$ ms and $\Delta x = 0.03125$ m (filled red triangles); PISO in the KEcons solver and a timestep $\Delta t = 20$ ms and $\Delta x = 0.03125$ m (red circles); icoFoam and a timestep $\Delta t = 20$ ms and $\Delta x = 0.03125$ m (red triangles).}

%{\color{red} ms is not written within $ms$. Also 20 ms, not 20ms}
\subsection{Non-periodic boundary verification}

\begin{figure}
\center
% left bot right top
\includegraphics[width=0.55\textwidth,trim={0cm 0cm 3cm 0cm},clip]{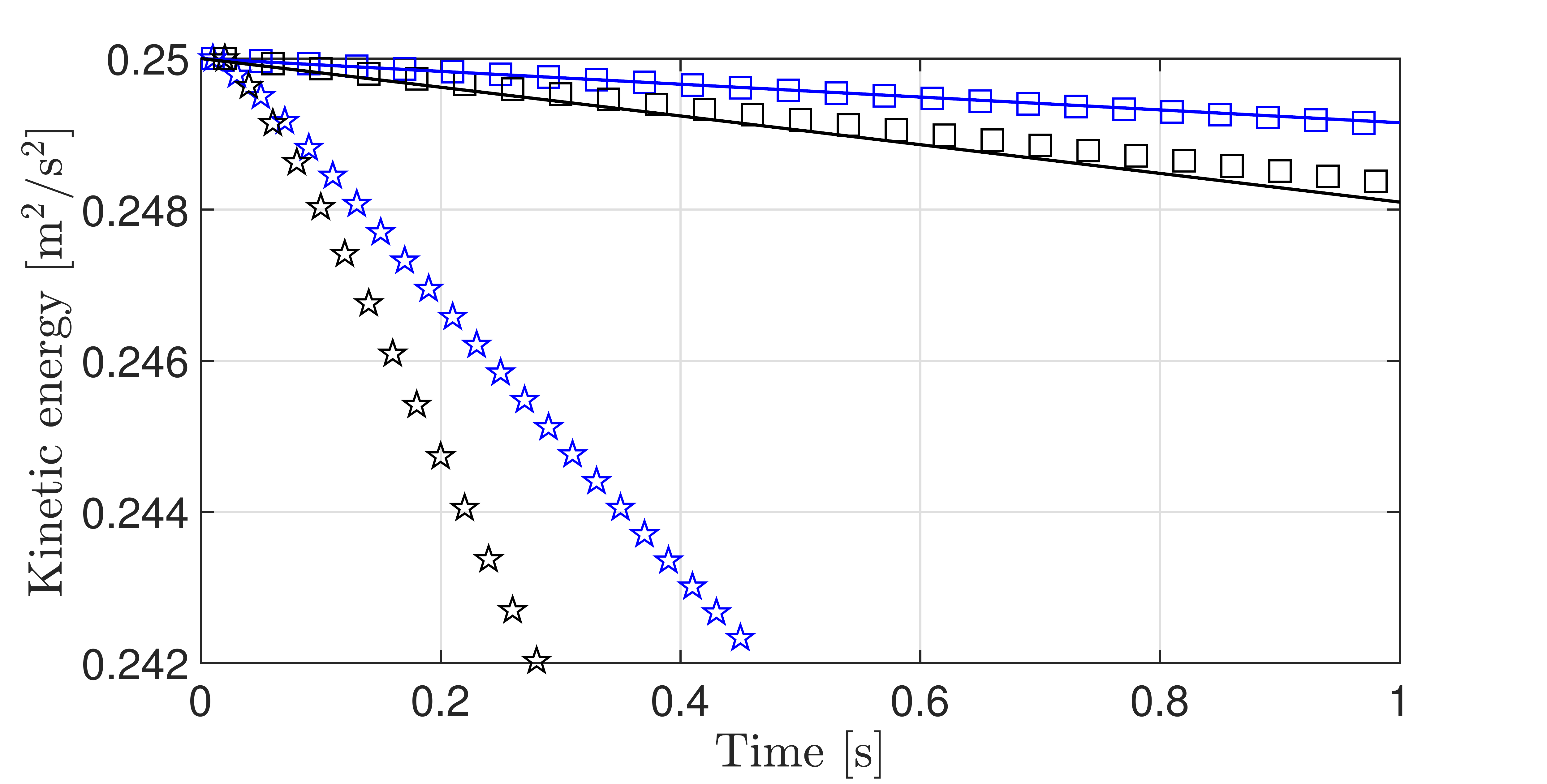}
\caption{Temporal convergence of KE. \textcolor{blue}{$\medsquare$} icoFoam with pressure correction $\Delta t = 10$ ms; \textcolor{black}{$\medsquare$} icoFoam with pressure correction and $\Delta t = 20$ ms; \textcolor{blue}{$\medstar$} icoFoam without pressure correction $\Delta t = 10$ ms; \textcolor{black}{$\medstar$} icoFoam without pressure correction $\Delta t = 20$ ms; \mythickline{blue} Ham et al. \cite{hamiac-kecons} $\Delta t = 10$ ms; \mythickline{black} Ham et al. \cite{hamiac-kecons} $\Delta t = 20$ ms.}
\label{fig:tgvInertIco}
\end{figure}

The construction KEcons solver involved several modifications which led to improved energy conservation properties of the OpenFOAM solver. One could ask which one these modifications is the most critical. In Fig.~\ref{fig:tgvInertIco}, the baseline icoFoam solver using pressure gradient corrections is compared to the KEcons solver and the baseline icoFoam solver. It appears that most of the improvements between the described formulation and the baseline OpenFOAM solver are due to the pressure scheme. In order to observe the improvements due to the formulation of the convective scheme as well as the time-staggering, a different test problem is studied. Similar to the discussion in Sec.~\ref{sec:motiv}, the new formulation is compared to the \verb|pisoFoam| formulation using linear schemes and the \verb|corrected| scheme for pressure. Therefore, the changes observed for the velocity field will reflect the influence of the convective term in the new momentum equation formulation. The changes observed in the mixture fraction field will reflect the improvements due also to the convective term formulation and time-staggering implementation. 

\begin{figure}
\center
% left bot right top
\includegraphics[width=0.45\textwidth,trim={0cm 0cm 11cm 0cm},clip]{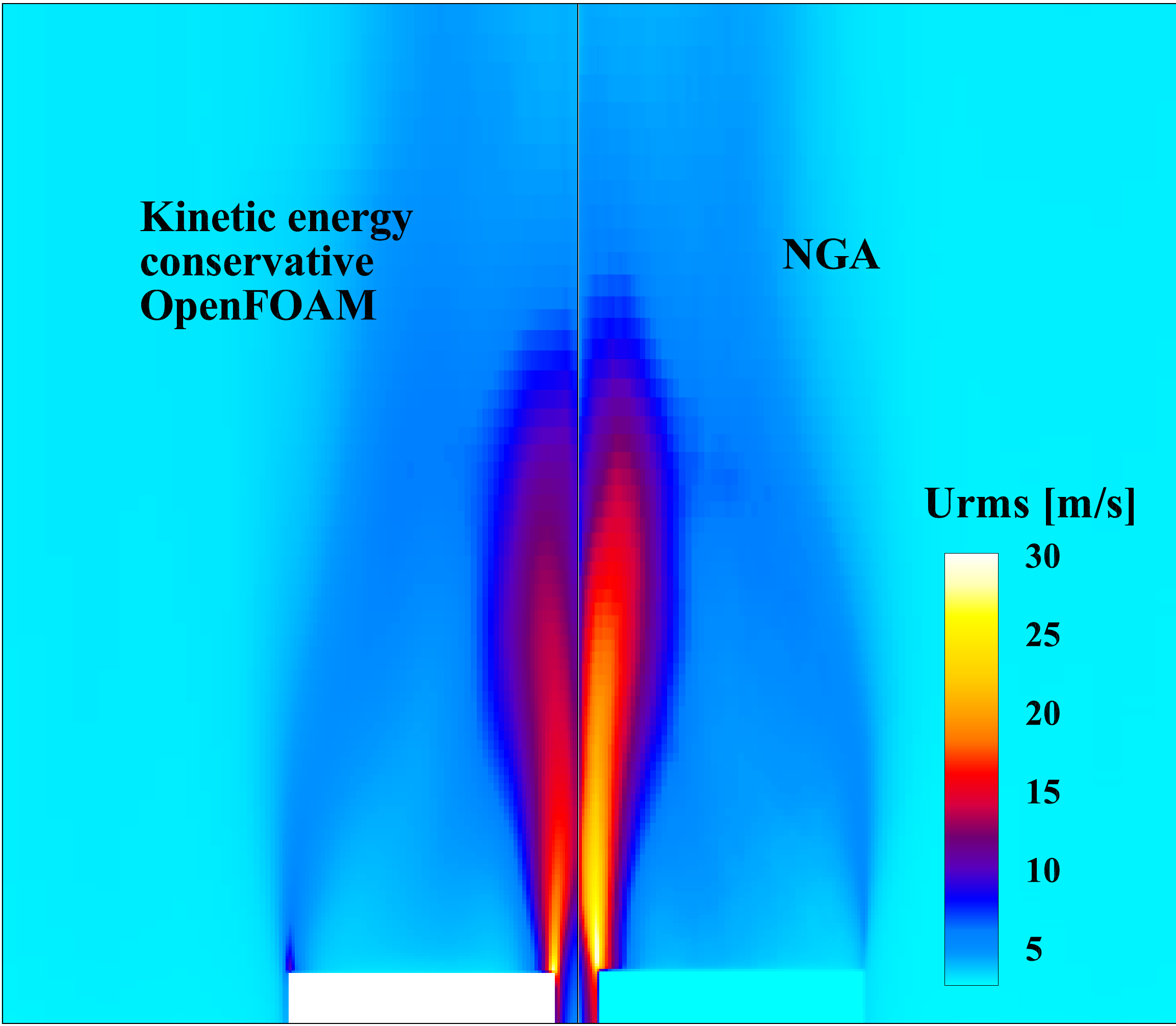}
\includegraphics[width=0.45\textwidth,trim={0cm 0cm 11cm 0cm},clip]{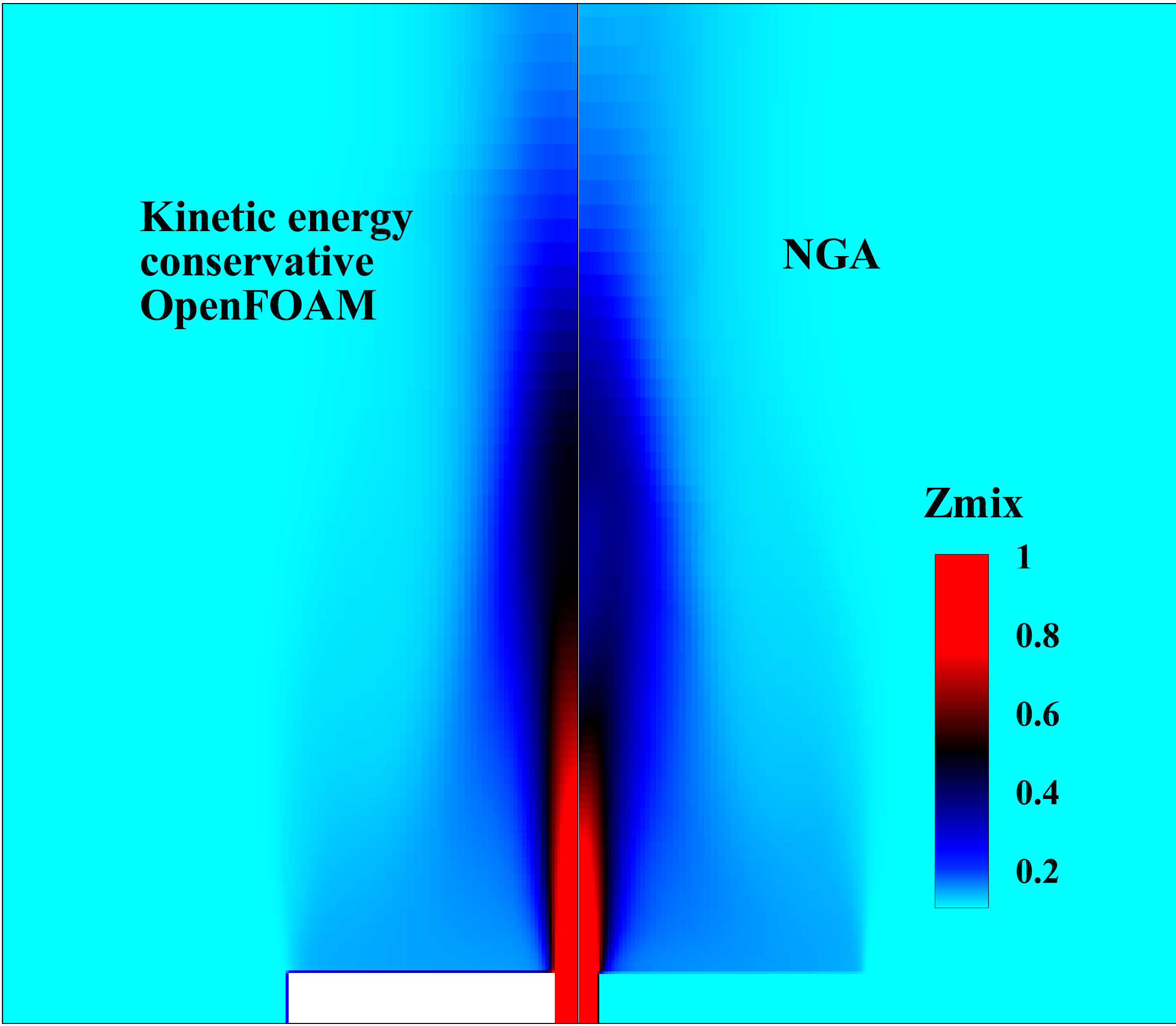}
\caption{Comparison of the axial RMS velocity between the KEcons solver and NGA (left) and comparison of the mean mixture fraction between the KEcons solver and NGA (right)}
\label{fig:MeanCompcons}
\end{figure}

\begin{figure}
\center
% left bot right top
\includegraphics[width=0.49\textwidth,trim={3cm 0cm 4cm 1cm},clip]{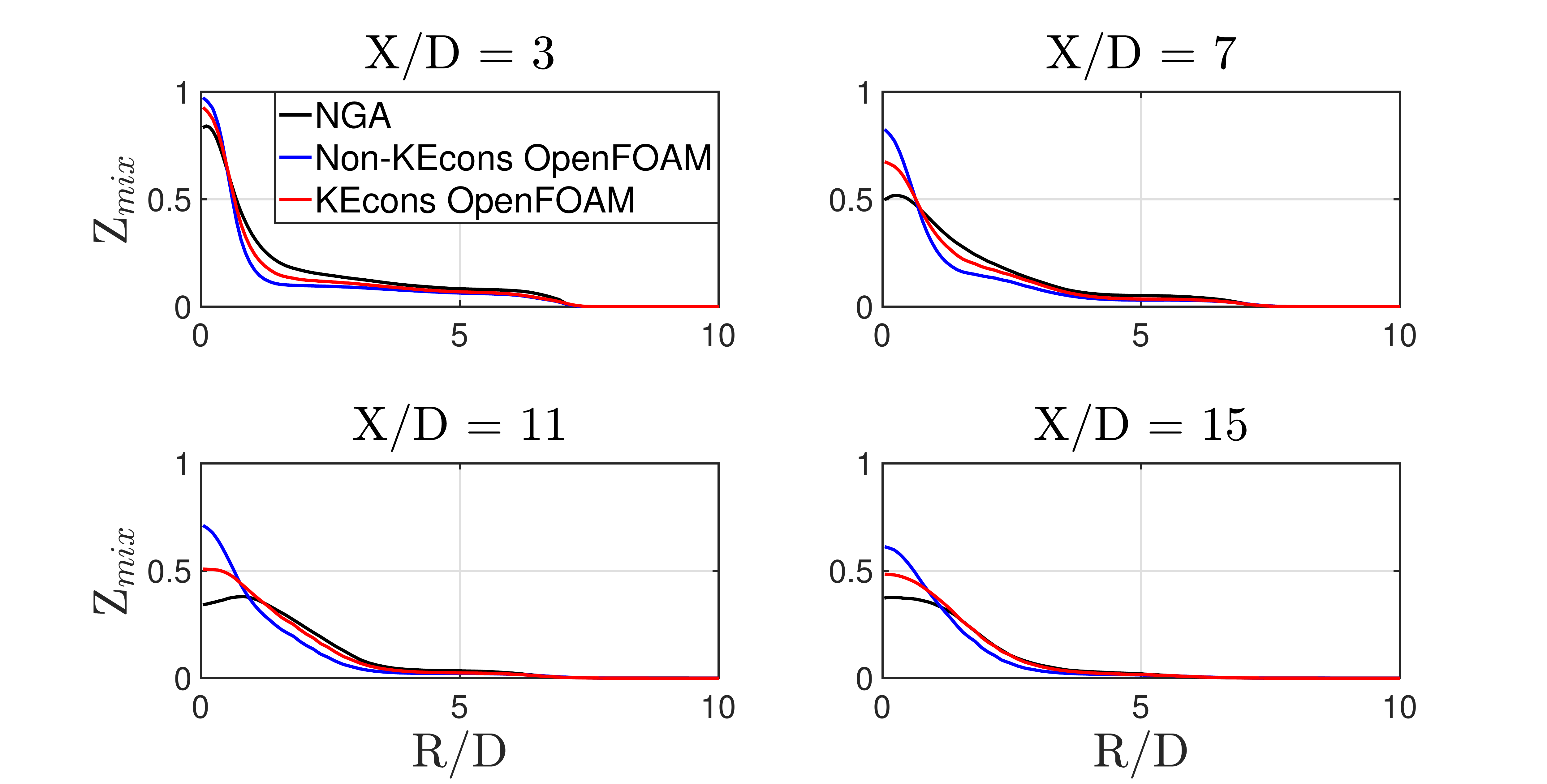}
\includegraphics[width=0.49\textwidth,trim={3cm 0cm 4cm 1cm},clip]{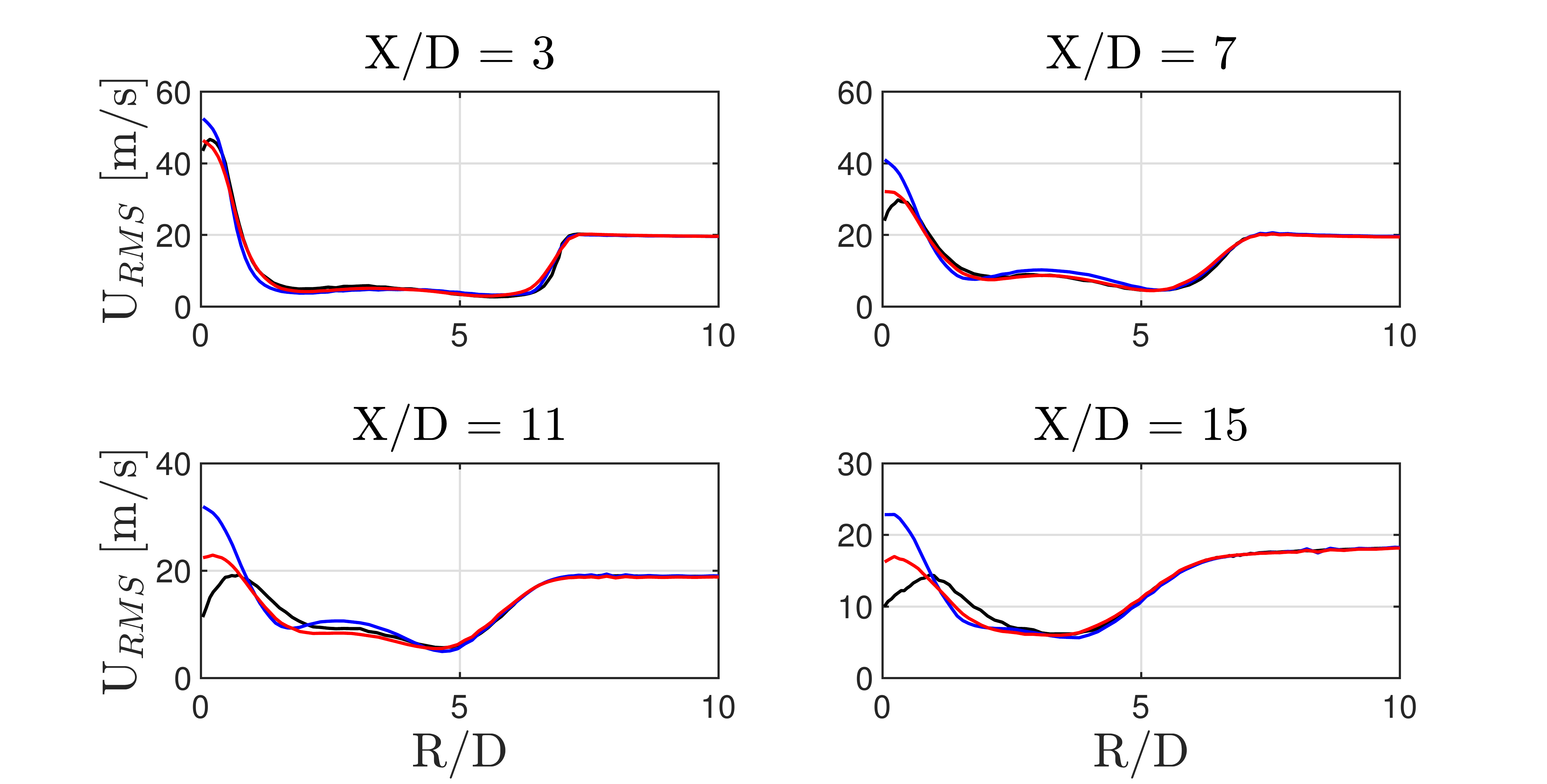}
\caption{1-D line plot comparison between NGA, the KEcons solver and the non-KEcons solver for mixture fraction (left) and axial RMS velocity (right)}
\label{fig:1dbb}
\end{figure}

The mean mixture fraction field and the RMS axial velocity are plotted against the NGA results in Fig.~\ref{fig:MeanCompcons}. The modified low-Ma solver shows visibly improved results in terms of the core jet length, as compared to the non-conservative solver (Fig.~\ref{fig:MeanComp}). This is consistent with the improvements seen in the TG vortex cases (Sec.~\ref{sec:resultstg}). The radial profiles of these statistics are plotted in Fig.~\ref{fig:1dbb}. Overall, the new solver produces results closer to the structured grid NGA solver, with the maximum improvement near the centerline. %The centerline difference mixture fraction difference with NGA has been divided by a factor 2 with the KEcons solver. %While the magnitude of the RMS velocity are closer to NGA, the improvement is not as significant as in the mixture fraction case. Most of the improvement can be seen in terms of the distribution of the RMS velocity in the radial direction. 

%{\color{red} Where do you define KECons solver as a name? Clarify or change captions and text accordingly.}

\subsection{Variable Density TG Vortex Case}
\label{sec:vardensTGV}

\label{sec:verifvardens}
%\subsection{Variable density Taylor-Green vortices \cite{shunn-MMS}}

%{\color{red} This contains viscous term - mention that KE will not be conserved, along with non-zero dilatation. We need to get the right evolution of KE}

The TG case can be modified to include density changes by including an additional transport equation. This manufactured method solution (MMS) technique corresponds to problem 3 in \cite{shunn-MMS}. The relevant parameters (see \cite{shunn-MMS} for details) are: $\rho_0 = 5; \rho_1 = 1; k = 2; \omega =2; u_F = 0; v_F = 0; \mu = 0.001$. Unlike the steady Taylor-Green vortex case in Sec.~\ref{sec:verifconsdens}, the flow field oscillates at the same rate as the density field. Similar to the case studied in Sec.~\ref{sec:veriflowmach}, the density field is directly obtained from a transported scalar field (analytical solution available in \cite{shunn-MMS}), for which a source term is provided. No additional source terms are used for the momentum or mass conservation equations. The analytical velocity field is now given by the following set of equations:
\begin{equation}
    u_x(x,y,t) =  \frac{\rho_1 - \rho_0}{\rho(x,y,t)} (\frac{-\omega}{4k}) cos(\pi k x) sin(\pi k y) sin(\pi \omega t),
\end{equation}
and
\begin{equation}
    u_y(x,y,t) =  \frac{\rho_1 - \rho_0}{\rho(x,y,t)} (\frac{-\omega}{4k}) sin(\pi k x) cos(\pi k y) sin(\pi \omega t).
\end{equation}

\begin{figure}
\center
% left bot right top
%\includegraphics[width=0.336\textwidth,trim={10cm 0cm 10cm 0cm},clip]{./Figures/2DTGVardensstructbis-eps-converted-to.pdf}
%\includegraphics[width=0.49\textwidth,trim={10cm 0cm 2cm 0cm},clip]{./Figures/2DTGVardensunstructbis-eps-converted-to.pdf}
\includegraphics[width=0.327\textwidth,trim={10cm 0cm 10cm 0cm},clip]{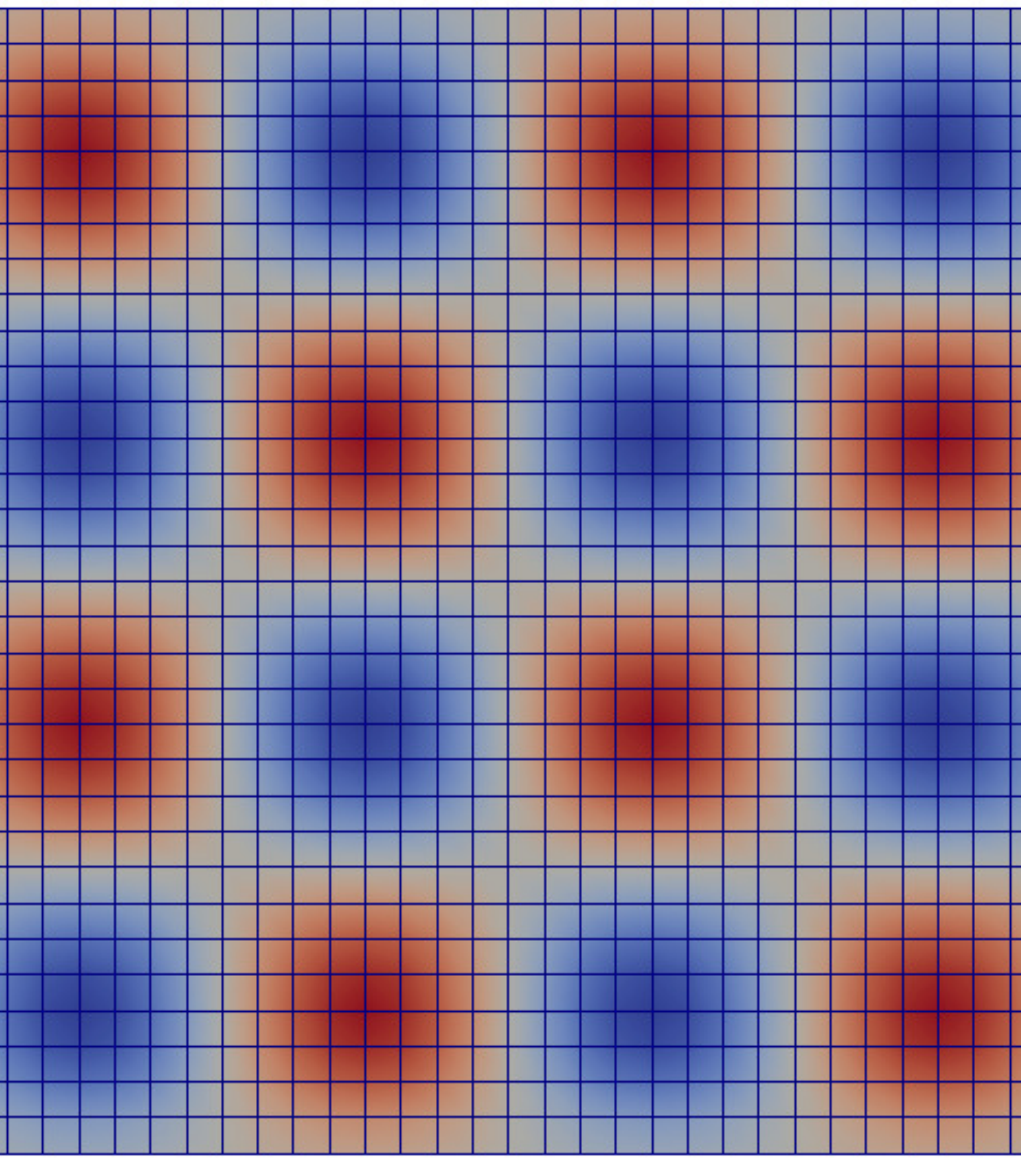}
\includegraphics[width=0.49\textwidth,trim={10cm 0cm 1.25cm 0cm},clip]{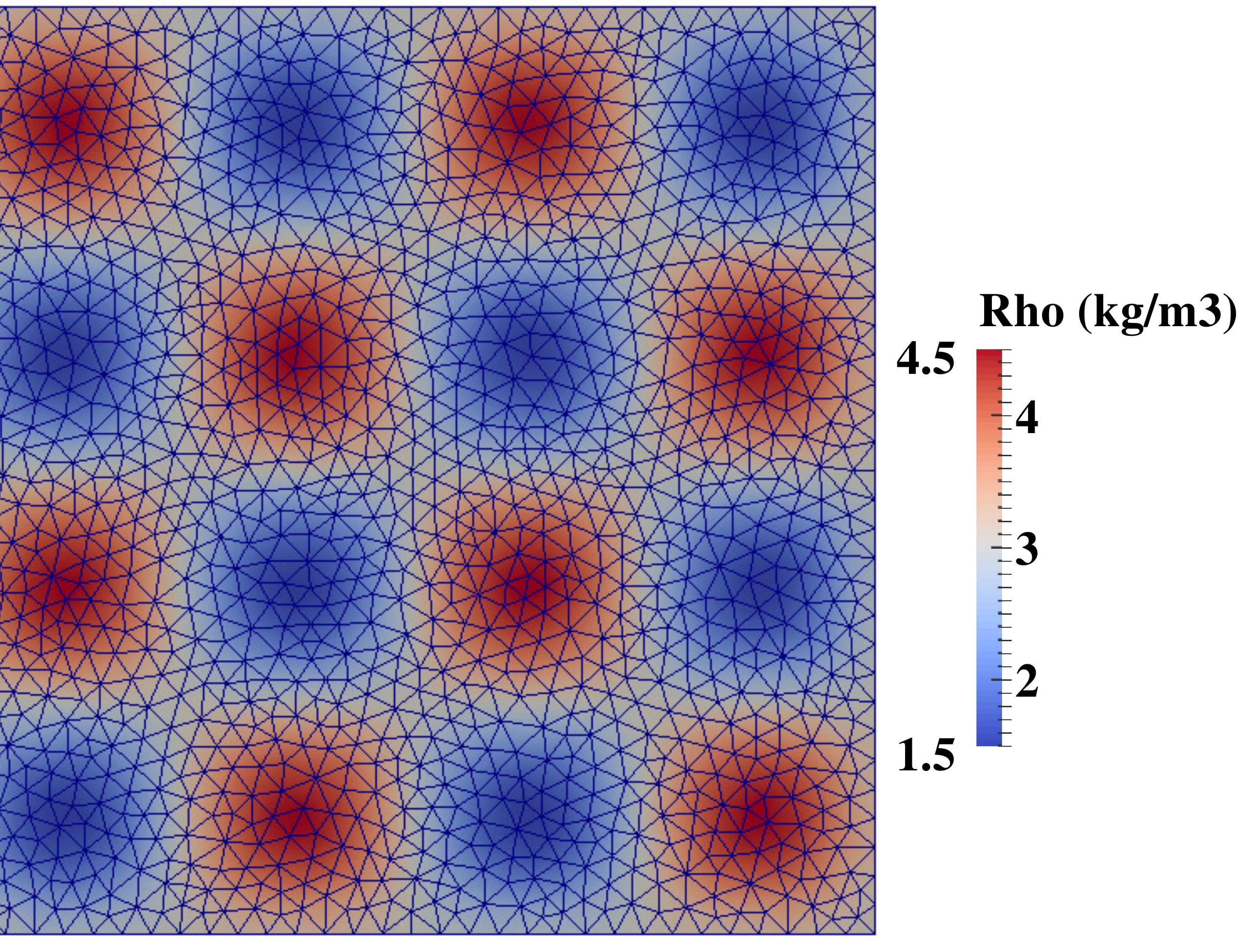}
\caption{Density field at t=0.1s of the 2D variable density MMS configuration using with the structured mesh (left) and the unstructured mesh with $\Delta x  = 0.0625$ m (right) cases.}
\label{fig:ill2dtgvardens}
\end{figure}

For the solver to be accurate, it is important to capture the coupling between the density changes and the velocity field. Since the momentum equation does not have a direct forcing term, its temporal variations are a result of the density changes alone. For these tests, the maximum convective CFL number is held constant at 0.15 in a fashion similar to \cite{shunn-MMS}. An instantaneous density field is shown in Fig.~\ref{fig:ill2dtgvardens}. It was first found that the variable density solver using the hybrid formulation of the pressure correction and the non-KEcons solver were prone to instabilities.

\begin{figure}
\center
% left bot right top
\includegraphics[width=0.55\textwidth,trim={1.5cm 0cm 4cm 0cm},clip]{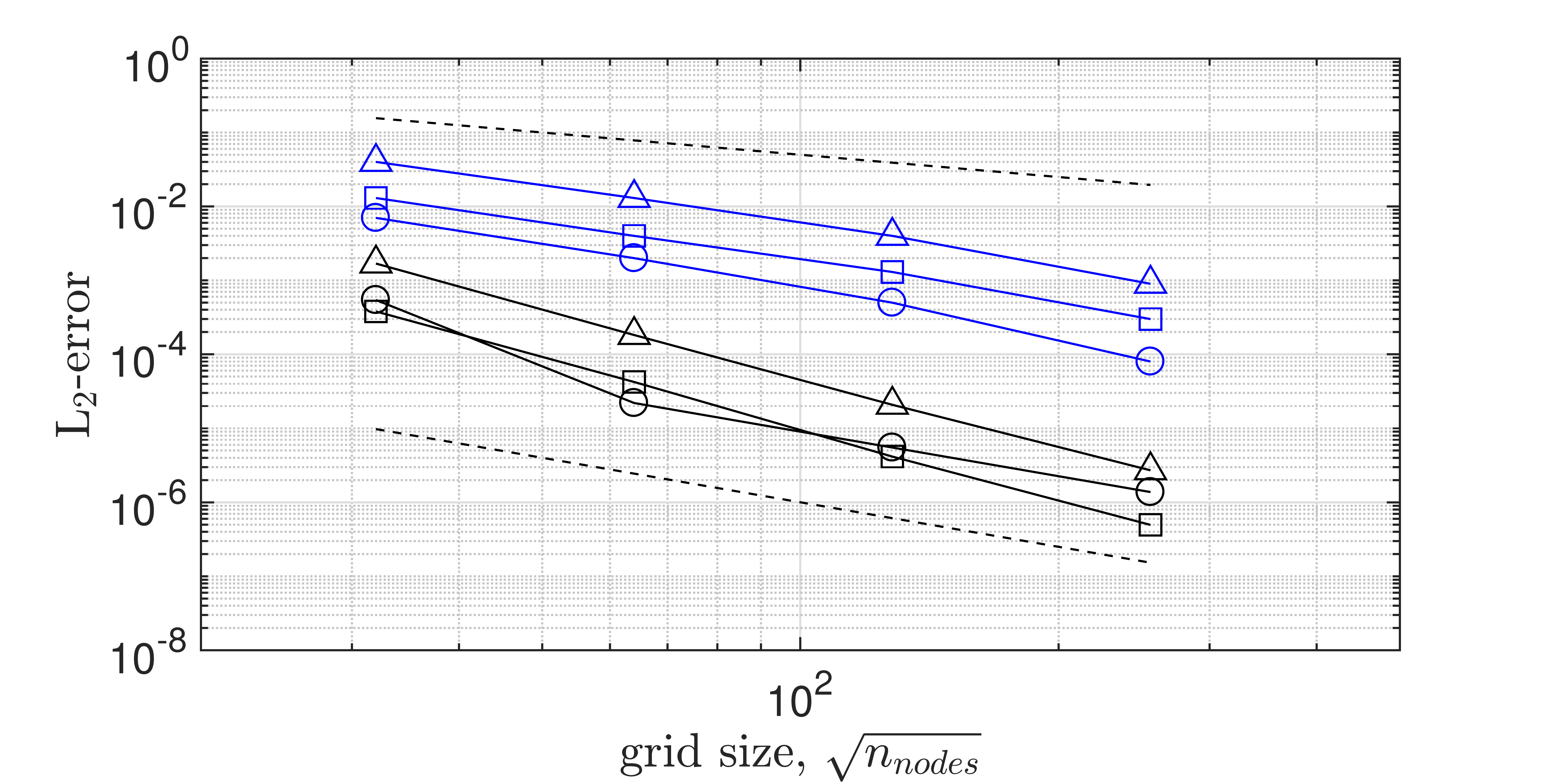}
\caption{Spatial convergence of the error between the numerical and the analytical MMS solution for the structured mesh case, compared with the results of Shunn et al. \cite{shunn-MMS}. \mybarredcircle{black} velocity convergence; \mybarredcircle{blue} velocity convergence in Shunn et al. \cite{shunn-MMS}; \mybarredsquare{black} scalar convergence, \mybarredsquare{blue} scalar convergence in Shunn et al. \cite{shunn-MMS}; \mybarredtriangle{black} density convergence, \mybarredtriangle{blue} density convergence in Shunn et al. \cite{shunn-MMS}.}
\label{fig:2dspaceconvhex}
\end{figure}

\begin{figure}
\center
% left bot right top
\includegraphics[width=0.55\textwidth,trim={1.5cm 0cm 4cm 0cm},clip]{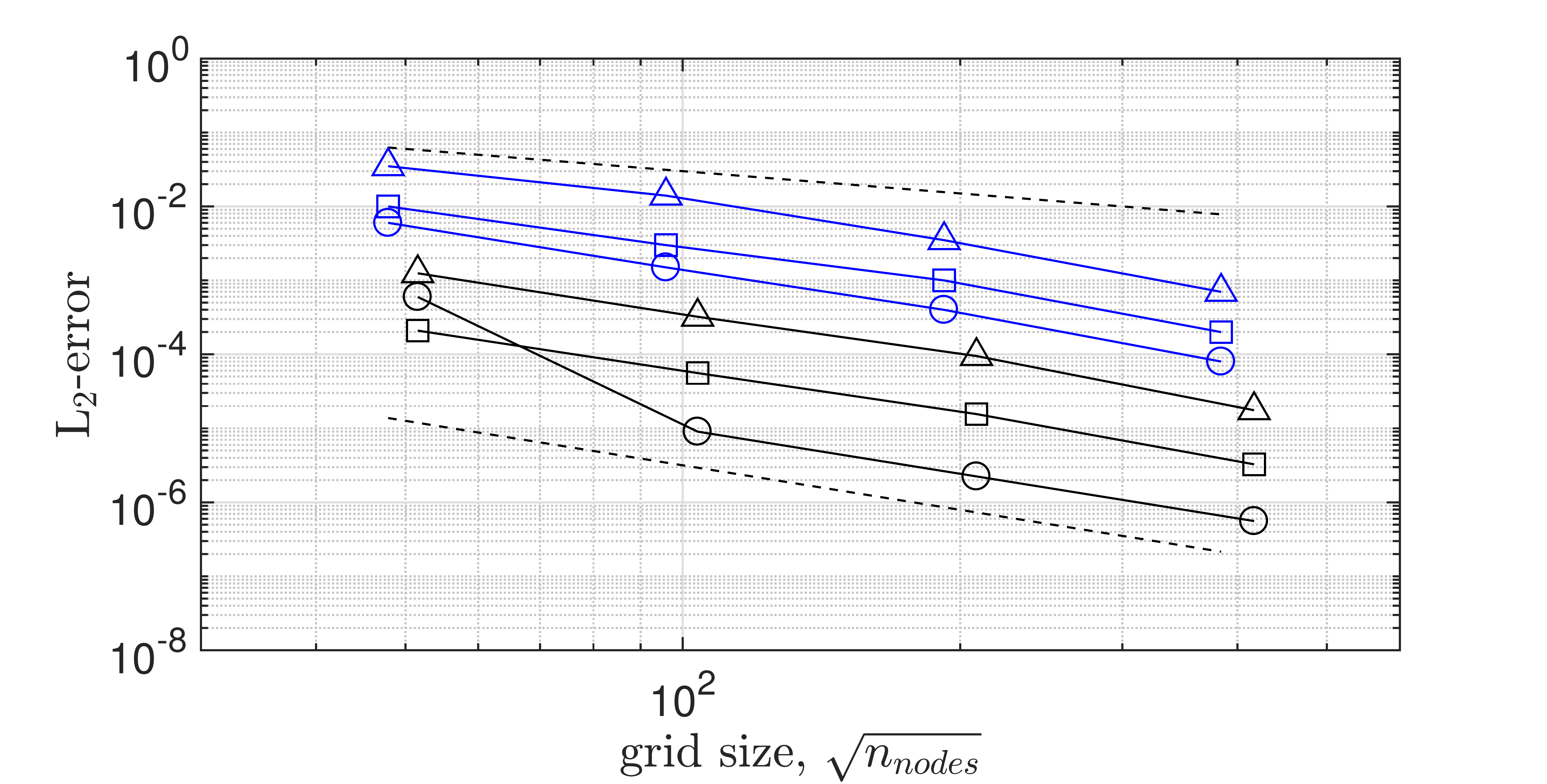}
\caption{Spatial convergence of the error between the numerical and the analytical MMS solution for the unstructured mesh case, compared with the results of Shunn et al. \cite{shunn-MMS}. \mybarredcircle{black} velocity convergence; \mybarredcircle{blue} velocity convergence in Shunn et al. \cite{shunn-MMS}; \mybarredsquare{black} scalar convergence, \mybarredsquare{blue} scalar convergence in Shunn et al. \cite{shunn-MMS}; \mybarredtriangle{black} density convergence, \mybarredtriangle{blue} density convergence in Shunn et al. \cite{shunn-MMS}.}
%\caption{Spatial convergence of the error between the numerical and the analytical MMS solution for the unstructured mesh case, compared with the results of Shunn et al. \cite{shunn-MMS}. Velocity convergence (black circles), velocity convergence in Shunn et al.(blue circles); Scalar convergence (black squares), scalar convergence in Shunn et al.(blue squares); Density convergence (black triangles), density convergence in Shunn et al.(blue triangles).}
\label{fig:2dspaceconvtri}
\end{figure}

Using the new low-Mach number solver, second order accuracy in space could be obtained as shown in Fig.~\ref{fig:2dspaceconvhex}. This is consistent with tests of similar collocated numerical scheme presented in \cite{shunn-MMS}. The same configuration is simulated using a 2D grid composed of triangular cells. Similar to the constant density TG case, this tetrahedral mesh based simulation is more applicable to practical OpenFOAM computations. The spatial convergence remains second order (Fig.~\ref{fig:2dspaceconvtri}), which is encouraging for practical application. 

%It is important to note that the method of Shunn et al.\,\cite{shunn-MMS} generally produces lower errors for each mesh size, although the convergence rate is similar to that of the solver implemented in the current work. 

The results are displayed next to the ones generated by Shunn et al. \cite{shunn-MMS} who used non-zero convective velocity parameters $u_F$ and $v_F$. The interpolation methods are also likely to be different since Shunn et al. \cite{shunn-MMS} use a node-based code, meaning that variables are stored at the nodes rather than the cell centers. The discretization of the scalar source terms also plays a large role in the solver accuracy especially in MMS procedure. Depending on the time at which the source terms have been computed ($n$ or $n+1/2$), significant accuracy differences can be observed. One should therefore not infer any conclusions from the magnitude of errors between both solvers but rather focus on the rate of convergence for the error, which appears to be 2$^{nd}$ order in space.

\begin{figure}
\center
% left bot right top
\includegraphics[width=0.49\textwidth,trim={0cm 0cm 0cm 0cm},clip]{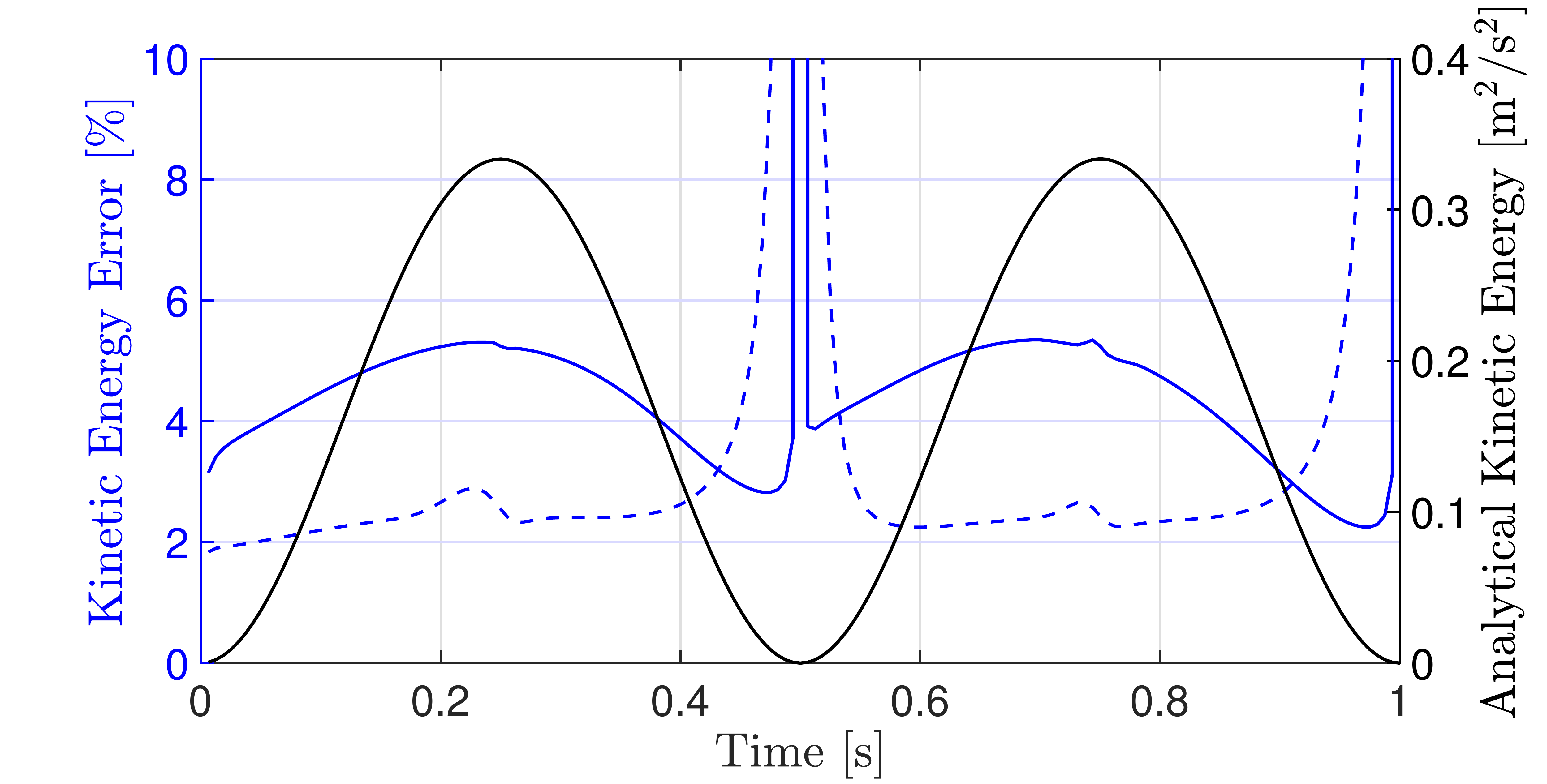}
\includegraphics[width=0.49\textwidth,trim={0cm 0cm 0cm 0cm},clip]{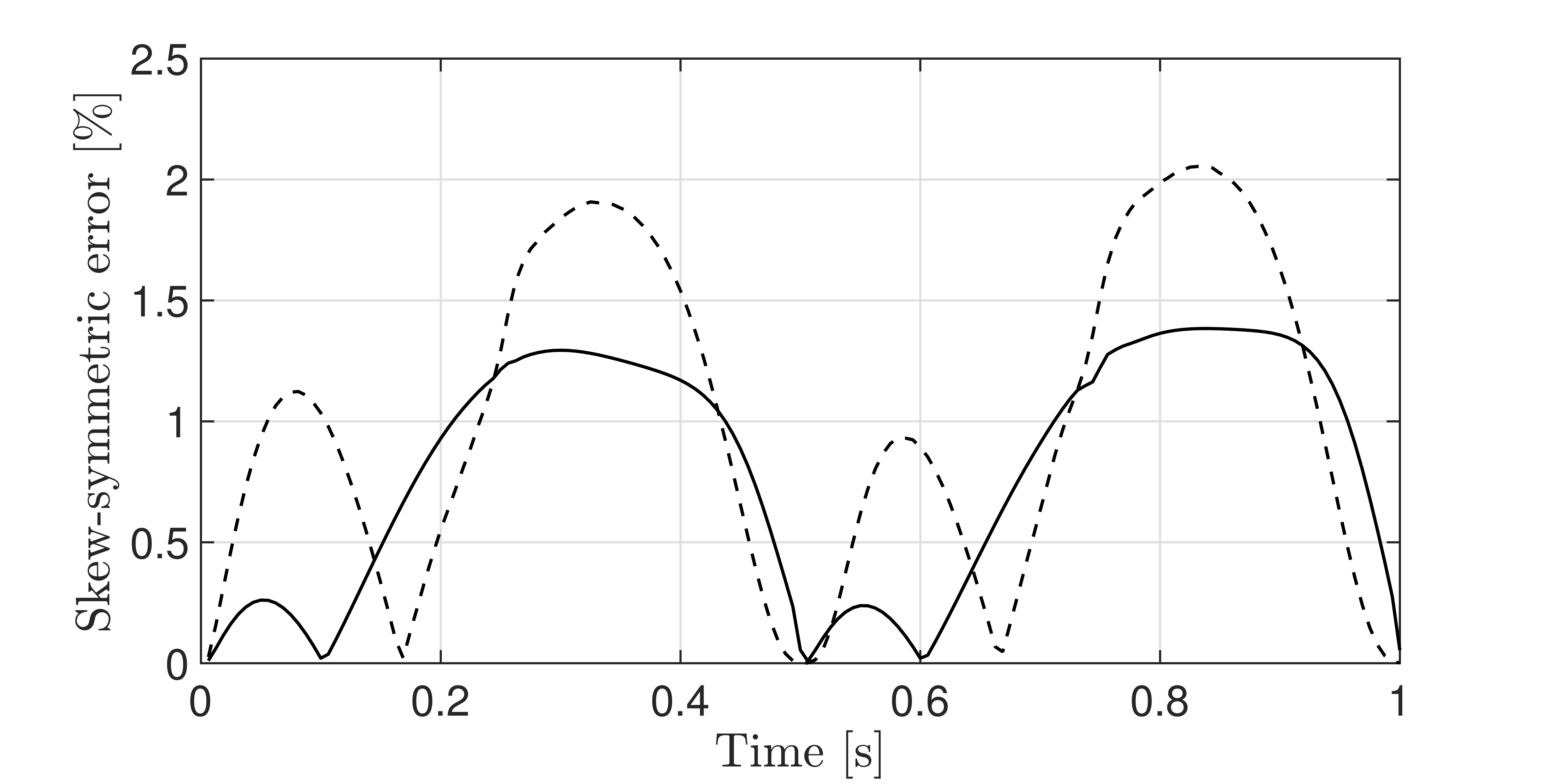}
\caption{Left: global KE of the 2D analytical MMS solution (\mythickline{black}) as a function of time plotted along with percentage error in KE for the square mesh case (\mythickline{blue}) and the triangular mesh case (\mythickdashedline{blue}). Right: skew-symmetric error contribution to the total KE error for the structured (\mythickline{black}) and the triangular mesh (\mythickdashedline{black}).}
\label{fig:skewsymerr}
\end{figure}

Since the test case contains viscosity, and non-zero dilatation, KE will not be conserved. Instead, the solver should minimize the difference between the analytical and computed KE values. In order to verify KE conservation, the dot product of the time and convective terms of the momentum equation with $\boldsymbol{\hat{u}}$ is computed, and is effectively $\frac{\overline{\rho}^{t,n+1} (u^2)^{n+1} - \overline{\rho}^{t,n} (u^2)^{n}}{\Delta t}$. Any error comes from an inaccurate estimation of $u^{n+1}$ (explicit term of the momentum convection) or from a lack of primary mass conservation in the continuity equation. In Fig.~\ref{fig:skewsymerr}, the contribution of this error is plotted along with the total KE error for the 1024 cells case using a timestep of $6.25 ms$. This error is called the skew-symmetric error, similarly to Sec.~\ref{sec:vardensTGV}. It can be observed that most of the KE error does not come from the skew-symmetric error but from other components of the equations solved. This gives confidence in the ability of the new methodology to ensure energy conservation for the time derivative and the momentum convection terms.

%\begin{figure}[H]
%\center
% left bot right top
%\includegraphics[width=0.9\textwidth,trim={0cm 0cm 0cm 0cm},clip]{./Figures/HexMesh32.png}
%\caption{2D Hexahedral Mesh, 32}
%\label{fig:2dmeshhex}
%\end{figure}

%{\color{red} Plot KE error in \% so it is clear that error is small}

\subsection{Complex geometry with density model and non-mass conservative flow}

% Alex's Notes
% DLR Configuration and Mesh
% Table on flow parameters
% Contour plot of velocity ??
% Figures of comparison for different axial location for Velocity, RMS Vel and Soot VF between cons and non-cons schemes

In order to demonstrate the solver performance in a practical geometry, combustion and soot formation in a pressured model aircraft combustor is simulated. The flow configuration is based on the experimental set up at DLR \cite{geigle2014soot}. Here, the new solver is used to simulate combustor operation at 3 bar pressure.

\begin{figure}[htb]
\center
% left bot right top
\includegraphics[width=0.40\textwidth,trim={0cm 0cm 0cm 0cm},clip]{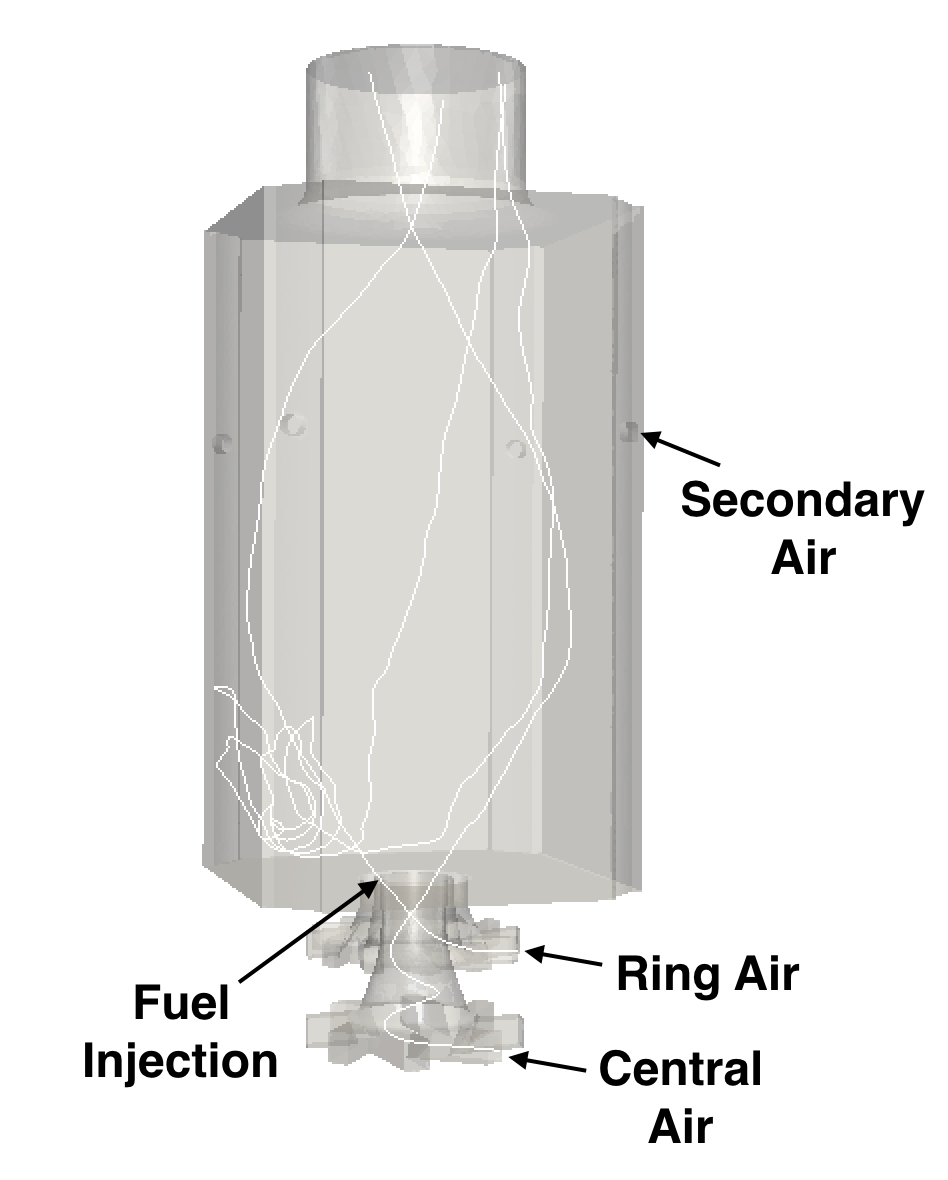}
\caption{DLR combustor geometry with air inlets marked. White lines are particle trajectories from Lagrangian approach originating from the inlets, showing streamlines of velocity in the combustor.}
\label{fig:DLRgeom}
\end{figure}

The combustor is similar to the one used by Geigle et al.~\cite{geigle2015investigation} and Koo et al.~\cite{koodlr}. The combustor geometry is shown in Fig.~\ref{fig:DLRgeom} with typical streamline in the combustor from inlet to outlet. It was designed for operation at 10 kW/bar power and installed with large optical access for simultaneous laser acquisition of velocity, temperature, species mass fractions and soot volume fraction. It has a cross-sectional area of 68 $\times$ 68 mm$^2$, with a height of 120 mm. The inflow consists of three concentric nozzles: two room temperature air inlets with swirling velocity and 60 annular straight channel fuel ($C_2H_4$) inlet in between the two air flows with a size of 0.5 $\times$ 0.4 mm$^2$ each. A single constricted exit of diameter 40 mm removes the combustion products. At 80 mm height, four additional air ducts of 5 mm diameter inject secondary air into the combustor radially, meeting at the combustor central axis, forming a stagnation point. Due to the high pressure, combined with locally rich fuel-air conditions, significant soot formation is observed. Estimation of soot profiles is the main target of these computations. Due to the extreme sensitivity of soot formation to local thermochemical conditions, as well as the trajectory of the fluid particles within the combustor, the use of temporally accurate LES solvers become important. In this study, the specific case of 3 bar operation with 460.3 slpm of primary air, 39.3 slpm of ethylene fuel and 187.4 slpm of secondary oxidation air is considered. The global equivalence ratio is approximately 0.86, which is considerably below the sooting limit. Hence, any soot particle observed is generated due to local inefficiencies in mixing and oxidation of the fuel.

\begin{figure}[htb]
\center
% left bot right top
\includegraphics[width=0.50\textwidth,trim={0cm 0cm 0cm 0cm},clip]{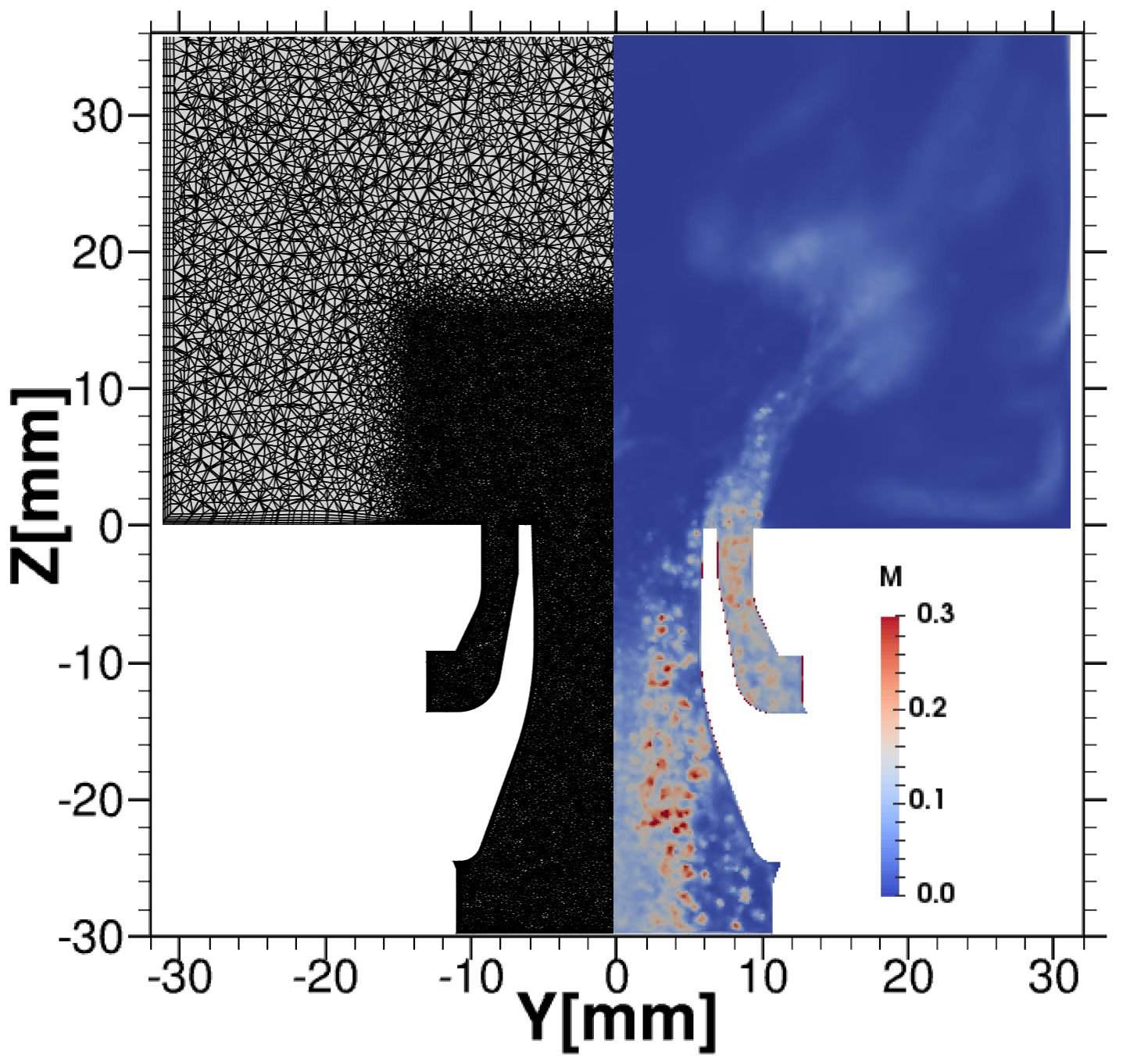}
\caption{Mesh refinement near the inlet and wall plotted along with $M$ field.}
\label{fig:DLRconfig}
\end{figure}

The computational domain consists of approximately 12 million tetrahedral cells, with grid refinement applied near the inlets and the near-wall region (Fig.~\ref{fig:DLRconfig}). In LES, it is necessary to resolve the large-scale structures, and a metric for such refinement is the Pope criterion \cite{popebook}. A fractional energy $M$, defined as the ratio betwen the sub-filter KE and the total KE is used to determine the resolution adequacy. Since the sub-filter KE is not directly available from the resolved fields, a model is used to estimate this quantity. The cell-size is refined until this ratio is below 0.30 everywhere in the domain. Note that even with this requirement, there might be instantaneous $M$ values that are above the cut-off threshold. Figure~\ref{fig:DLRconfig} shows the $M$ field plotted for the final grid used in this study.

The low-Mach number LES equations are solved along with a flamelet-based model for turbulent combustion, and moments-based model for soot particle evolution \cite{mueller2012model}. This requires additional transport equations for mixture fraction, progress variable, enthalpy, and soot moments. details of the model can be found in \cite{koodlr,mueller_pof}.

The LES governing equations are obtained by Favre-filtering the momentum and scalars including soot moments. Unclosed terms in the subfilter flux are closed with the dynamic SGS model \cite{gpmc91}. Filtered chemistry reaction terms for the enthalpy equation are closed using the presumed-PDF approach described in Mueller et al.~\cite{mueller2012model}. For these tests, the timestep is held constant at 5 $\times$ 10$^{-7}$ s. The LES simulation was performed on 2048 cores with time-averaged data taken after 10 flow-through times, totaling approximately 80 wall-clock hours.

\begin{figure}[h]
\center
% left bot right top
\includegraphics[width=1.0\textwidth,trim={0cm 0cm 0cm 0cm},clip]{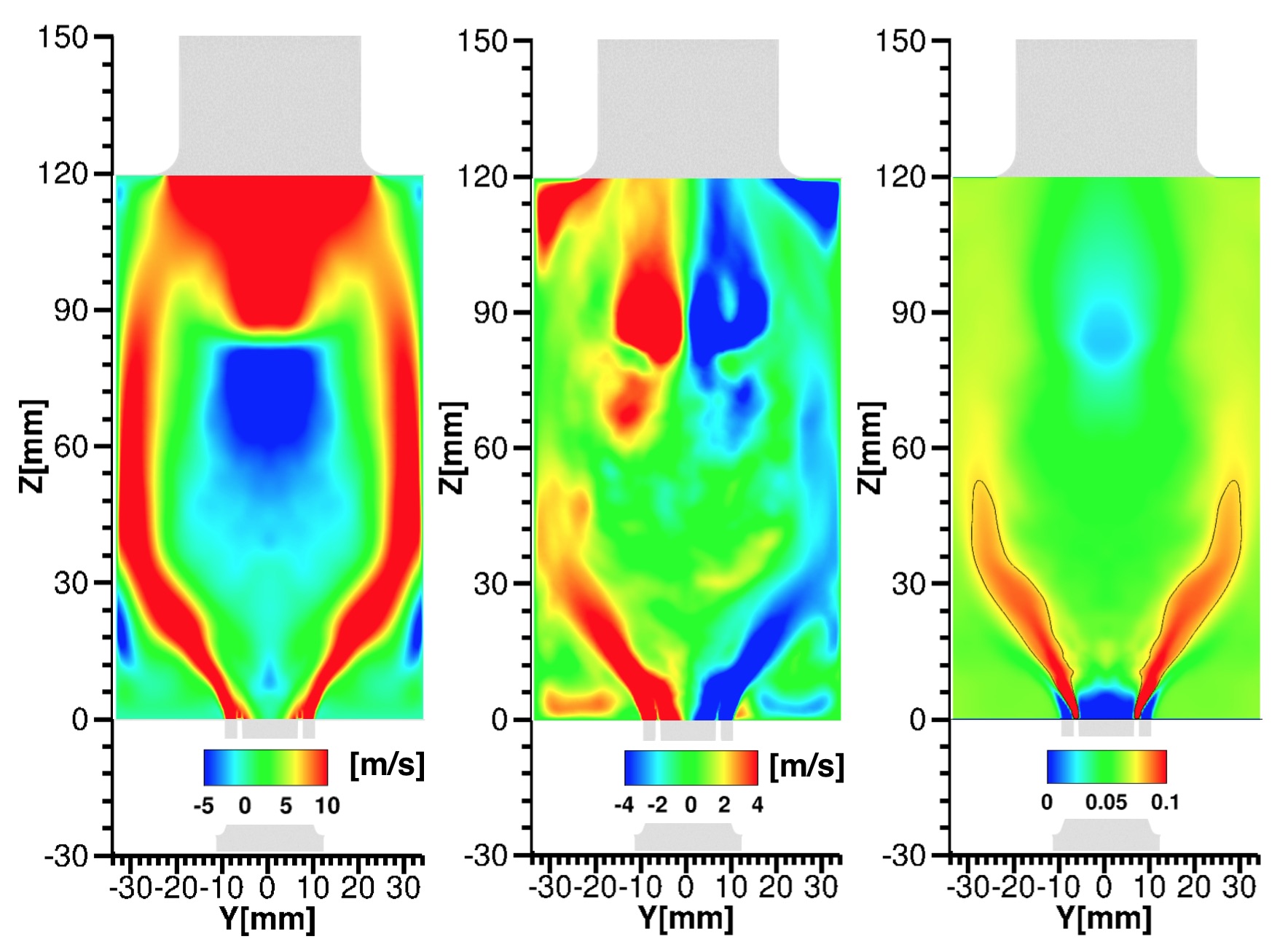}
\caption{Mean axial velocity (left), mean tangential velocity (center), and mixture fraction with stoichiometric line (right).}
\label{fig:DLR_Contour}
\end{figure}

Figure~\ref{fig:DLR_Contour} shows time-averaged two-dimensional images of axial velocity, tangential velocity and mixture fraction, obtained using the minimally-dissipative solver. It can be seen that the flow structure consists of a large inner recirculation zone (marked by negative axial velocity component), as well as a smaller but persistent outer recirculation zone. The size and strength of these recirculation zones determine the level of mixing, and the tendency of the combustor to form soot particles.

\begin{figure}[h]
\center
% left bot right top
\includegraphics[width=1.0\textwidth,trim={0cm 0cm 0cm 0cm},clip]{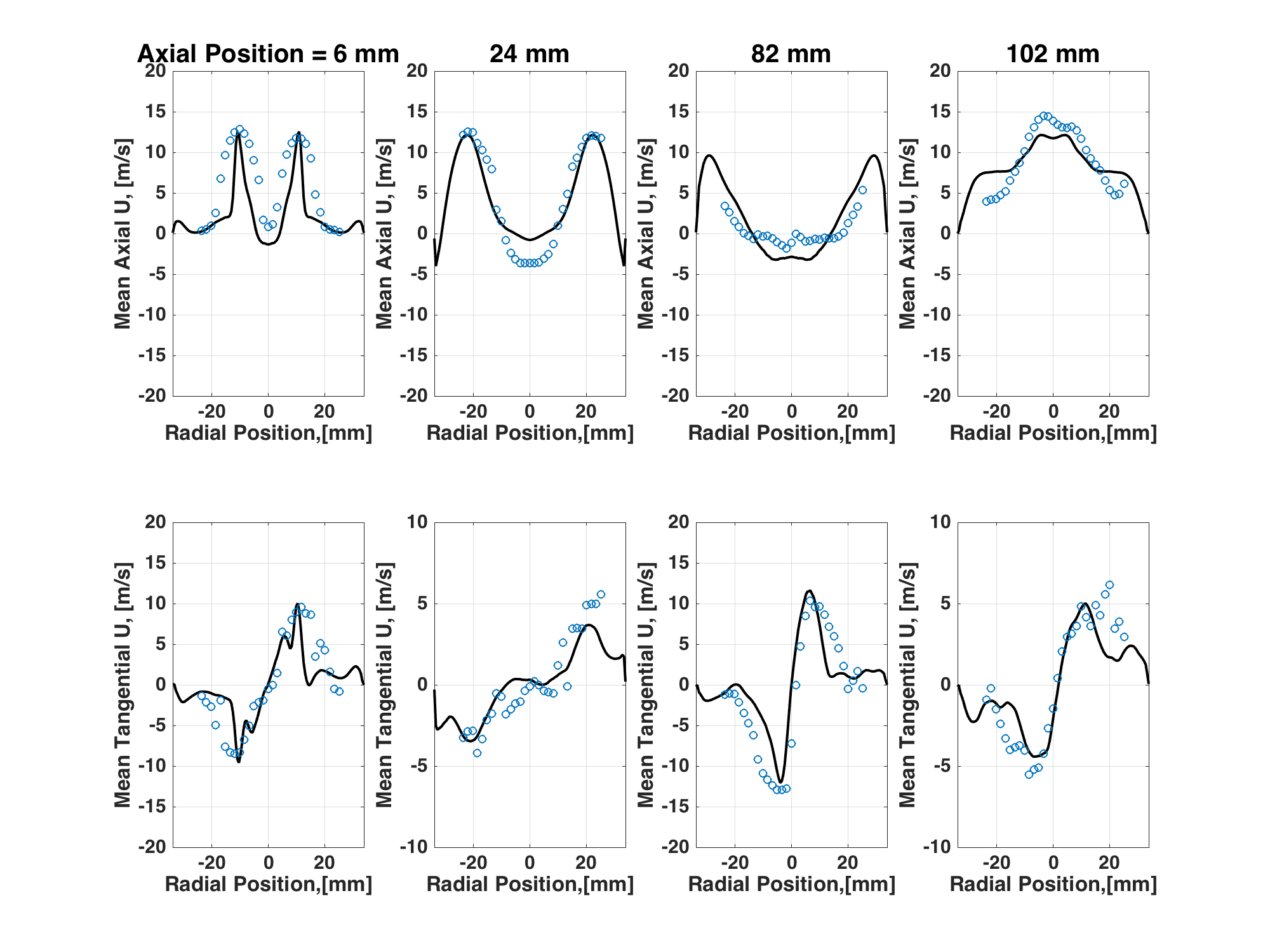}
\caption{Time-averaged axial (top) and tangential (bottom) velocity profiles at selected axial locations. Simulation (\mythickline{black}) and experimental data (\mycircle{blueviolet}).}
\label{fig:DLRMeanU}
\end{figure}

Figure~\ref{fig:DLRMeanU} shows time-averaged velocity statistics compared against experimental data. The time-averaged axial velocity is well-captured by the solver, including the presence of the high-velocity regions along the fuel injection path. Further, the low-velocity recirculation region is also captured accurately. Similarly, the mean tangential velocity shows very good agreement with the experimental data. In particular, the swirl component that determines the flame stabilization mechanism is well represented by the LES solver.

\begin{figure}[h]
\center
% left bot right top
\includegraphics[width=1.0\textwidth,trim={0cm 0cm 0cm 0cm},clip]{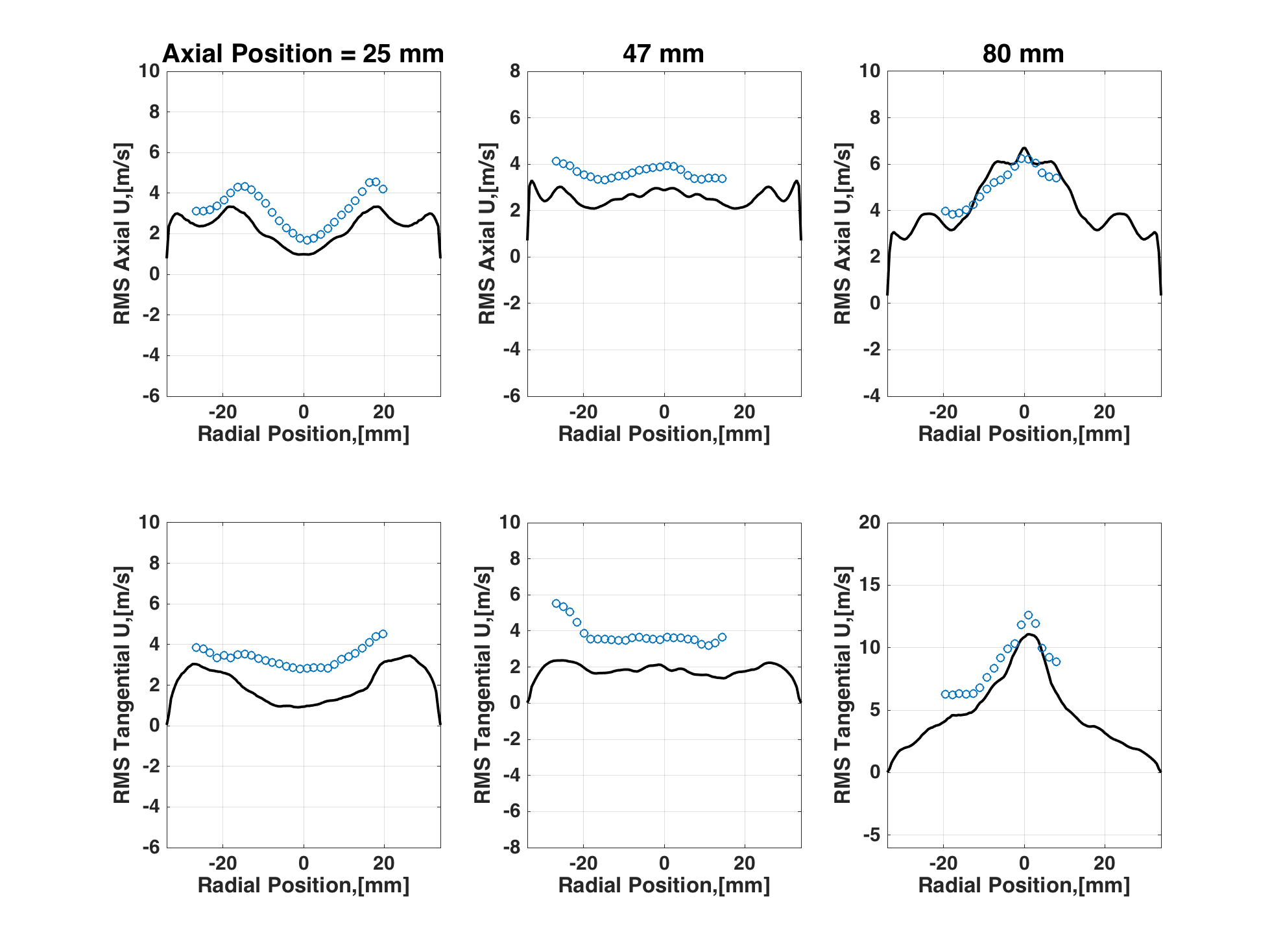}
\caption{RMS axial (top) and tangential (bottom) velocity profiles at selected axial locations. Simulation (\mythickline{black}) and experimental data (\mycircle{blueviolet}).}
\label{fig:DLRRMSU}
\end{figure}

Perhaps, of more interest is the root-mean square (RMS) velocity, which is directly tied to the ability of the LES solver to maintain turbulent fluctuations from dissipating through numerical inaccuracies. Figure~\ref{fig:DLRRMSU} shows that at all axial locations considered, the RMS velocity of the both the axial and tangential components are well captured. This provides confidence that the method is not dissipative even for complex geometries, and the turbulent structures remain accessible on unstructured grids as well.

\section{Conclusion}

A minimally-dissipative low-Mach number solver has been implemented in OpenFOAM. It was established that the default hybrid pressure correction approach leads to significant errors in mass conservation in a discrete sense, and was numerically unstable for certain flow conditions. The skew-symmetric formulation of the low-Mach number solver ensured discrete conservation of mass, while reducing dissipation of quadratic quantities. For variable density and reacting flows problems, a formulation for coupling scalar transport equations was also implemented. Verification cases demonstrated that the new solver significantly reduces KE dissipation even for highly skewed meshes. Further, the method is shown to be second-order accurate for variable density flow problems in structured and unstructured mesh cases. 

Since LES computations are highly sensitive to numerical errors, ensuring such primary and secondary conservation properties will help isolate modeling and numerical errors, leading to more useful model validation exercises. The algorithm developed will benefit from formulations for a suitable pressure gradient treatment on collocated meshes. In conclusion, the current study establishes a reliable and open source solver for LES with applications in complex turbulent reacting flows.

%{\color{red} Finish when everything is in place.}

\newpage
%\bibliography{bibtex_database}
%\bibliographystyle{aiaa}
%\newpage

\appendix

\section{Distinction between PISO and fractional timestep method}
\label{sec:pisoFts}

The fractional timestep \cite{kimmoin-projection} and PISO \cite{issa-piso} algorithm follow the same idea, which relies on enforcing mass conservation with a pressure correction term. Below, the distinction between the two approaches is described.

The fractional timestep method is the most intuitive method but the least-implicit approach, in the sense that not all terms of the momentum equation are updated using the same velocity vector. First, the velocity at the new timestep is guessed (and is denoted by $\boldsymbol{u}^g$) using the most updated estimate for the pressure term. Then a discrete momentum equation is written using  $\boldsymbol{u}^g$ everywhere except in the time derivative term. The viscous term is dropped without changing the purpose of the discussion.

\begin{equation}
    \label{eq:predictor}
    \frac{\rho^{n+1} \boldsymbol{u}^g - \rho^n \boldsymbol{u}^n}{\Delta t} + \boldsymbol{\mathcal{C}}^g = -\nabla p^n.
\end{equation}

The convection and dissipation terms are written implicitly, leading to an equation for $\boldsymbol{u}^g$, which is an estimated velocity field. Using $\boldsymbol{u}^g$ in the convection term, the fractional timestep procedure can be progressed as:

\begin{equation}
    \frac{\rho^{n+1} \boldsymbol{u}^* - \rho^n \boldsymbol{u}^n}{\Delta t} + \boldsymbol{\mathcal{C}}^g = 0.
\end{equation}
This equation is used to obtain $\boldsymbol{u}^*$ and subsequently, a pressure equation is formulated as
\begin{equation}
\label{eq:ftsmom}
    \frac{\rho^{n+1} \boldsymbol{u}^{n+1} - \rho^{n+1} \boldsymbol{u}^*}{\Delta t} =  -\nabla p^{n+1}.
\end{equation}
The above equation can be used to solver for pressure, as well as the velocity field $\boldsymbol{u}^{n+1}$.

The PISO procedure follows the same idea but splits the derivative operator between a part treated implicitly and a part treated explicitly. The first step of the method is exactly similar to Eq.~\ref{eq:predictor}. The following correction step is formulated differently. Each operator (convection, diffusion, time) is treated as a sparse square matrix of size equal to the number of cells. This matrix is split between components that will be treated implicitly ($A$) and explicitly ($H$).

The pressure correction procedure is initialized as 
\begin{equation}
(A_{time}+A_{conv}) [\boldsymbol{u}^*] + H_{conv} [\boldsymbol{u}^g]  = 0,
\end{equation}
where $A_{time}$ denotes the implicit part of the time derivative operator,  $A_{conv}$ and $H_{conv}$ denote respectively the implicit and explicit part of the convection operator, and $[\boldsymbol{u}]$ denotes the full velocity field.

The fractional velocity $\boldsymbol{u}^*$ is then corrected into a mass conservative velocity  $\boldsymbol{\overline{u}}$ using a pressure correction $[\nabla p_1]$ field.
\begin{equation}
(A_{time}+A_{conv}) ([\boldsymbol{\overline{u}}]-[\boldsymbol{u}^*])  = -[\nabla p_1].
\end{equation}
In order to achieve second order accuracy in time, another layer of correction is needed. Therefore next, $\boldsymbol{\overline{u}}$ is used as a guess in the exact same way $\boldsymbol{u}^g$ was used.
\begin{equation}
(A_{time}+A_{conv}) [\boldsymbol{u}^{**}] + H_{conv} [\boldsymbol{\overline{u}}]  = 0,
\end{equation}
\begin{equation}
\label{eq:pisomom}
(A_{time}+A_{conv}) ([\boldsymbol{u}^{n+1}]-[\boldsymbol{u}^{**}])  = -[\nabla p_2].
\end{equation}

Comparing Eq.~\ref{eq:pisomom} and Eq.~\ref{eq:ftsmom}, it is clear that the momentum equations are advanced differently in the PISO method and the fractional timestep method. The difference lies in the implicit treatment of some terms and the additional pressure correction layer in the PISO method.

\section{Application to non-mass-conserving systems}
\label{sec:nonmasscons}

Many reacting flow applications involve exchange of mass between different phases, especially in the context of dispersed-phase flows \cite{raman-dispersedphase,vervisch-weaklyspray,heye_proc}. In such cases, the continuity equation contains a source term that represents the mass added or removed from the continuous gas-phase. Here, the applicability of the low-Mach number minimally-dissipative solver to such systems is investigated.

%\subsubsection{Derivation of the KE equation for non mass-conservative solvers}

Assuming that mass addition only affects the continuity equation, the modified mass conservation equation can be written as:
\begin{equation}
 \frac{\partial \rho}{\partial t} + \nabla \cdot \rho \boldsymbol{u} = \dot{S},
\end{equation}
where $\dot{S}$ denotes the mass exchange source term. For the sake of simplicity, pressure and viscosity are not included without any loss of generality. The continuous momentum equation can be written as
\begin{equation}
\label{eq:spraymom}
\frac{\partial \rho \boldsymbol{u}}{\partial t} + \nabla \cdot (\rho \boldsymbol{u} \boldsymbol{u}) = 0.
\end{equation}
Multiplying by $\boldsymbol{u}$, the KE equation can be written as 
\begin{equation}
\label{eq:sprayKe}
\frac{\partial \rho \frac{u^2}{2}}{\partial t} + \nabla \cdot (\rho \frac{u^2}{2} \boldsymbol{u}) + \frac{u^2}{2} (\frac{\partial \rho}{\partial t} + \nabla \cdot (\rho \boldsymbol{u})) = 0.
\end{equation}
Based on the continuity equation, the last term is non-zero and equal to $\dot{S}$. Following Morinishi \cite{morinishi-skew}, the momentum equation is still written in discrete conservative form as \cite[Eq.128]{morinishi-skew} and becomes:
\begin{equation}
 \frac{\delta \overline{\rho}^t \boldsymbol{u}}{\delta t} + \frac{\delta \overline{\boldsymbol{\phi}_f}^t \overline{\boldsymbol{\hat{u}}}^x}{\delta x} + \frac{\boldsymbol{\hat{u}}}{2}(\frac{\delta \overline{\boldsymbol{\phi}_f}^t}{\delta x} + \frac{\delta \overline{\rho}^t}{\delta t}).
\end{equation} 

The skewed-symmetric form varies from the divergence form, only by the last term on the RHS. Its contribution to the conservation of KE is then similar to the continuous equation and is \textit{a fortiori} proportional to the evaporation rate. %This property is not specific to skew-symmetric scheme and the same relation can be obtained using the method outlined in \cite{pierce_thesis}. The residual term will have the following form:

\section{Pressure Poisson equation}
\label{sec:appPoisson}

The pressure Poisson equation formulation is derived here for both the fractional timestep and PISO schemes.

In the fractional timestep method, the momentum equation in discrete form is given by:
\[ \frac{\overline{\rho}^{n+1} \boldsymbol{u}^* - \overline{\rho}^{n} \boldsymbol{u}^{n}}{\Delta t} + \boldsymbol{\mathcal{C}}^{n+1/2} = \boldsymbol{\mathcal{D}}^{n+1/2},\],
while the continuity equation is written as:
\[\frac{\overline{\rho}^{n+1} \boldsymbol{u}^{n+1} - \overline{\rho}^{n+1} \boldsymbol{u}^*}{\Delta t} = -\nabla p^{n+1/2}.\]
Applying the divergence operator to the above equation leads to 
\[ (\frac{\partial \rho}{\partial t})^{n+1} + \nabla \cdot (\boldsymbol{\phi_f}^*)= \Delta t \nabla \cdot(\nabla p^{n+1/2}).\]

In the PISO formulation, the derivation is not as straightforward since it involves the splitting operators (see~\ref{sec:pisoFts}). In what follows, the brackets $[.]$ denote fields as opposed to cell values. Each pressure correction results in the same following relation:

\[ A ([\boldsymbol{u}^{n+1}] - [\boldsymbol{u}^{*}])= -[\nabla p^{n+1/2}].\]

Multiplying by $\rho^{n+1}$ leads to
\[A ([\overline{\rho}^{n+1}] [\boldsymbol{u}^{n+1}] - [\overline{\rho}^{n+1}] [\boldsymbol{u}^{*}])= -[\overline{\rho}^{n+1}] [\nabla p^{n+1/2}].\]
Applying the divergence operator and rearranging results in
\[ (\frac{\partial [\rho]}{\partial t})^{n+1} + \nabla \cdot ([\boldsymbol{\phi_f}^*])= \nabla \cdot(A^{-1} [\overline{\rho}^{n+1}] [\nabla p^{n+1/2}]).\]

Recall that $\rho$ is defined at timesteps $n+\frac{3}{2}$ and $n+\frac{1}{2}$. Hence $\overline{\rho}^t$ is defined at timestep $n+1$. Noting that $\overline{\rho}^t U^{*} = \phi^*$, Eq.~\ref{eq:pressequform} is obtained.

%% References


\bibliographystyle{elsarticle-num}

\bibliography{bibtex_database}

%TO IMPROVE

%improve open foam in intro 
%explain that we ar eusing a modified version of icoFOam --> OK
%Write up the equations in continuous form at the beginning --> OK
%time staggered plot : explain the time staggering with a cartoon ---> OK
%Some of Alex's remark are not taken into account yet
%{\bf Alex: We can explicitly show this difference between linear and midpoint method using %the bluff body case.}
%{\bf Malik: Ok, I'll keep this in mind}
%{\bf Alex: Perhaps instead of plotting Kin energy in the y-axis, we could plot percentage of Kin Energy dissipated}
%{\bf Malik: Ok for plots that don't contain comparison with Ham}

\end{document}